\documentclass[journal,comsoc]{IEEEtran}

\usepackage[backend=bibtex,
     bibstyle=ieee,
     citestyle=numeric-comp,
     sortcites=true,
     maxnames=2,
     maxbibnames=12,
     minnames=1,
     mincrossrefs=99,
     uniquelist=false,
     defernumbers=true
]{biblatex}

\makeatletter
\newcounter{IEEE@bibentries}
\renewcommand\IEEEtriggeratref[1]{%
  \renewbibmacro{finentry}{%
    \stepcounter{IEEE@bibentries}%
    \ifthenelse{\equal{\value{IEEE@bibentries}}{#1}}
    {\finentry\@IEEEtriggercmd}
    {\finentry}%
  }%
}
\makeatother

\usepackage{amsmath,amssymb,amsfonts}
\usepackage{newtxmath}
\usepackage{booktabs}
\usepackage{tabularx}
\usepackage{csquotes}

\usepackage{tikz}
\usepackage{tikzpeople}
\usetikzlibrary{trees}
\usetikzlibrary{positioning}
\usetikzlibrary{arrows.meta,arrows}
\usetikzlibrary{decorations.pathreplacing}
\usepackage[edges]{forest}
\forestset{qtree/.style={for tree={parent anchor=south,
           child anchor=north,align=center,inner sep=0pt}}}

\newcommand{\extraprotect}[1]{\unexpanded{\unexpanded{#1}}}

\usepackage{cleveref}
\crefname{figure}{\figurename}{Tables}
\crefname{table}{\tablename}{Figures}

\usepackage{xspace}
\newcommand*{\eg}{e.g.\@\xspace}
\newcommand*{\ie}{i.e.\@\xspace}

\definecolor{sron0}{HTML}{332288}
\definecolor{sron1}{HTML}{88CCEE}
\definecolor{sron2}{HTML}{117733}
\definecolor{sron3}{HTML}{DDCC77}
\definecolor{sron4}{HTML}{CC6677}
\definecolor{sron5}{HTML}{AA4499}

\usepackage{subfig}

\begin{document}

\title{IPFS and Friends: A Qualitative Comparison of\\Next Generation Peer-to-Peer Data Networks}

\author{Erik~Daniel and~Florian~Tschorsch%
\thanks{Erik Daniel and Florian Tschorsch are with the Department of
Distributed Security Infrastructures at Technische Universit\"at Berlin, 10587~Berlin, Germany;
e-mail: \mbox{erik.daniel@tu-berlin.de} and \mbox{florian.tschorsch@tu-berlin.de}}%
}

\maketitle
%\copyrightnotice

\begin{abstract}
Decentralized, distributed storage offers a way to reduce the impact of data silos
as often fostered by centralized cloud storage.
While the intentions of this trend are not new,
the topic gained traction due to technological advancements,
most notably blockchain networks.
As a consequence, we observe that a new generation of peer-to-peer data networks emerges.
In this survey paper, we therefore provide a technical overview
of the next generation data networks.
We use select data networks to introduce general concepts and to emphasize new developments.
Specifically, we provide a deeper outline of the Interplanetary File System and
a general overview of Swarm, the Hypercore Protocol, SAFE, Storj, and Arweave.
We identify common building blocks and provide a qualitative comparison.
From the overview, we derive future challenges and research goals concerning data networks.
\end{abstract}


\section{Introduction}
Nowadays, users store and share data
by using cloud storage providers in one way or another.
Cloud storages are organized centrally,
where the storage infrastructure is typically
owned and managed by a single logical entity.
Such cloud storage providers are responsible for
storing, locating, providing, and securing data.
While cloud storage can have many economical and technical advantages,
it also raises a series of concerns.
The centralized control and governance leads to data silos
that may affect accessibility, availability, and confidentiality.
Data access might, for example, be subject to censorship.
At the same time, data silos pose a valuable target for breaches
and acquiring data for sale, which risk security and privacy.
In general, users lose their self-determined control
and delegate it to a cloud provider.

\begin{figure}
    \begin{tikzpicture}
      \footnotesize
      \node[outer sep=0,inner sep=0] (t) {};
      \node[right=68mm of t,outer sep=0,inner sep=0] (t_end) {};
      \node[outer sep=0,inner sep=0,right=4mm of t_end] (time1) {};
      \node[right=4mm of time1] (time2) {};
  
      \draw[thick] (t) -- ([xshift=8mm]t_end);
      \draw[dotted,thick] (t_end) -- (time1);
      \draw[->,thick] (time1) -- (time2);
  
      \draw[-,thick] ($(t.east)!0.0!(t_end)$) -- ([yshift=2mm]$(t.east)!0.0!(t_end)$)
        node[above] (f1) {Napster};
      \draw[-,thick] ($(t.east)!0.05!(t_end)$) -- ([yshift=-5mm]$(t.east)!0.05!(t_end)$)
        node[below]{Gnutella};
      \draw[-,thick] ($(t.east)!0.075!(t_end)$) -- ([yshift=7mm]$(t.east)!0.075!(t_end)$)
        node[above]{Freenet};
      \draw[-,thick] ($(t.east)!0.17!(t_end)$) -- ([yshift=2mm]$(t.east)!0.17!(t_end)$)
        node[above]{Chord};
      \draw[-,thick] ($(t.east)!0.18!(t_end)$) -- ([yshift=-5mm]$(t.east)!0.18!(t_end)$)
        node[right,pos=0.9]{PAST};
      \draw[-,thick] ($(t.east)!0.95!(t_end)$) -- ([yshift=10mm]$(t.east)!0.95!(t_end)$)
        node[above] (s1) {IPFS};
      \draw[-,thick] ($(t.east)!0.98!(t_end)$) -- ([yshift=-15mm]$(t.east)!0.98!(t_end)$)
        node[below] (s2) {Storj v1};
  
      \node[diamond,fill=black,scale=0.6] (bittorrent) at ($(t.east)!0.145!(t_end)$) {};
      \draw[-,thick] (bittorrent.center) -- ([yshift=-10mm]$(bittorrent)$)
        node[below,align=center] (f2) {\textbf{BitTorrent} \\ \emph{Chunk Exchange}};
  
      \node[diamond,fill=black,scale=0.6] (kademlia) at ($(t.east)!0.24!(t_end)$) {};
      \draw[-,thick] (kademlia.center) -- ([yshift=10mm]$(kademlia)$)
        node[above, align=center] (f3) {\textbf{Kademlia} \\ \emph{Peer discovery}};
  
      \node[diamond,fill=black,scale=0.6] (bitcoin) at ($(t.east)!0.59!(t_end)$) {};
      \draw[-,thick] (bitcoin.center) -- ([yshift=10mm]$(bitcoin)$)
        node[above,align=center](t1){\textbf{Bitcoin} \\ \emph{Incentives}};
  
      \node[diamond,fill=black,scale=0.6] (ccn) at ($(t.east)!0.67!(t_end)$) {};
      \draw[-,thick] (ccn.center) -- ([yshift=-10mm]$(ccn)$)
        node[below,align=center] (t2) {\textbf{Content-Centric} \\ \textbf{Networking} \\ \emph{Content Addressing}};
  
      \node [fill=sron0,rectangle,opacity=0.25,anchor=west,
        yshift=-2pt,xshift=-15pt,minimum height=115pt,minimum width=90pt,
        label={[align=center]below:First Generation}] at (t) {};
      \node [fill=sron1,rectangle,opacity=0.25,anchor=west,
        yshift=-2pt,xshift=80pt,minimum height=115pt,minimum width=85pt,
        label={[align=center]below:Transition Phase}] {};
      \node [draw=sron2,ultra thick,fill=sron2,rectangle,opacity=0.25,anchor=west,
        yshift=-2pt,xshift=170pt,minimum height=115pt,minimum width=55pt,
        label={[align=center,font=\bfseries]below:Next Generation}] {};
  
    \end{tikzpicture}
    \caption{Precursor technologies of the next generations of P2P data networks.}
    \vspace{-1em}
    \label{fig:precursor}
  \end{figure}
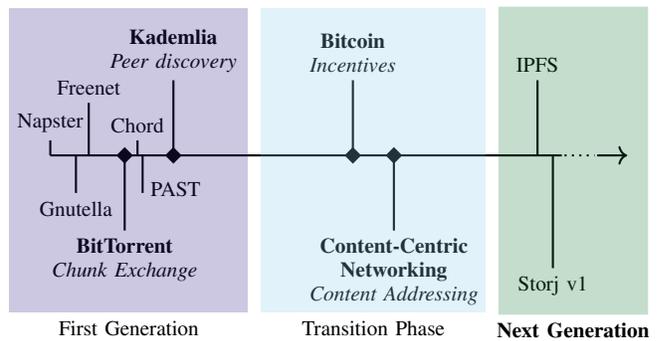

One direction to break free from data silos and to reduce trust assumptions
are \emph{peer-to-peer data networks}.
The term summarizes a family of data storage approaches
that build upon a peer-to-peer~(P2P) network and
include aspects of data storage, replication, distribution, and exchange.
As typical for P2P networks, peers interact directly, build an overlay network,
share resources, and can make autonomous local decisions.
Consequentially, P2P data networks strive
to jointly manage and share storage.
While the main goals and principles of P2P networks did not change over the last two decades,
P2P networks evolved over time, improving the usability and functionality.
In \cref{fig:precursor}, we illustrate the development
from the first generation to the next generation of data networks.

\subsection{First Generation of Data Networks}
There are many different older P2P networks
that can be classified as data networks as well.
The popularity of P2P technologies emerged in 1999
with the audio file sharing network Napster,
closely followed by Gnutella
for sharing all types of files~\cite{saroiu2001measurement}.
Napster and Gnutella marked the beginning
and were followed by many other P2P networks focusing on
specialized application areas or novel network structures.
For example, Freenet~\cite{clarke2000freenet}
realizes anonymous storage and retrieval.
Chord~\cite{stoica2001chord}, CAN~\cite{ratnasamy2001scalable},
and Pastry~\cite{rowstron2001pastry} provide protocols
to maintain a structured overlay network topology.
In particular, BitTorrent~\cite{cohen2003incentives} received a lot of attention
from both users and the research community.
BitTorrent introduced an incentive mechanism to achieve Pareto efficiency,
trying to improve network utilization achieving a higher level of robustness.
We consider networks such as Napster, Gnutella, Freenet, BitTorrent, and many more
as first generation P2P data networks,
which primarily focus on file sharing.

\citeauthor{androutsellis2004survey}~\cite{androutsellis2004survey}
provide a 2004 state of the art overview of P2P content distribution technologies
providing a broad overview of the previous generation.
Other previous works also provide closer looks at the previous generation
with a closer focus on specific P2P data networks
(\eg, FreeNet and Past)~\cite{hasan2005survey, ashraf2019comparative}
or decentralized files systems in general
(\eg, Google FS and Hadoop Distributed FS)~\cite{thanh2008taxonomy}.

The advancements in P2P technologies
and the popularity of this first generation of data networks,
affected the areas of distributed file systems~\cite{hasan2005survey}
and content distribution technologies~\cite{androutsellis2004survey}.
This trend also falls under the umbrella of data networks in general
and P2P data networks in particular.

\subsection{Transistion Phase}
One component which seemed to be missing in P2P file sharing systems
was a way to improve long-term storage and availability of files.
With the introduction of Bitcoin~\cite{nakamoto2009bitcoin} in~2008,
the P2P idea in general and the joint data replication in particular
gained new traction.
Distributed ledger technologies provide availability, integrity, and
byzantine fault tolerance in a distributed system.
In particular, cryptocurrencies showed their potential
as a monetary incentive mechanism in a decentralized environment.
These and additional trends and developments,
\eg, Kademlia~\cite{maymounkov2002kademlia} and
information-centric networking~\cite{ahlgren2012survey},
lead to the invention of what we denote as
the next generation of P2P data networks.

\subsection{Next Generation of Data Networks}
Starting with the introduction of IPFS~\cite{benet2014ipfs} in 2014,
we define the next generation of data networks
as systems and concepts for \emph{decentralized sharing and storing of data},
which appeared in the last decade.
We provide a technical overview of
this next generation of P2P data networks.
In contrast to the existing literature,
we provide a comparative overview of next generation data networks,
\ie, P2P data networks.
We focus on storage and content sharing
independent of the utilization of a blockchain.

In this paper, we show how these new systems are built,
how they utilize the gained knowledge from previous systems,
as well as new developments and advancements over the last decade.
We identify building blocks, similarities, and trends of these systems.
While some of the systems are building blocks themselves for
other applications, \eg, decentralized applications~(DApps),
we focus on two main system aspects:
\emph{content distribution} and \emph{distributed storage}.
Furthermore, we provide insights in the incentive mechanisms,
deployed for retrieving or storing files, or both.
Since many new data networks were developed,
we cannot provide a full overview of all data networks.
Instead, we focus on a few select systems
with sophisticated or unique mechanisms, different use cases,
and different degree of content and user privacy.
Our overview focuses on concepts and abstracts from implementation details
to extract general insights.
Yet, it should be noted that the systems are prone to change
due to ongoing development.
Our survey paper makes use of a wide range of sources,
including peer-reviewed papers, white papers as well as
documentations, specifications, and source code.

Specifically, we focus on IPFS~\cite{benet2014ipfs},
Swarm~\cite{tron2020swarmbook}, the Hypercore Protocol~\cite{ogden2018dat},
SAFE~\cite{lambert2014safe},
Storj~\cite{storjlabs2018storj}, and
Arweave~\cite{arweave2019williams}.
In particular, the InterPlanetary File System~(IPFS) has gained popularity
as storage layer for blockchains~\cite{ali2017iot,norvill2018ipfs,wang2018blockchain,steichen2018blockchain,khatal2020fileshare,hoang2020privacy,xu2019healthchain}
and was subject of a series of studies~\cite{patsakis2019hydras,shen2019understanding,muralidharan2019interplanetary,ascigil2019towards,nyaletey2019blockipfs,heinisuo2019asterism,prunster2020total,henningsen2020mapping,henningsen2020crawling,de2021accelerating,guidi2021data}.
Furthermore, we put our overview of these systems in context
to preceding systems and research directions,
namely BitTorrent, information-centric networking, and blockchains.
By contrasting precursor systems, we sketch the evolution
of data networks and are able to profoundly discuss advancements
of the next generation.

Based on this overview, we extract the building blocks
and some unique aspects of P2P data networks.
While all systems allow distributed content sharing and storage,
they seem to focus on either of the aspects.
That is, each system aims to serve a slightly different purpose
with different requirements and points of focus.
This leads to different design decisions in network organization,
file look up, degree of decentralization, redundancy, and privacy.
For example, Storj aims for a distributed cloud storage
while the Hypercore protocol focuses on distributing large datasets.
Similarly, IPFS aims to replace client-server structure of the web
and therefore needs a stronger focus on data look up
than BitTorrent where mainly each file is located in its own overlay network.
At the same time, we found many similarities in the approach of building data networks,
for example, using Kademlia to structure the network or finding peers,
split files into pieces, or incentivizing different tasks to increase functionality.

Other research on next generation data networks
particularly focus on the interaction with blockchains.
\citeauthor{huang2020blockchain}~\cite{huang2020blockchain}
mainly cover IPFS and Swarm, and
\citeauthor{benisi2020blockchain}~\cite{benisi2020blockchain}
discuss these technologies with an even stronger focus on the blockchain aspects.
\citeauthor{casino2019immutability}~\cite{casino2019immutability} take a closer look
at the immutability of decentralized storage and
its consequences and possible threats.
Some data networks, however, make a clear decision against the usage of blockchains,
due to scalability or latency problems.
In our survey paper, we therefore take a broader perspective on data networks,
looking at the design decisions of data networks beyond blockchains.

A more general view on recent P2P networks is given
by~\mbox{\citeauthor{naik2020next}}~\cite{naik2020next}.
They describe next level P2P networks,
the evolution of classic networks like BitTorrent and Chord,
and discuss performance aspects under churn.
It should be noted that, their definition of next level network is different
from our next generation definition as they define IPFS as a \enquote{classic P2P network}.
We instead argue that P2P data networks evolved over time,
incorporating ideas from newly established fields,
for example, explicit incentive mechanisms.

The remainder is structured as follows:
The survey transitions from a system view,
over a component view to a research perspective on data networks.
As part of the system view, we first provide
background information of technological precursors of data networks~(\Cref{sec:basics}).
Subsequently, we introduce \enquote{IPFS and Friends}
and provide a detailed technical overview of the next generation of data networks~(\Cref{sec:ipfs} and \Cref{sec:datanetworks}).
Lastly, we mention related systems and concepts~(\Cref{sec:honorable}).
As part of the component view,
we derive the building blocks of data networks and share insights gained from the technical overview~(\Cref{sec:buildingblocks}).
Finally, we transition to a research perspective
and identify research areas and open challenges~(\Cref{sec:goals}).
\Cref{sec:conclusion} concludes this survey.

\section{Technological Precursors}\label{sec:basics}
It has been more than two decades since the first appearance of P2P data networks.
During this time the technology evolved and influenced the development of new networks.
We observe that there are basically three \enquote{eras} of P2P data networks:
It started with P2P file sharing and networks such as BitTorrent and Kademlia in 1999--2002,
which we consider the first generation.
This era is followed by a \enquote{transition phase}, where new ideas
such as information-centric networking and cryptocurrencies emerged.
Approximately since 2014 with the invention of IPFS,
we see a new generation of P2P data network gaining traction.
For a better understanding of and appreciation for the influences,
we provide an introduction to important \enquote{precursor} technologies
that paved the ground,
namely, BitTorrent, Kademlia, information-centric networking, self-certifying names, and blockchains.

\subsection{BitTorrent}
The BitTorrent protocol~\cite{cohen2003incentives} is a P2P file sharing protocol.
It has an incentive structure controlling the download behavior,
attempting to achieve fair resource consumption.
The goal of BitTorrent is to provide a more efficient way to distribute files
compared to using a single server.
This is achieved by utilizing the fact
that files are replicated with each download,
making the file distribution self-scalable.

Files are exchanged in torrents.
In general, each torrent is a P2P overlay network responsible for one file.
To exchange a file with the BitTorrent protocol a \texttt{.torrent} file,
containing meta-data of the file and a contact point, a tracker, is created.
It is also possible to define multiple files in a \texttt{.torrent} file.
The torrent file needs to be made available, \eg, on a web server, before the file can be shared.
The tracker serves as a bootstrapping node for the torrent.
Peers that have complete files are called seeders.
Peers that still miss chunks are called leechers.
Leechers request chunks and serve simultaneously as
download points for already downloaded chunks.

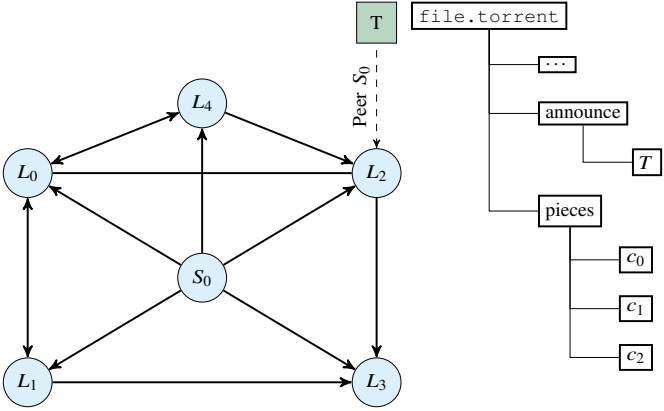
\begin{figure}
  \footnotesize
  \centering
  \resizebox{\columnwidth}{!}{
  \begin{tikzpicture}[%
    grow via three points={one child at (0.5,-0.7) and
    two children at (0.5,-0.7) and (0.5,-1.4)},
    edge from parent path={(\tikzparentnode.south) |- (\tikzchildnode.west)},auto,>=stealth']

    \tikzstyle{peer}=[circle, draw, thin,fill=sron1!30, minimum width=7mm]

    \node[peer] (s0) at (0,0) {$S_{0}$};
    \node[peer] (l1) at (-2.5,1.5) {$L_{0}$};
    \node[peer] (l2) at (-2.5,-1.5) {$L_{1}$};
    \node[peer] (l3) at (2.5,1.5) {$L_{2}$};
    \node[peer] (l4) at (2.5,-1.5) {$L_{3}$};
    \node[peer] (l5) at (0,2.5) {$L_{4}$};

    \node[rectangle, draw, thin,fill=sron2!30,anchor=south,yshift=15mm,inner sep=2mm]
      (tracker) at (l3.north) {T};

    \tikzstyle{link}=[thick]

    \path[->,link] (s0) edge (l1);
    \path[->,link] (s0) edge (l2);
    \path[->,link] (s0) edge (l3);
    \path[->,link] (s0) edge (l4);
    \path[->,link] (s0) edge (l5);
    \path[<->,link] (l1) edge (l2);
    \path[<->,link] (l1) edge (l5);
    \path[link] (l1) edge (l3);
    \path[->,link] (l2) edge (l4);
    \path[->,link] (l3) edge (l4);
    \path[<-,link] (l3) edge (l5);

    \path[dashed,<-] (l3) edge node[sloped, above] {Peer $S_{0}$} (tracker);

    \tikzstyle{every node}=[draw=black,thick,anchor=west,xshift=2mm]
        \node[anchor=north west] (d) at (tracker.north east) {\texttt{file.torrent}}

            child { node {\dots}}
            child { node {announce}
                child { node {$T$}}}
            child[missing] {}
            child { node {pieces}
                child { node {$c_{0}$}}
                child { node {$c_{1}$}}
                child { node {$c_{2}$}}};

  \end{tikzpicture}}
    \caption{Conceptional overview of BitTorrent.}
    %\vspace{-1em}
    \label{fig:bittorrent}
\end{figure}

A conceptual overview of how BitTorrent deals with files can be seen in \cref{fig:bittorrent}.
The roles and their interaction are as follows: a peer gets the \texttt{.torrent} file,
contacts the tracker~$T$ listed in the \texttt{.torrent} file, gets a list of peers,
connects to the peers and becomes a leecher.
In the figure, the peer~$S_{0}$ serves as a seed of the file and
the peers~$L_{i}$ represent the leechers requesting the different chunks.
As illustrated for the \texttt{.torrent} file, the file is split into chunks~$c_{j}$.
After a leecher successfully acquired all chunks, it becomes a new seed.
Seed~$S_{0}$ and leechers build the torrent network for the file.
Other files are distributed in different torrent networks with possibly different peers.

Instead of the presented centralized trackers, there are also trackerless torrents.
In a trackerless torrent, seeds are found with a distributed hash table~(DHT).
The client derives the key from the torrent file and the DHT returns
a list of available peers for the torrent.
The BitTorrent client can use a predetermined node or
a node provided by the torrent file for bootstrapping the DHT.

The feature that made BitTorrent unique (and probably successful)
is the explicit incentivization of peers to exchange data,
which are implemented in the file sharing strategies rarest piece first and tit-for-tat.
Rarest piece first describes the chunk selection of BitTorrent.
It ensures a minimization of chunk overlap,
making file exchange more robust against node churn.
The chunks that are most uncommon in the network are preferably selected for download.
Tit-for-tat describes the bandwidth resource allocation mechanism.
In BitTorrent peers decide to whom they upload data based on the downloaded data from a peer.
This should prevent leechers from only downloading without providing any resources to others.

BitTorrent is well researched~\cite{pouwelse2005bittorrent,bharambe2006analyzing,xia2010survey}
and has proven its test of time.
Despite its age, it is still actively used by millions of people~\cite{ramanathan2020quantifying}
for sharing files and also serves as a role model for newer peer-to-peer file distribution systems.
In addition, the BitTorrent Foundation and Tron Foundation
developed BitTorrent Token (BTT)~\cite{bittorrentfoundation2019bittorrentbtt},
which serves as a blockchain-based incentive layer
to increase the availability and persistence of files.
The new incentive structure extends tit-for-tat by bid data.
The bid data determines BTT/byte rate, which the peer pays for continued service.
In exchange for the payment, the peer is unchoked and eligible to receive data.
The exchange of tokens is handled by a payment channel.

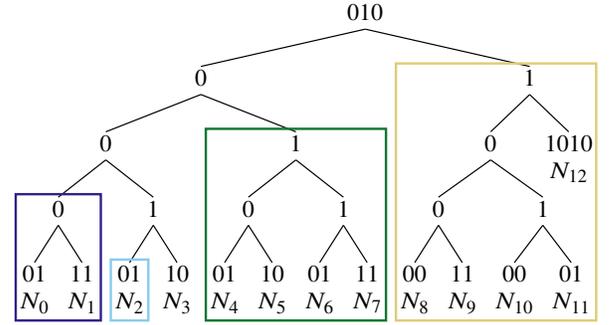
\begin{figure}
\footnotesize
\centering
\resizebox{0.9\columnwidth}{!}{
\begin{forest}, qtree,
    [010
        [0,
         [0
          [0,tikz={\node [draw=sron0,thick, rectangle,inner sep=1, fit to=tree]{};}
           [01 \\ $N_{0}$]
           [11 \\ $N_{1}$]
          ]
          [1
           [01 \\ $N_{2}$,tikz={\node [draw=sron1,thick, inner sep=1, rectangle,fit to=tree]{};}]
           [10 \\ $N_{3}$]
          ]
         ]
         [1,tikz={\node [draw=sron2,inner sep=1, thick,rectangle, fit to=tree]{};}
          [0
           [01 \\ $N_{4}$]
           [10 \\ $N_{5}$]
          ]
          [1
           [01 \\ $N_{6}$]
           [11 \\ $N_{7}$]
          ]
         ]
        ]
        [1,tikz={\node [draw=sron3,rectangle, thick, inner sep=1, fit to=tree]{};}
         [0
          [0
           [00 \\ $N_{8}$]
           [11 \\ $N_{9}$]
          ]
          [1
           [00 \\ $N_{10}$]
           [01 \\ $N_{11}$]
          ]
         ]
         [1010 \\ $N_{12}$]
        ]
     ]
\end{forest}
}
    \caption{Kademlia tree with 13 nodes and random ids. Highlighting the buckets for $N_{3}$}
    \label{fig:kadtree}
\end{figure}

\subsection{Kademlia}
From today's perspective,
Kademlia~\cite{maymounkov2002kademlia} is probably the most widely used DHT.
As we will see later, the majority of P2P data networks builds upon Kademlia
one way or another.
Kademlia also influenced protocols for P2P file exchange like BitTorrent,
which enables trackerless torrents by using a Kademlia-based DHT~\cite{bep005}.

In general, Kademlia can be classified as a structured overlay network,
which specifies \emph{how} to structure and maintain the P2P network.
To this end, peers are assigned an identity, which determines its position
and consequently its neighbors.
For the neighbor selection, an XOR metric is used.
The advantage of the XOR metric is that it is symmetric and unidirectional.
Depending on their XOR distance nodes are sorted into $k$-buckets.
The buckets are arranged as a binary tree, where the shortest prefix determines the bucket.
If a new node belongs to a bucket which contains $k$~nodes including itself,
the bucket gets split into smaller buckets,
otherwise the new node is dropped.
An exemplary Kademlia tree with 8\,bit identifiers is shown in \cref{fig:kadtree}.

\subsection{Information-Centric Networking}
Another precursor worth mentioning is information-centric networking~(ICN).
Even though ICN is not a P2P data network, some of its ideas
and concepts are at least similar to some data networks.
Contrary to P2P data networks, ICN proposes to change the network layer.
The routing and flow of packets should change from point-to-point location search
to requesting content directly from the network.
As an example, let us assume we wanted to retrieve some data, \eg, a website,
and we know that this website is available at \texttt{example.com}.
First, we request the location of the host of the site via DNS, \ie, the IP address.
Afterwards, we establish a connection to retrieve the website.
In ICN, we would request the data directly
and would not address the host where the data is located.
Any node storing the website could provide the data immediately.

\citeauthor{jacobson2009networking}~\cite{jacobson2009networking} proposed
content-centric networking, where these content requests are interest packets.
Owner(s) of the content can then directly answer the interest
packet with data packets containing the content.
This requires other mechanisms for flow control,
routing, and security on an infrastructure level.
Interest packets are broadcasted and peers sharing interest in data can share resources.
There are multiple projects dealing with ICN,
\eg, Named Data Networking~\cite{zhang2014named}~(NDN).
With Ntorrent~\cite{mastorakis2017ntorrent}, \citeauthor{mastorakis2017ntorrent}
propose an extension of NDN to implement a BitTorrent-like mechanism in NDN.
Further general information on ICN can be found in~\cite{ahlgren2012survey}.
Due to the content-centric nature of data networks,
they could be broadly interpreted as overlay implementations of ICN.

\subsection{Self-Certifying Names}
The change from a host-focused to a content-focused
communication introduces new security problems.
Furthermore, as caching becomes a major feature of the network
specific threats need to be considered, \eg,
cache poisoning or Denial-of-Service attacks against the cache.
More broadly, security issues of ICN in general include content authentication,
authorization and access control, and privacy~\cite{rfc7945}.

Currently, the major focus of security research lies on authentication.
Due to extensive use of caching, the data provider is not necessarily
the original source (data owner) of an object anymore.
This requires mechanisms that enable the recipient to
assess validity (integrity), provenance (content origin), and relevance of an object.

One way to ensure validity and relevance are self-certifying names.
Self-certifying names can be enabled with the usage of hash pointers
(or more generically content hashes) to reference content.
The content of a file is used as input of cryptographic hash function,
\eg, SHA-3~\cite{dworkin2015sha}.
The resulting digest can then be used to identify the content and
the client can verify the integrity of the file locally.
The cryptographic properties of the hash function,
most importantly pre-image and collision resistance,
ensure that nobody can replace or modify the input data without changing its digest.
In this case, the name provides integrity and relevance, however,
it remains questionable who is responsible for verifying the objects,
\eg, client and/or intermediates.
Furthermore, a self-certifying name by itself cannot provide provenance or
proof an object's origin.
Cryptographic signatures can guarantee the authenticity of an objects origin,
but require a public key infrastructure or a web of trust to verify signatures.
While this allows verifying the authenticity of an object's origin,
it is still possible to send malformed objects,
which therefore requires mechanisms to secure integrity.
The longevity of content through caching requires careful key management
to protect against compromised cryptographic credentials.

Access control has similar problems: once data is released
it is difficult to limit access or revoke the publication.
Encryption could limit access but might require out-of-band key distribution.
Further insights on security, privacy, access control and other challenges in ICN
can be found in~\cite{tourani2017security,rfc7927}.

\subsection{Blockchain}
The introduction of Bitcoin~\cite{nakamoto2009bitcoin} in 2008 enabled
new possibilities for distributed applications.
Bitcoin is an ingenious, intricate combination of ideas from the areas of
linked timestamping, digital cash, P2P networks, byzantine fault tolerance, and
cryptography~\cite{narayanan17bitcoinpedigree,tschorsch2016bitcoin}.
One of the key innovations that Bitcoin brought forward was
an open consensus algorithm that actively incentivizes peers to be compliant.
Therefore, it uses the notion of coins, generated in the process, \ie, mining.

While the term blockchain typically refers to an entire system and its protocols,
it also refers to a particular data structure, similar to a hash chain or hash tree.
That is, a blockchain orders blocks
that are linked to their predecessor with a cryptographic hash.
This linked data structure ensures the integrity of the blockchain data,
\eg, transactions.
The blockchain's consistency is secured by a consensus algorithm,
\eg, in Bitcoin the Nakamoto consensus.
For more details on Bitcoin and blockchains, we refer to~\cite{tschorsch2016bitcoin}.

Since blockchains suffer from problems such as scalability,
different designs have been developed to mitigate these problems.
The different designs opened a new category,
which is referred to as Distributed Ledger Technologies~(DLT).
DLTs provide distributed, byzantine fault tolerant, immutable, and ordered logs.
Unfortunately, the feasibility of a purely DLT-based data network is limited,
due to a series of scalability problems and limited on-chain storage capacity%
~\cite{gervais16blockchain-sec-perf,bonneau15challenges}.
Moreover, storing large amounts of data in a blockchain
that was designed as medium of exchange and store of value,
\ie, cryptocurrencies such as Bitcoin, leads to high transactions fees.
However, research and development of DLTs shows the feasibility of blockchain-based data networks,
\eg, Arweave (cf. \Cref{sec:arweave}).

In general, however, cryptocurrencies allowing decentralized payments
can be used in P2P data networks as an incentive structure.
As we will elaborate in the following,
such an incentive structure can increase the robustness and availability of data networks
and therefore address weaknesses of previous generations.

\section{Interplanetary File System (IPFS)}\label{sec:ipfs}
The Interplanetary File System (IPFS)~\cite{benet2014ipfs} is
a bundle of subprotocols and a project driven by Protocol Labs.
IPFS aims to improve the web's efficiency
and to make the web more decentralized and resilient.
IPFS uses content-based addressing, where content is not addressed via a location
but via its content.
The way IPFS stores and addresses data with its deduplication properties,
allows efficient storage of data.

Through IPFS, it is possible to store and share files in a decentralized way,
increasing censorship-resistance for its content.
IPFS can be used to deploy websites building a distributed web.
It is used as a storage service complementing blockchains,
enabling different applications on top of IPFS.

Since IPFS uses content-based addressing, it focuses mainly on immutable data.
IPFS however supports updatable addresses for content
by integrating the InterPlanetary Name System (IPNS).
IPNS allows to link a name (hash of a public key)
with the content identifier of a file.
IPNS entries are signed by the private key and
can arbitrarily be (re)published to the network (default $4\,h$).
Each peer maintains its own LRU cache of resolved entries (default 128 entries).
An IPNS entry has a specific lifetime,
after which it is removed from cache (default $24\,h$).
By changing the mapping of fixed names to content identifiers,
file updates can be realized.
Please note, content identifiers are unique and file specific.

In addition, IPFS employs its own incentive layer,
\ie, Filecoin~\cite{protocollabs2017filecoin},
to ensure the availability of files in the network.
Yet, IPFS works independently from Filecoin and vice-versa.
This is a prime example of how a cryptocurrency can be integrated
to incentivize peers.

\begin{figure*}
\centering
    \subfloat[File $F$]{
    \begin{forest} qtree, forked edges,
    [\extraprotect{$F = ((a,b),(c,d))$}, no edge
        [\extraprotect{$(a,b)$}
         [$a$
          [$a_{0}$]
          [$a_{1}$]
          [$a_{2}$]
         ]
         [$b$
          [$b_{0}$]
          [$b_{1}$]
          [$b_{2}$]
         ]
        ]
        [\extraprotect{$(c,d)$}
         [$c$
          [$c_{0}$]
          [$c_{1}$]
          [$c_{2}$]
         ]
         [$d$
          [$d_{0}$]
          [$d_{1}$]
         ]
        ]
    ]
    \end{forest}}
    \hspace{1cm}
    \subfloat[File $F'$]{
    \begin{forest} qtree, forked edges,
    [\extraprotect{$F' = ((a,b),(c,d'),(e,f))$}, sron4, no edge
        [\extraprotect{$(a,b)$}
         [$a$
          [$a_{0}$]
          [$a_{1}$]
          [$a_{2}$]
         ]
         [$b$
          [$b_{0}$]
          [$b_{1}$]
          [$b_{2}$]
         ]
        ]
        [\extraprotect{$(c,d')$}, font=\bfseries, sron4
         [$c$
          [$c_{0}$]
          [$c_{1}$]
          [$c_{2}$]
         ]
         [$d'$, font=\bfseries, sron4
          [$d_{0}$]
          [$d_{1}$]
          [$d_{2}$, sron4]
         ]
        ]
        [\extraprotect{$(e,f)$}, for tree={, font=\bfseries, sron4}
         [$e$
          [$e_{0}$]
          [$e_{1}$]
          [$e_{2}$]
         ]
         [$f$
          [$f_{0}$]
          [$f_{1}$]
          [$f_{2}$]
         ]
        ]
    ]
    \end{forest}}
    \caption{Simplified IPFS file structure visualizing Merkle DAGs of CIDs and the concept of deduplication.}
    \label{fig:fipfsswarm}
\end{figure*}
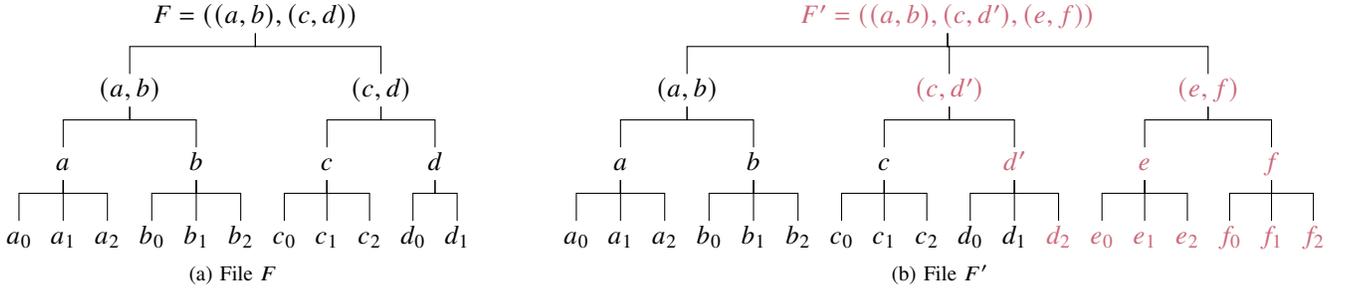

\subsection{General Functionality}
IPFS uses the modular P2P networking stack \emph{libp2p}~\cite{gitlibp2p}.
In fact, \emph{libp2p} came into existence from developing IPFS.
In IPFS nodes are identified by a node id.
The node id is the hash of their public key.
For joining the network, the IPFS development team deployed some bootstrap nodes.
By contacting these nodes, a peer can learn new peers.
The peers with which a node is connected, denoted as its swarm.
Peers can be found via a Kademlia-based DHT and local node discovery.
The communication between connections can be encrypted.
While IPFS uses Kademlia, its connections are not completely determined by Kademlia.
In IPFS, a node establishes a connection to newly discovered nodes,
trying to put them into buckets~\cite{henningsen2020mapping}.
Idle connections are trimmed once a \texttt{HighWater} threshold
is achieved (default 900) until the \texttt{LowWater} threshold is reached (default 600).
A connection is considered idle, if it is not explicitly protected
and existed longer than a grace period (default $20\,s$).
Protected connections are, for example, currently actively used connections,
explicitly added peers, or peers required for DHT functionality.

An object in IPFS (file, list, tree, commit) is split into chunks or blocks.
Each block is identifiable by a content identifier~(CID),
which can be created based on a recipe from the content.
From these blocks, a Merkle directed acyclic graph~(DAG) is created.
Merkle DAGs are similar to Merkle trees.
Each node of the Merkle DAG has an identifier
determined by the node's content.
In contrast, however, Merkle DAGs are not required to be balanced,
a node may carry a payload, and a node can have multiple parents.
The root of the Merkle DAG can be used to retrieve the file.
IPFS employs block deduplication: each stored block has a different CID.
This facilitates file versioning,
where newer file versions share a lot of blocks with older versions.
In this case, only the differences between the versions need to be stored instead
of two complete Merkle DAGs.
The blocks have an added wrapper specifying the UNIXFS type of the block.

In order to illustrate the Merkle DAGs and the mechanism of deduplication,
let us assume that our survey paper and an earlier draft of the paper
are stored on IPFS.
\cref{fig:fipfsswarm} shows a simplified representation of the Merkle DAGs of the two files.
Each node represents a chunk and the label represents the node CID, the content hash.
The DAG is created from bottom to top,
since the intermediate node's CID depends on its descendants.
The actual data is located in the leaves.
In the final version, we assume additional information was appended to the content,
which results in a different root node and additional nodes.
Therefore, in our example, $F$ is the root CID of the draft and
$F'$ the root of the finished survey.

The blocks themselves are stored on devices or providers.
The DHT serves as a look-up for data providers.
As in Kademlia, nodes with node ids closest to the CID
store the information about the content providers.
A provider can announce that it is storing specific blocks.
Possession of blocks is reannounced in a configurable time frame (default $12\,h$).

\subsubsection{Bitswap}
The actual exchange of blocks is handled by the \emph{Bitswap} Protocol.
Each node has a \enquote{want}, \enquote{have}, and \enquote{do not want} list.
The different lists contain CIDs which the node wants/has or does not want.
CIDs on a do not want list are not even cached and simply dropped on receive.
A node sends the CIDs on its want list to its swarm.
Neighbors in possession of this block send the block
and a recipe for creating the CID.
The node can then verify the content by building the CID from the recipe.
If no neighbor possesses a wanted CID, IPFS performs a DHT lookup.
After a successful DHT lookup, a node possessing the CID
is added to the swarm and send the want list.

For a peer to download, a file it needs to know the root CID.
After acquiring the CID of an object's Merkle DAG root,
it can put this root CID on the want list
and the previously described Bitswap/DHT takes over.
The root block gives information about its nodes,
resulting in new CIDs which have to be requested.
Subsequent CID requests are not send to all neighbors.
The neighbors answering the root CID are prioritized
and are grouped in a session.
Since version 0.5, Bitswap sends a \texttt{WANT-HAVE} message for subsequent requests
to multiple peers in the session and to one peer an optimistic \texttt{WANT-BLOCK} message.
The \texttt{WANT-HAVE} message asks if the peer possesses the block
and \texttt{WANT-BLOCK} messages request the block directly.
If a block is received, other pending request
can be canceled with a \texttt{CANCEL} message~\cite{de2021accelerating}.
Previously, neighbors were asked for the block simultaneously,
resulting in possibly receiving a block multiple times.
Once all leaves of the tree are acquired, the file is locally available.
Files are not uploaded to the network only possession is announced.

\begin{figure*}
\scriptsize
\resizebox{\textwidth}{!}{
    \begin{tikzpicture}[node distance=15mm,auto,>=stealth']
    \node[bob, minimum height=10mm] (auth) {Author};

    \node (net) [right=25mm of auth] {
        \resizebox{65mm}{!}{\begin{tikzpicture}
        \tikzstyle{peer}=[circle, draw, thin,fill=cyan!20, minimum width=8mm]

        \node [peer] (a) at (-6.75, 3) {$N_{0}$};
        \node [peer] (c) at (-4.25, 4.75) {$N_{1}$};
        \node [peer] (e) at (-5, 2.0) {$N_{2}$};
        \node [peer, very thick] (f) at (-7.25, 1.25) {$N_{3}$};
        \node [peer] (i) at (-3, 0.0) {$N_{4}$};
        \node [peer] (j) at (-3.5,3.75) {$N_{5}$};
        \node [peer] (m) at (-1.25, 3.75) {$N_{6}$};
        \node [peer] (o) at (-0.75, 0.15) {$N_{7}$};
        \node [peer] (p) at (1.0, 3.25) {$N_{8}$};
        \node [peer] (s) at (4, 4) {$N_{9}$};
        \node [peer] (t) at (3.5, 2.75) {$N_{10}$};
        \node [peer,very thick] (u) at (4.75, 1.5) {$N_{11}$};
        \node [peer] (z) at (0, 2.0) {$N_{12}$};

        \path[thick] (a) edge (c);
        \path[thick] (a) edge (f);
        \path[thick] (c) edge (e);
        \path[thick] (e) edge (f);

        \path[thick] (i) edge (j);
        \path[thick] (i) edge (o);
        \path[thick] (j) edge (m);
        \path[thick] (m) edge (o);

        \path[thick] (p) edge (s);
        \path[thick] (p) edge (u);
        \path[thick] (p) edge (z);
        \path[thick] (s) edge (t);
        \path[thick] (s) edge (z);
        \path[thick] (t) edge (u);
        \path[thick] (t) edge (z);
        \path[thick] (u) edge (z);

        \path[thick] (a) edge (j);
        \path[thick] (a) edge (p);
        \path[thick] (c) edge (m);
        \path[thick] (c) edge (s);
        \path[thick] (e) edge (o);
        \path[thick] (e) edge (z);
        \path[thick] (f) edge (i);
        \path[thick] (f) edge (u);

        \path[thick] (i) edge (t);
        \path[thick] (j) edge (z);
        \path[thick] (m) edge (p);
        \path[thick] (o) edge (u);

        \path[thick] (z) edge (a);
        \path[thick] (z) edge (i);
        \path[thick] (z) edge (o);
        \path[thick] (z) edge (m);
        \path[thick] (z) edge (f);

    \end{tikzpicture}}};

    \draw [decorate,decoration={brace,amplitude=10pt,mirror,raise=4pt}]
        (net.south west) -- (net.south east);

    \node[alice, minimum height=10mm, right=25mm of net] (cauth) {Reviewer};
    \node[left=1mm of auth] (f1) {\includegraphics[height=10mm]{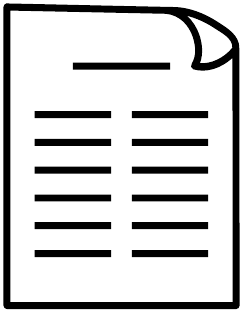}};

    \node (bs) [below=6mm of net,label={[yshift=0mm,align=left,text width=30mm]right:Bitswap transfer. Assuming only one neighbor has the data.}]{\resizebox{60mm}{!}{
            \begin{tikzpicture}[node distance=35mm,auto,>=stealth']
                \tikzstyle{label}=[above,font=\tt,scale=1]
                \scriptsize

                \node[charlie, minimum height=5mm] (b) {$N_{i}$};
                \node[bob, minimum height=5mm, right= of b] (a) {$N_{3}$};
                \node[alice, minimum height=5mm,right= of a] (c) {$N_{11}$};

                \node[below=40mm of c.south] (c_ground) {};
                \node[below=40mm of a.south] (a_ground) {};
                \node[below=40mm of b.south] (b_ground) {};
                \draw[thick] ([yshift=-3mm]a.south) -- (a_ground);
                \draw[thick] ([yshift=-3mm]b.south) -- (b_ground);
                \draw[thick] ([yshift=-3mm]c.south) -- (c_ground);
                \draw[->,thick] ($(c.south)!0.15!(c_ground)$) -- node[label,midway]{WANT-HAVE F} ($(a.south)!0.15!(a_ground)$);
                \draw[->,thick] ($(c.south)!0.17!(c_ground)$) -- ($(b.south)!0.17!(b_ground)$);
                \draw[<-,thick] ($(c.south)!0.27!(c_ground)$) -- node[label,midway]{HAVE F} ($(a.south)!0.27!(a_ground)$);
                \draw[<-,thick] ($(c.south)!0.37!(c_ground)$) -- node[label,near end]{IDONT\_HAVE F} ($(b.south)!0.37!(b_ground)$);
                \draw[->,thick] ($(c.south)!0.52!(c_ground)$) -- node[label,midway]{WANT-BLOCK F} ($(a.south)!0.52!(a_ground)$);
                \draw[<-,thick] ($(c.south)!0.62!(c_ground)$) -- node[label,midway]{BLOCK F} ($(a.south)!0.62!(a_ground)$);
                \draw[->,thick] ($(c.south)!0.72!(c_ground)$) -- node[label,midway]{WANT-BLOCK (a,b),(c,d)} ($(a.south)!0.72!(a_ground)$);
                \draw[<-,thick] ($(c.south)!0.82!(c_ground)$) -- node[label,midway]{BLOCK (a,b),(c,d)} ($(a.south)!0.82!(a_ground)$);
                \draw[<-,thick] ($(c.south)!0.82!(c_ground)$) -- node[below,midway]{\vdots} ($(a.south)!0.82!(a_ground)$);
            \end{tikzpicture}}};

    \node (filetree) [below=of f1,label={[yshift=0mm]below:File DAG}] {\resizebox{33mm}{!}{
        \begin{forest} qtree, forked edges
            [\extraprotect{$\mathbf{F} = ((a,b),(c,d))$}, no edge
                [\extraprotect{$(a,b)$}
                [$a$
                [$a_{0}$]
                [$a_{1}$]
                [$a_{2}$]
                ]
                [$b$
                [$b_{0}$]
                [$b_{1}$]
                [$b_{2}$]
                ]
                ]
                [\extraprotect{$(c,d)$}
                [$c$
                [$c_{0}$]
                [$c_{1}$]
                [$c_{2}$]
                ]
                [$d$
                [$d_{0}$]
                [$d_{1}$]
                ]
                ]
            ]
        \end{forest}}};

        \node[right=1mm of cauth] (f2) {\includegraphics[height=10mm]{survey}};

        \draw[->] ([yshift=0mm]f1.south) -- ([yshift=0mm]filetree.north) node[right,midway] {1. Create DAG};
        \draw[dashed,->] ([yshift=0.5mm]auth.north) |- ([yshift=4mm]net.north) -| ([yshift=0.5mm]cauth.north) node[above,pos=0] {2. Share root CID -- F};

        \draw[->] ([yshift=0mm]cauth.west) -- ([yshift=0mm]net.east) node[above,midway] {3. Request blocks};
        \draw[->] ([yshift=0mm]auth.east) -- ([yshift=0mm]net.west) node[above,midway] {4. Provide blocks};
        \draw[->] ([yshift=-10mm]net.east) -| (f2.south) node[above,pos=0.15] {5. Blocks received};

    \end{tikzpicture}}
    \caption{Conceptual overview of IPFS.}
    \label{fig:ipfs}
\end{figure*}

A typical file exchange using IPFS is illustrated in \cref{fig:ipfs}.
Here, the author of a survey paper uses IPFS to exchange it with a reviewer.
The author with the node identity~$N_{3}$ provides the survey via IPFS,
which splits it into a Merkle DAG (see also~\cref{fig:fipfsswarm}).
The author shares the resulting root CID~$F$ of the DAG with the Reviewer via an out-of-band channel.
The reviewer, whose node has the identity~$N_{11}$, requests $F$ from the network.
Bitswap deals with the exchange asking neighbors for the blocks, subsequently requesting the DAG.
Since nobody except for the author of the survey can answer the reviewer's request,
the author eventually provides the file to the reviewer using Bitswap.
When the reviewer acquired all blocks, she assembles the file and can read the survey.

\subsubsection{Availability}
IPFS does not have any implicit mechanisms for repairing and
maintaining files or ensuring redundancy and availability in the network.
In our previous example,
only the author~$N_{3}$ and the reviewer~$N_{11}$
hold all blocks of the survey.
There is no active replication due to the protocol.
Files can, however, be \enquote{pinned}
to prevent a node from deleting blocks locally.
Otherwise, content is only cached and
can be deleted via garbage collection at any point in time.
Furthermore, files cannot be intentionally deleted in other nodes,
deletes always happen locally only.
For a file to disappear, it needs to be removed from every cache
and every pinning node.

For storage guarantees Filecoin~\cite{protocollabs2017filecoin} exists.
Filecoin employs a storage and retrieval
market for storing and retrieving files.
The storage market is responsible for storing data,
\eg, match clients to storage miners and record their deals on the ledger,
reward/punish the storage miners, and verify the continuous storage.
The retrieval market is responsible for retrieving files.
Retrieval miners serve data in exchange for Filecoin
and only need the data at the time of serving the request.
To ensure that both parties, clients and retrieval miners,
can cooperate and are compensated,
data retrieval is ensured with payment channels,
data is sent in small pieces and compensated with micro-payments
before the next piece is sent.

While the storage and retrieval market handle their tasks slightly different,
the main principle is the same.
There are three different orders: bid, ask, and deal.
The bid order is a service request of the client
that it wants to store or retrieve files.
The ask order is a service offer from a storage or retrieval node announcing
storage or retrieval conditions.
The deal order is the statement closing the deal between bid and ask orders.
Orders are stored in the Orderbook.
For the storage market the Orderbook is stored on-chain,
to ensure informed decision making of clients and inform the market of the trends.
The orders are added in clear revealing information.
The Orderbook of the retrieval market is off-chain to increase retrieval speed.

The execution of deals is maintained using a distributed ledger
with Proof-of-Replication~(PoRep) and Proof-of-Space-Time~(PoST).
The PoRep is a proof that a storage node replicated data,
ensuring that the data is stored on separate physical storage.
PoST proves the continuous storage over time.
For more information on the different proofs,
we refer to the tech report~\cite{benet2017porep}.

\subsection{Features and Extensions}
IPFS supports multiple transport/network protocols,
or cryptographic hash functions to increase its adaptability.
This is possible through the usage of multi-address and multi-hash data structures.
Multi-address is a path structure for encoding addressing information.
They allow a peer to announce its contact information (\eg, IPv4 and IPv6),
transport protocol (\eg, TCP and UDP) and port.
Multi-hash is used to provide multiple different hash functions.
The digest value is prepended with the digest length, and the hash function type.
Multi-hashes are used for the node id and part of the CID.

The CID in IPFS is used for identifying blocks.
A CID is a cryptographic hash of its content with added meta data.
The meta data includes the used hashing algorithm and its length (multi-hash),
the encoding format (InterPlanetary Linked Data) and the version.
More specifically, the multi-hash prepended with encoding information
is InterPlanetary Linked Data~(IPLD),
and IPLD prepended with version information is the actual IPFS CID.

While IPFS itself has no mechanism to ensure redundancy/availability,
IPFS Cluster allows the creation and administration of
an additional overlay network of nodes, separate from the IPFS main network.
IPFS Cluster helps to ensure data redundancy and data allocation in a defined swarm.
The cluster manages pinned data, maintains a configured amount of replicas,
repinning content when necessary,
and considers free storage space while selecting nodes for pinning data.
IPFS Cluster needs a running IPFS node.
IPFS Cluster uses \emph{libp2p} for its networking layer.

IPFS Cluster ensures horizontal scalability of files without any incentives.
It can be used by a content provider to increase availability
without relying on caching in the network.
Filecoin can be used to incentivize others to store files.

\subsection{Use Cases}
In the following, we provide a brief overview of a few representative areas
to showcase some use cases of IPFS.
Please note, however, that this is by far not an exhaustive list
and also not the focus of this paper.
Nevertheless, it offers insights into current use cases and potential
of data networks in general and IPFS in particular.

IPFS is often combined with blockchains in many areas concerning data exchange.
In this case, the blockchain can fulfill various purposes.
The blockchain can provide integrity, authenticity, or serve as a pointer to the data.
Application areas include medical data~\cite{xu2019healthchain,kumar2020distributed},
tracking agricultural products~\cite{hao2018safe},
or in general data from the Internet of things (IoT)~\cite{ali2017iot}.
The blockchain can also serve as a mechanism to provide access control
to IPFS data~\cite{wang2018blockchain,steichen2018blockchain,battah2020blockchain},
or an audit trail for the data~\cite{nyaletey2019blockipfs}.
Data networks in general can also be used to improve storage issues of blockchains,
\eg, by off-chaining transaction data~\cite{norvill2018ipfs}.
Furthermore, some researchers propose new content sharing mechanism based
on IPFS and blockchains~\cite{chen2017improved,khatal2020fileshare,hoang2020privacy}.
Another proposed use case uses IPFS for making scientific papers publicly available
and a blockchain to provide a review mechanism matching reviewers, reviews, and papers~\cite{tenorio2019towards}.

However, there are also use cases for IPFS that do not involve a blockchain.
In compliance with its goal of decentralizing the Internet,
IPFS can be used for archiving websites in an InterPlanetary Wayback~\cite{alam2016interplanetary}.
Other use cases without a blockchain are
in combination with ICN as content delivery network~\cite{ascigil2019towards},
combining IPFS with scale-out network attached storage for Fog/Edge computing~\cite{confais2017object},
or for storing IoT data in combination with IPFS Cluster for increased availability~\cite{muralidharan2019interplanetary}.

Lastly, there is also the possibility for misuse of IPFS for malicious activities,
\eg, for ransomware as a service~\cite{karapapas2020ransomware} or
for the coordination of botnets~\cite{patsakis2019hydras}.

\subsection{Summary and Discussion}
IPFS combines ideas from BitTorrent, Kademlia, Git, and ICN
to create a new data network.
We believe that the integration of concepts
such as content addressing and deduplication
are promising as they have the potential to
improve retrieval times and storage overhead.

Furthermore, the different subprotocols like libp2p
for managing the peer discovery and connection handling,
and Bitswap for exchanging data are great developments,
that provide many opportunities for other P2P networks.
Additionally, optional mechanisms like Filecoin and IPFS Cluster
for improving the availability of files,
should provide additional use cases.

The wide support of different protocols increases the difficulty
to grasp the finer details of IPFS, though.
While encryption is supported in IPFS there are no additional mechanisms
for increasing the privacy of its participants.
The want and have list provide sensitive information about the participants.
As shown by \citeauthor{balduf2021monitoring}~\cite{balduf2021monitoring}
monitoring data requests in the network reveals information about its users.
Therefore, IPFS could have similar privacy problems to BitTorrent.
Furthermore, for good and bad it is not possible
to prevent replication or enforce deletion of content once released.

IPFS became a popular research topic.
In particular, IPFS itself and its performance, efficiency, and general usability
are subject of a series of papers~\cite{patsakis2019hydras,shen2019understanding,muralidharan2019interplanetary,ascigil2019towards,nyaletey2019blockipfs,heinisuo2019asterism,prunster2020total,henningsen2020mapping,henningsen2020crawling,de2021accelerating,guidi2021data}.
Please note that the referenced papers are pointers only.
More details on research concerning IPFS' functionality and
other data networks is discussed in a later section.

\section{Related P2P Data Networks}\label{sec:datanetworks}
Next to IPFS, many data networks are in development.
We give an overview of five other data networks,
pointing out their main concepts.
A summary and comparison of BitTorrent, IPFS,
and following data networks can be seen in \cref{tab:overview}.

\begin{table*}
\footnotesize
\centering
\caption{General overview of the different data networks.}
\newcolumntype{R}{>{\raggedright\arraybackslash}X}
\begin{tabularx}{\textwidth}{l R p{2.5cm}p{2cm}l}
\toprule
 \textbf{System} & \textbf{Main Goal and Distinct Feature} & \textbf{File Persistence} & \textbf{Token} & \textbf{Mutability} \\
\midrule
BitTorrent~\cite{cohen2003incentives} &
Efficient file distribution
utilizing tit-for-tat to provide Pareto optimality &
not guaranteed &
BitTorrent-Token~\cite{bittorrentfoundation2019bittorrentbtt} &
-- \\
%%%%%%%%%%%%%%%%%%%%%%%%%%%%%%%%%%%%%%%%%%%%%%%%%%%%%%%%%%%%%%
IPFS~\cite{benet2014ipfs,gitipfs} &
Decentralized web
achieving fast distribution through content addressing and wide compatibility &
not guaranteed &
Filecoin~\cite{protocollabs2017filecoin} &
IPNS \\
%%%%%%%%%%%%%%%%%%%%%%%%%%%%%%%%%%%%%%%%%%%%%%%%%%%%%%%%%%%%%%
Swarm~\cite{tron2020swarmbook,gitswarm} &
Decentralized storage and communication infrastructure
backed by a sophisticated Ethereum-based incentive mechanism &
not guaranteed &
Ethereum~\cite{wood2014ethereum} &
ENS, Feeds \\
%%%%%%%%%%%%%%%%%%%%%%%%%%%%%%%%%%%%%%%%%%%%%%%%%%%%%%%%%%%%%%
Hypercore~\cite{ogden2018dat,gitdat} &
Simple sharing of large mutable data objects (folder synchronization)
between selected peers &
not guaranteed &
-- &
yes \\
%%%%%%%%%%%%%%%%%%%%%%%%%%%%%%%%%%%%%%%%%%%%%%%%%%%%%%%%%%%%%%
SAFE~\cite{lambert2014safe,gitsafe} &
Autonomous data and communications network
using self-encryption and self-authentication
for improved decentralization and privacy &
public guaranteed,\newline private deletable &
Safecoin &
specific \\
%%%%%%%%%%%%%%%%%%%%%%%%%%%%%%%%%%%%%%%%%%%%%%%%%%%%%%%%%%%%%%
Storj~\cite{storjlabs2018storj,gitstorj} &
Decentralized cloud storage that
protects the data from Byzantine nodes with erasure codes and a reputation system &
determined lifetime,\newline deletable on request &
Centralized\newline Payments &
yes \\
%%%%%%%%%%%%%%%%%%%%%%%%%%%%%%%%%%%%%%%%%%%%%%%%%%%%%%%%%%%%%%
Arweave~\cite{arweave2019williams,gitarweave} &
Permanent storage in a blockchain-like structure
including content filtering &
blockweave &
Arweave token &
-- \\
\bottomrule
\end{tabularx}
\label{tab:overview}
\end{table*}

\subsection{Swarm}
Swarm~\cite{tron2020swarmbook} is a P2P distributed platform
for storing and delivering content developed by the Ethereum Foundation.
It provides censorship-resistance by not allowing any deletes,
as well as upload and forget properties.
Swarm is built for Ethereum~\cite{wood2014ethereum} and
therefore in some parts
depends on and shares design aspects of Ethereum.
The aim of Swarm is the provision of decentralized storage and streaming
functionality for the web3 stack~\cite{web3foundation},
a decentralized, censorship-resistant environment for sharing interactive content.
The Ethereum Foundation envisions Swarm as the
\enquote{hard disk of the world computer}.

Similar to IPFS, Swarm uses content-based addressing.
Contrary to IPFS, the content-based addressing in Swarm also decides the storage location.
To ensure availability, Swarm introduces areas of responsibility.
The area of responsibility are close neighbors of the node.
The nodes in an area of responsibility should provide chunk redundancy.
Mutability is supported through
versioning, keeping each version of the file.
Feeds, specially constructed and addressed chunks,
and the Ethereum Name Service~(ENS) are used for finding the mutated files.
ENS is a standard defined in the Ethereum Improvement Proposal 137~\cite{eip137}.
It provides the ability to translate addresses into human-readable names.
In contrast to IPNS, ENS is implemented as a smart contract
on the Ethereum blockchain.

\begin{figure}
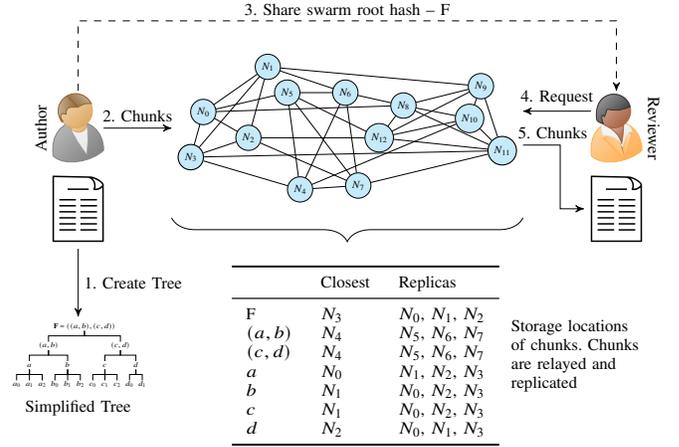

\scriptsize
\resizebox{\columnwidth}{!}{
    \begin{tikzpicture}[node distance=10mm,auto,>=stealth']
    \node[bob, minimum height=10mm] (auth) {};
    \node[rotate=90,anchor=south] at (auth.west) {Author};
    \node[minimum height=10mm, below=1mm of auth] (f1) {\includegraphics[height=10mm]{survey}};

    \node (net) [minimum height=5mm, right=of auth] {
        \resizebox{50mm}{!}{\begin{tikzpicture}
        \large
        \tikzstyle{peer}=[circle, draw, thin,fill=cyan!20, minimum width=8mm]

        \node [peer] (a) at (-6.75, 3) {$N_{0}$};
        \node [peer] (c) at (-4.25, 4.75) {$N_{1}$};
        \node [peer] (e) at (-5, 2.0) {$N_{2}$};
        \node [peer, very thick] (f) at (-7.25, 1.25) {$N_{3}$};
        \node [peer] (i) at (-3, 0.0) {$N_{4}$};
        \node [peer] (j) at (-3.5,3.75) {$N_{5}$};
        \node [peer] (m) at (-1.25, 3.75) {$N_{6}$};
        \node [peer] (o) at (-0.75, 0.15) {$N_{7}$};
        \node [peer] (p) at (1.0, 3.25) {$N_{8}$};
        \node [peer] (s) at (4, 4) {$N_{9}$};
        \node [peer] (t) at (3.5, 2.75) {$N_{10}$};
        \node [peer,very thick] (u) at (4.75, 1.5) {$N_{11}$};
        \node [peer] (z) at (0, 2.0) {$N_{12}$};

        \path[thick] (a) edge (c);
        \path (a) edge (e);
        \path[thick] (a) edge (f);
        \path[thick] (c) edge (e);
        \path (c) edge (f);
        \path[thick] (e) edge (f);

        \path[thick] (i) edge (j);
        \path (i) edge (m);
        \path[thick] (i) edge (o);
        \path[thick] (j) edge (m);
        \path (j) edge (o);
        \path[thick] (m) edge (o);

        \path[thick] (p) edge (s);
        \path (p) edge (t);
        \path[thick] (p) edge (u);
        \path[thick] (p) edge (z);
        \path[thick] (s) edge (t);
        \path (s) edge (u);
        \path[thick] (s) edge (z);
        \path[thick] (t) edge (u);
        \path[thick] (t) edge (z);
        \path[thick] (u) edge (z);

        \path[thick] (a) edge (j);
        \path[thick] (a) edge (p);
        \path[thick] (c) edge (m);
        \path[thick] (c) edge (s);
        \path[thick] (e) edge (o);
        \path[thick] (e) edge (z);
        \path[thick] (f) edge (i);
        \path[thick] (f) edge (u);

        \path[thick] (i) edge (t);
        \path[thick] (j) edge (z);
        \path[thick] (m) edge (p);
        \path[thick] (o) edge (u);

    \end{tikzpicture}}};

    \node[alice, minimum height=10mm, right=of net] (cauth) {};
    \node[rotate=270,anchor=south] at (cauth.east) {Reviewer};
    \node[minimum height=10mm, below=1mm of cauth] (f2) {\includegraphics[height=10mm]{survey}};

    \node (filetree) [below=of f1,label={[yshift=0mm]below:Simplified Tree}] {\resizebox{20mm}{!}{
        \begin{forest} qtree, forked edges
            [\extraprotect{$\mathbf{F} = ((a,b),(c,d))$}, no edge
                [\extraprotect{$(a,b)$}
                [$a$
                [$a_{0}$]
                [$a_{1}$]
                [$a_{2}$]
                ]
                [$b$
                [$b_{0}$]
                [$b_{1}$]
                [$b_{2}$]
                ]
                ]
                [\extraprotect{$(c,d)$}
                [$c$
                [$c_{0}$]
                [$c_{1}$]
                [$c_{2}$]
                ]
                [$d$
                [$d_{0}$]
                [$d_{1}$]
                ]
                ]
            ]
        \end{forest}}};

        \node (chunks) [right=of filetree,label={[yshift=0mm,align=left,text width=20mm]right:Storage locations of chunks. Chunks are relayed and replicated}] {\resizebox{!}{!}{
            \scriptsize
            \begin{tabular}{lll}
            \toprule
                        & Closest & Replicas \\
            \midrule
            F           & $N_{3}$ & $N_{0}$, $N_{1}$, $N_{2}$ \\
            $(a,b)$     & $N_{4}$ & $N_{5}$, $N_{6}$, $N_{7}$ \\
            $(c,d)$     & $N_{4}$ & $N_{5}$, $N_{6}$, $N_{7}$ \\
            $a$         & $N_{0}$ & $N_{1}$, $N_{2}$, $N_{3}$ \\
            $b$         & $N_{1}$ & $N_{0}$, $N_{2}$, $N_{3}$ \\
            $c$         & $N_{1}$ & $N_{0}$, $N_{2}$, $N_{3}$ \\
            $d$         & $N_{2}$ & $N_{0}$, $N_{1}$, $N_{3}$ \\
            \bottomrule
            \end{tabular}}};

        \draw [decorate,decoration={brace,amplitude=10pt,mirror,raise=4pt}] (net.south west) -- (net.south east);

        \draw[->] ([yshift=0mm]f1.south) -- ([yshift=0mm]filetree.north) node[right,midway] {1. Create Tree};
        \draw[->] ([yshift=0mm]auth.east) -- ([yshift=0mm]net.west) node[above,pos=0.5,align=left] {2. Chunks};

        \draw[dashed,->] ([yshift=0.5mm]auth.north) |- ([yshift=4mm]net.north) -| ([yshift=0.5mm]cauth.north) node[above,pos=0] {3. Share swarm root hash -- F};
        \draw[->] ([yshift=2.5mm]cauth.west) -- ([yshift=2.5mm]net.east) node[above,pos=0.7,midway,align=left] {4. Request};
        \draw[->] ([yshift=-2.5mm]net.east) -| ([xshift=-3.5mm]f2.west) node[above,pos=0.4] {5. Chunks} -- ([xshift=0mm]f2.west);

    \end{tikzpicture}}
    \caption{Conceptual overview of Swarm.}
    \label{fig:swarm}
\end{figure}

\cref{fig:swarm} provides a conceptual overview of Swarm,
where we continue to use the exchange of a survey paper between author and reviewer as running example.
Swarm splits a file, \ie, the survey into chunks which are arranged into a so-called \emph{Swarm hash}.
The Swarm hash is a combination of chunks arranged in a Merkle tree,
where leaf nodes represent input data and intermediate nodes are compositions
of reference to its child nodes.
The resulting chunks are uploaded to the network.
Swarm employs a Kademlia topology,
where the neighbors are determined by their identifiers distance.
It should be noted that additionally to the bucket connections,
Swarm relies on a nearest neighbor set, which are the remaining nodes of the neighborhood.
A neighborhood is basically the lowest amount of buckets that contain at least three other peers.
This nearest neighbor set is responsible for replication and is not necessarily symmetric.
As an example, in \cref{fig:swarm} the nearest neighbor for $N_{3}$ are
$N_{2}$, $N_{1}$ and $N_{0}$, while for $N_{12}$,
the neighbors are $N_{8}$, $N_{9}$, $N_{10}$ and $N_{11}$ (cf.~\cref{fig:kadtree}).
The uploaded chunks are relayed, stored, and replicated at the location closest to their address.
To retrieve the survey, the swarm root hash is necessary.
The network relays requests according to the content address.

To ensure compliant node behavior, Swarm provides an incentive layer.
The incentive structure is based on SWAP, SWEAR, and SWINDLE.
The SWarm Accounting Protocol~(SWAP) handles the balancing of data exchange between nodes.
Each node maintains local accounting information.
A peer basically buys a chunk on credit without interest from the serving node.
The price for chunks is negotiable between the peers.
Requests are served until certain imbalance thresholds,
\ie chunks are served lopsidedly and the debt becomes too high.
After the first threshold is reached,
the node expects a settlement of the debt for further service.
If the second threshold is reached,
due to not settling the debt,
the node is disconnected.
The debt can be settled with cheques, which can be interpreted as a simple one-way payment channel.
SWarm Enforcement And Registration (SWEAR) and
Secured With INsurance Deposit Litigation and Escrow (SWINDLE)
shall ensure persistence of content.
Furthermore, Swarm's incentive structure has postage stamps,
which provide a mechanism against junk uploads and
also a lottery mechanism to incentivize the continued storage of chunks.

Postage stamps can be acquired in batches from a smart contract.
The postage stamps are attached to an uploaded chunk and signed by the owner of the stamp.
This serves as a proof of payment for uploading chunks.
Stamp usage can only be monitored by relaying or storing nodes.
This allows reusage/overusage of stamps.
To reduce the risk of overusing stamps,
the stamps are only for certain prefix collisions,
limiting stamps to chunks in certain storage areas.

The postage stamps are used in a lottery.
The lottery provides value to chunks preventing early deletes of chunks.
Through the lottery, storage nodes can gain part of the initial costs for the stamp.
In the lottery an address area is chosen.
Nodes in the proximity of the area can apply for reward.
By applying, nodes testify possession of chunks in the area.
The nodes define a price for storing a chunk.
After proving possession of the chunks, the node with the cheapest prize gets the reward.

\paragraph*{Discussion}
Swarm provides sophisticated incentive concepts.
Settling unbalanced retrieval with cheques provides a faster and cheaper way
to settle discrepancies than relying on blockchain transactions.
The postage stamps with the lottery give an additional incentive for storing chunks.
Additionally, while it does cost to upload content,
nodes can earn the cost by actively serving chunks to participants.
However, postage stamps link a user to uploaded content.
While Swarm provides a certain degree of sender anonymity,
the upload pseudonymity might limit available content.

Considering the determined storage locations by the
Distributed Immutable Store for Chunks~(DISC),
the network might face storage problems.
Feeds, which can provide user-defined space in the network,
in the form of recovery feeds and pinning, might be able to mitigate
these disadvantages.

Overall, Swarm clearly depends on the Ethereum ecosystem.
While it is advantageous for the incentive structure,
since Ethereum is actively developed and has a broad user base,
it also requires users to depend on Ethereum.
While having this potentially large user base,
research of use cases or research investigating Swarm's mechanism is rare.
The connection of Swarm and Ethereum could be one reason for the lack of research,
since Swarm seems less complete than IPFS and
Ethereum itself still maintains many research opportunities.

\subsection{Hypercore Protocol/Dat}
The Hypercore Protocol~\cite{ogden2018dat,datproto} (formerly Dat Protocol)
supports incremental versioning of the content and meta data similar to Git.
The Hypercore Protocol consists of multiple sub-components.
While strictly speaking Hypercore is one of the sub-components,
for simplicity we use the term to reference the Hypercore Protocol in general.
In Hypercore, data is stored in a directory like structure and
similar to BitTorrent each directory is dealt with its own network.
The protocol supports different storage modes,
where each node can decide which data of a directory
and which versions of the data it wants to store.
Furthermore, the protocol supports subscription to live changes of all/any files in a directory.
All communication in the protocol is encrypted.
In order to find and read the data,
it is necessary to know a specific read key.

The protocol is designed to share large amounts of mutable data.
The motivation for creating the protocol was to
prevent link rot and content drift of scientific literature.
The protocol allows sharing of only part of the data with random access.

Hypercore can be understood as sharing a folder.
Files in a folder can be modified, added, and deleted.
This also includes and allows mutable files.

\begin{figure}
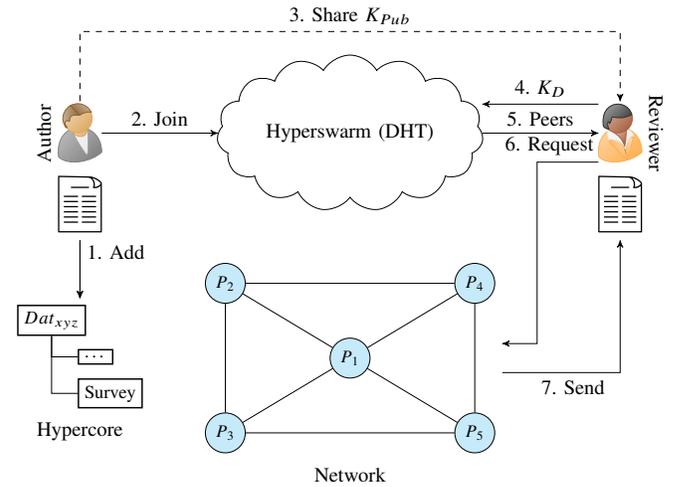

\resizebox{\columnwidth}{!}{
    \begin{tikzpicture}[node distance=20mm,auto,>=stealth']

    \node[bob, minimum height=10mm] (auth) {};
    \node[rotate=90,anchor=south] at (auth.west) {Author};
    \node[minimum height=10mm, below=1mm of auth] (f1) {\includegraphics[height=10mm]{survey}};

    \node (net0) [cloud, draw,cloud puffs=15,cloud puff arc=120, aspect=2, inner ysep=1em, right= of auth] {Hyperswarm (DHT)};
    \node (net1) [below=10mm of net0,label={[yshift=0mm]below:Network}] {
        \resizebox{50mm}{!}{\begin{tikzpicture}
        \tikzstyle{peer}=[circle, draw, thin,fill=cyan!20, minimum width=8mm]
        \node[peer] (s0) at (0,0) {$P_{1}$};
        \node[peer] (l1) at (-2.5,1.5) {$P_{2}$};
        \node[peer] (l2) at (-2.5,-1.5) {$P_{3}$};
        \node[peer] (l3) at (2.5,1.5) {$P_{4}$};
        \node[peer] (l4) at (2.5,-1.5) {$P_{5}$};

        \path (s0) edge (l1);
        \path (s0) edge (l2);
        \path (s0) edge (l3);
        \path (s0) edge (l4);
        \path (l1) edge (l2);
        \path (l1) edge (l3);
        \path (l2) edge (l4);
        \path (l3) edge (l4);

    \end{tikzpicture}}};

    \node[alice, minimum height=10mm, right=of net0] (cauth) {};
    \node[rotate=270,anchor=south] at (cauth.east) {Reviewer};
    \node[minimum height=10mm, below=1mm of cauth] (f2) {\includegraphics[height=10mm]{survey}};

    \node (filetree) [scale=0.9,below=10mm of f1,label={[yshift=0mm]below:Hypercore}] {
        \begin{tikzpicture}[%
        grow via three points={one child at (0.5,-0.7) and
        two children at (0.5,-0.7) and (0.5,-1.4)},
        edge from parent path={(\tikzparentnode.south) |- (\tikzchildnode.west)}]
        \tikzstyle{every node}=[draw=black,thick,anchor=west]
        \node (d) {$Dat_{xyz}$}
            child { node {\dots}}
            child { node {Survey}};
        \end{tikzpicture}};

    \draw[->] ([yshift=0mm]f1.south) -- ([yshift=0mm]filetree.north);

    \draw[->] ([yshift=0mm]f1.south) -- ([yshift=0mm]filetree.north) node[right, pos=0.2,align=left] {1. Add};
    \draw[->] ([yshift=0mm]auth.east) -- ([yshift=0mm]net0.west) node[above, midway] {2. Join};
    \draw[dashed,->] ([yshift=0.5mm]auth.north) |- ([yshift=4mm]net0.north) -| ([yshift=0.5mm]cauth.north) node[above,pos=0] {3. Share $K_{Pub}$};
    \draw[->] ([yshift=5mm]cauth.west) -- ([yshift=5mm]net0.east) node[above, midway] {4. $K_{D}$};
    \draw[<-] ([yshift=0mm]cauth.west) -- ([yshift=0mm]net0.east) node[above, midway] {5. Peers};
    \draw[->] ([yshift=-5mm,xshift=0mm]cauth.west) -| ([xshift=6mm,yshift=2.5mm]net1.east) node[above, pos=0.4] {6. Request} -- ([yshift=2.5mm]net1.east);
    \draw[->] ([yshift=-2.5mm]net1.east) -| ([yshift=0mm]f2.south) node[below, pos=0.3] {7. Send};
    \end{tikzpicture}}
    \caption{Conceptual overview of Hypercore.}
    \vspace{-1.5em}
    \label{fig:dat}
\end{figure}

A conceptual overview of Hypercore can be seen in \cref{fig:dat}.
For peer discovery, Hypercore uses Hyperswarm,
a Kademlia-based DHT.
If the author wants to share the survey using the Hypercore Protocol,
the author needs to create a Hypercore and add the survey.
To be able to be found by Hyperswarm, it is necessary to join the Hyperswarm overlay network.
By sharing the public key~$K_{Pub}$, the reviewer can calculate the discovery key $K_{D}$ and
after finding peers and joining the data network decrypt the messages.
Once the other overlay network is joined
the unstructured network of volunteers can share the data and the survey can be retrieved.

\paragraph*{Discussion}
Hypercore allows sharing of data by exchanging a public key.
It is possible to acquire a specific version
and only specific regions of the data.
This makes it simple, especially for large datasets, and allows mutable data.
The protocol natively concentrates on sharing collection of files,
which broadens the usability of the protocol.

Due to the encryption and a discovery key, the protocol ensures confidentiality.
A public key allows the calculation of the discovery key,
but it is not possible to reverse the public key.
This prevents others from reading the data.
A downside of Hypercore is the lack of
additional authentication mechanisms beyond the public key,
which prevents additional fine-grained access control.
Furthermore, it still leaks meta data since the discovery key is only a pseudonym.

Hypercore has no incentive structure for replicating data
and the data persistence relies on its participants.
Research utilizing or analyzing Hypercore/Dat is rare.
While the protocol seems well developed and usable,
research seems to focus on IPFS, instead.

\subsection{Secure Access For Everyone (SAFE)}
The Secure Access For Everyone~(SAFE) network~\cite{lambert2014safe,safeprimer}
is designed to be a fully autonomous decentralized data and communication network.
Even authentication follows a self-authentication~\cite{irvine2010self} mechanism,
which does not rely on any centralized component.
The main goal of SAFE is to provide a network which everyone can join and use
to store, view, and publish data without leaving trace of their activity on the machine.
This would allow participants to publish content with low risks of persecution.

SAFE supports three different data types: Map, Sequence, and Blob.
The data can be further divided into public and private data.
Map and sequence are Conflict-free Replicated Data Types,
which is important in case of mutable data to ensure consistency.
The Blob is for immutable data.
All data in the SAFE network is encrypted, even public data.
The used encryption algorithm is self-encryption~\cite{irvine2010selfencrypt},
which uses the file itself to encrypt the file.
That is, a file is split into at least three fixed size chunks.
Each chunk is hashed and encrypted with the hash of the previous chunk,
\ie, $n-1$ where $n$ is the current chunk.
Afterwards, the encrypted chunk gets obfuscated with
the concatenated hashes of the original chunks.
In case of SAFE, the obfuscated chunks are stored in the network.
For decryption, a data map is created during the encryption process.
The data map contains information about the file and maps
the hash of obfuscated chunks to the hash of the real chunks.
For public data, the decryption keys are provided by the network.
While private data can be deleted, public data should be permanent.
Therefore mutable data can only be private.
A Name Resolution System allows human-readable addresses for retrieving data.

The network itself is organized by XOR addresses according to a Kademlia-based DHT.
Furthermore, the network is split into sections.
When a new vault wants to join the network,
the new vault needs to prove, that it can provide the required resources
and is then randomly assigned a XOR address and therefore to a section.
The sections are maintained dynamically.
According to the amount of vaults in the network,
sections are split and the vaults are reassigned to new sections.
Sections that grow too small are prioritized by getting new nodes
or can request relocation of nodes to balance section size.
Changing the section increases the vault's node age.
Node age is a measurement of trust, can be lost and must then be re-earned.
Only a certain amount of nodes can make decisions in a section, the elders.
Elders are the oldest nodes in the section.
The elders can vote on accepting or denying events in a section
and a certain quorum of elders has to approve and group sign the decision
for it to become valid.
Events in a network section are, for example,
joining/leaving of a node or storing a chunk.
The authenticity of the elders is ensured by a SectionProofChain,
which holds the elders' group signatures and is a sequence of public keys proving the validity of a section.
The sequence is updated and signed everytime the group of elders changes.

\begin{figure}
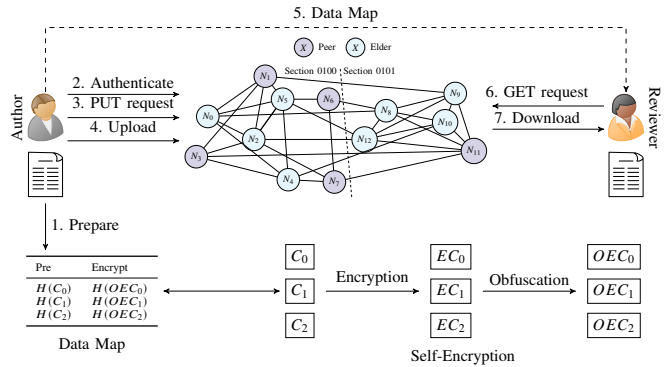

\resizebox{\columnwidth}{!}{
    \begin{tikzpicture}[node distance=25mm,auto,>=stealth']

    \node[bob, minimum height=10mm] (auth) {};
    \node[rotate=90,anchor=south] at (auth.west) {Author};
    \node[minimum height=10mm, below=1mm of auth] (f1) {\includegraphics[height=10mm]{survey}};

    \node (net) [right=of auth] {
        \resizebox{65mm}{!}{\begin{tikzpicture}\large
        \tikzstyle{peer}=[circle, draw, thin,fill=sron0!20, minimum width=8mm]
        \tikzstyle{elder}=[circle, draw, thin,fill=sron1!20, minimum width=8mm]

        \node [peer,label={[right,xshift=5mm,yshift=-4mm]Peer}] (sa) at (-2.5, 6) {$X$};
        \node [elder, label={[right,xshift=5mm,yshift=-4mm]Elder}] (st) at (-0.25, 6) {$X$};

        \node[anchor=west] at (-0.5,5) {Section 0101};
        \node[anchor=east] at (-0.5,5) {Section 0100};
        \path[dashed,thick] (0,-0.5) edge (-0.5,5.1);

        \node [elder] (a) at (-6.75, 3) {$N_{0}$};
        \node [peer] (c) at (-4.25, 4.75) {$N_{1}$};
        \node [elder] (e) at (-4.75, 2.0) {$N_{2}$};
        \node [peer] (f) at (-7.25, 1.25) {$N_{3}$};
        \node [elder] (i) at (-3.25, 0.25) {$N_{4}$};
        \node [elder] (j) at (-3.5,3.75) {$N_{5}$};
        \node [peer] (m) at (-1.5, 3.75) {$N_{6}$};
        \node [peer] (o) at (-1.25, 0.15) {$N_{7}$};
        \node [elder] (p) at (1.0, 3.25) {$N_{8}$};
        \node [elder] (s) at (4, 4) {$N_{9}$};
        \node [elder] (t) at (3.5, 2.75) {$N_{10}$};
        \node [peer] (u) at (4.75, 1.5) {$N_{11}$};
        \node [elder] (z) at (0, 2.0) {$N_{12}$};

        \path (a) edge (c);
        \path (a) edge (e);
        \path (a) edge (j);
        \path (a) edge (i);
        \path (c) edge (e);
        \path (c) edge (f);
        \path (c) edge (j);
        \path (e) edge (f);
        \path (e) edge (j);
        \path (e) edge (j);

        \path (i) edge (j);
        \path (i) edge (o);
        \path (j) edge (m);
        \path (m) edge (o);

        \path (p) edge (s);
        \path (p) edge (t);
        \path (p) edge (u);
        \path (p) edge (z);
        \path (s) edge (t);
        \path (s) edge (u);
        \path (s) edge (z);
        \path (t) edge (u);
        \path (t) edge (z);
        \path (u) edge (z);

        \path (a) edge (p);
        \path (c) edge (s);
        \path (e) edge (o);
        \path (e) edge (z);
        \path (f) edge (i);
        \path (f) edge (u);

        \path (i) edge (t);
        \path (j) edge (z);
        \path (m) edge (p);
        \path (o) edge (u);

    \end{tikzpicture}}};

    \node[alice, minimum height=10mm, right=of net] (cauth) {};
    \node[rotate=270,anchor=south] at (cauth.east) {Reviewer};
    \node[minimum height=10mm, below=1mm of cauth] (f2) {\includegraphics[height=10mm]{survey}};

    \node (map) [below=10mm of f1,xshift=10mm,label={[yshift=0mm]below:Data Map}] {
            \scriptsize
            \begin{tabular}{ll}
            \toprule
            Pre & Encrypt \\
            \midrule
            $H(C_{0})$ & $H(OEC_{0})$ \\
            $H(C_{1})$ & $H(OEC_{1})$ \\
            $H(C_{2})$ & $H(OEC_{2})$ \\
            \bottomrule
            \end{tabular}};

    \node (selfe) [right=of map,label={[yshift=0mm]below:Self-Encryption}] {
        \begin{tikzpicture}[node distance=25mm]
            \node [draw, rectangle, minimum width=4mm] (c1) {$C_{0}$};
            \node [draw, rectangle, minimum width=4mm, below=2mm of c1] (c2) {$C_{1}$};
            \node [draw, rectangle, minimum width=4mm, below=2mm of c2] (c3) {$C_{2}$};

            \node [draw, rectangle, minimum width=4mm, right=of c1] (c4) {$EC_{0}$};
            \node [draw, rectangle, minimum width=4mm, below=2mm of c4] (c5) {$EC_{1}$};
            \node [draw, rectangle, minimum width=4mm, below=2mm of c5] (c6) {$EC_{2}$};

            \node [draw, rectangle, minimum width=4mm, right=of c4] (c7) {$OEC_{0}$};
            \node [draw, rectangle, minimum width=4mm, below=2mm of c7] (c8) {$OEC_{1}$};
            \node [draw, rectangle, minimum width=4mm, below=2mm of c8] (c9) {$OEC_{2}$};

            \draw[->] ([xshift=2.5mm]c2.east) -- ([xshift=-2.5mm]c5.west) node[above, pos=0.5] {Encryption};
            \draw[->] ([xshift=2.5mm]c5.east) -- ([xshift=-2.5mm]c8.west) node[above, pos=0.5] {Obfuscation};

        \end{tikzpicture}};

    \draw[<->] (selfe -| map.east) -- ([yshift=0mm]selfe.west) node[left, midway] {};

    \draw[->] ([yshift=0mm]f1.south) -- (map.north -| f1.south) node[right, midway,align=left] {1. Prepare};
    \draw[->] ([yshift=5mm,xshift=1mm]auth.east) -- ([yshift=5mm]net.west) node[above, midway,align=left] {2. Authenticate};
    \draw[->] ([yshift=0mm,xshift=1mm]auth.east) -- ([yshift=0mm]net.west) node[above, midway,align=left] {3. PUT request};
    \draw[->] ([yshift=-5mm,xshift=1mm]auth.east) -- ([yshift=-5mm]net.west) node[above, midway,align=left] {4. Upload};

    \draw[dashed,->] ([yshift=0.5mm]auth.north) |- ([yshift=1mm]net.north) -| ([yshift=0.5mm]cauth.north) node[above,pos=0] {5. Data Map};

   \draw[->] ([yshift=2.5mm,xshift=-1mm]cauth.west) -- ([yshift=2.5mm]net.east) node[above, pos=0.6,align=left] {6. GET request};
    \draw[<-] ([yshift=-2.5mm,xshift=-1mm]cauth.west) -- ([yshift=-2.5mm]net.east) node[above, pos=0.6,align=left] {7. Download};

    \end{tikzpicture}}
    \caption{Conceptual overview of SAFE.}
    \label{fig:safe}
\end{figure}

A conceptual overview of the SAFE network can be seen in \cref{fig:safe}.
Considering our running example,
the survey is split into chunks, self-encrypted, and used to generate a data map.
After the self-authentication process, a \emph{PUT} request is sent to the network.
When the elders in the section responsible for storing chunks agree, the data is stored.
For downloading the file, the data map is required.
The data map is used for \emph{GET} requests to acquire the obfuscated encrypted chunks.
For downloading public data, authentication is not necessary.
After acquiring the chunks the file can be recreated with the help of the data map.

In the SAFE network, storing data is charged with
the network's own currency, \ie, Safecoin.
The Safecoin balance of the clients is monitored by client managers
and approved/rejected with the help of SAFE's consensus mechanisms.
Nodes can earn Safecoin by farming, \ie, providing content to requesters.

\paragraph*{Discussion}
The self-authentication, self-encryption, and the network organization
give the user a high degree of control over their data.
The absence of central components reduce single points of failure.
Furthermore, privacy and to a certain degree anonymity are key features of the SAFE network.
The network requires authentication for storing data only.
Retrieving data is mediated via a client-selected proxy,
which provides pseudonymous communication.
Safecoin is intended to provide an incentive layer
which ensures the availability and reliability of the network.

\citeauthor{paul2014security}~\cite{paul2014security} provide a first security analysis of SAFE in 2014,
concerning confidentiality, integrity and availability as well as possible attacks.
In 2015, \citeauthor{jacob2015security}~\cite{jacob2015security} analyzed the security of the network
with respect to authenticity, integrity, confidentiality, availability, and anonymity.
The authors explained how the self-authentication and the decentralized nature
could be potentially exploited to reveal personal data of single entities.

SAFE is in development since 2006 and considers recent research and developments,
but remains (at the time of writing) in its alpha phase.
We feel that SAFE has a potential to establish the topic of anonymity
as a distinct feature when compared to the other data networks.

\subsection{Storj}
Storj~\cite{storjlabs2018storj} is a P2P storage network.
In the following, we refer to version 3.0.
It concentrates on high durability of data, low latency,
and high security and privacy for stored data.
End-to-end encryption for communication, file locations, and files is supported.
For the high durability of files or in other words better availability of files in the network,
Storj uses erasure codes.
Furthermore, low bandwidth consumption is also one main design goal.
The protocol assumes object size of $4\,MB$ or more,
while lower object sizes are supported the storage process could be less efficient.
In Storj, decentralization is interpreted as no single operator is
solely responsible for the operation of the system.
In a decentralized system, trust and Byzantine failure assumptions are important.
Storj assumes no altruistic, always good behaving nodes,
a majority of rational nodes, behaving only malicious when they profit,
and a minority of Byzantine malicious nodes.

Storj aims to be a decentralized cloud storage.
\citeauthor{storjlabs2018storj} wants to provide an alternative
to centralized storage providers.
For this purpose, Storj provides compatibility with Amazon~S3
application programming interface to increase the general
acceptance and ease the migration for new user.
Since Storj provides cloud storage, user are allowed
to store and retrieve data as well as delete, move, and copy data.

The Storj network consists of three node types, satellite, storage, and uplink nodes.
The satellite nodes administrate the storage process and maintenance of files.
The encryption of meta data and even file paths adds an additional protection of meta data.
Uplink nodes are end users, who want to store and access files.
Storage nodes store the data.
Storage and uplink nodes choose with which Satellite nodes to cooperate.
This results in a network similar to BitTorrent where satellites become central parts.

\begin{figure}
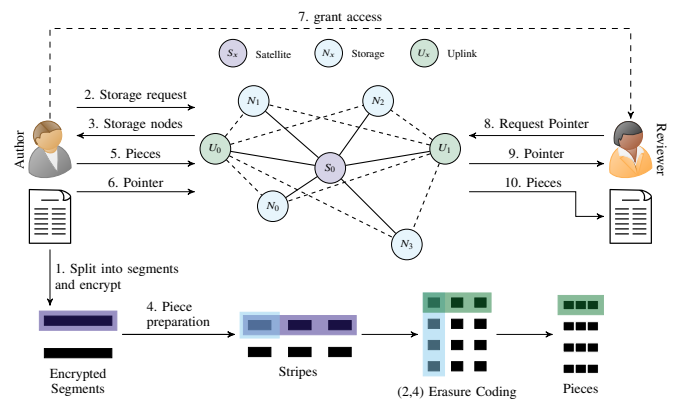

\scriptsize
\resizebox{\columnwidth}{!}{
    \begin{tikzpicture}[node distance=22mm,auto,>=stealth']
    \node[bob, minimum height=10mm] (auth) {};
    \node[rotate=90,anchor=south] at (auth.west) {Author};
    \node[minimum height=10mm, below=1mm of auth] (f1) {\includegraphics[height=10mm]{survey}};

    \node (net) [right=of auth] {
        \resizebox{50mm}{!}{\begin{tikzpicture}\footnotesize
        \tikzstyle{satellite}=[circle, draw, thin,fill=sron0!20, minimum width=8mm]
        \tikzstyle{storage}=[circle, draw, thin,fill=sron1!20, minimum width=8mm]
        \tikzstyle{uplink}=[circle, draw, thin,fill=sron2!20, minimum width=8mm]

        \node [satellite,scale=0.9,label={[right,xshift=5mm,yshift=-4mm]Satellite}] (sa) at (-2.5, 5) {$S_{x}$};
        \node [storage,scale=0.9,label={[right,xshift=5mm,yshift=-4mm]Storage}] (st) at (0, 5) {$N_{x}$};
        \node [uplink,scale=0.9,label={[right,xshift=5mm,yshift=-4mm]Uplink}] (up) at (2.5, 5) {$U_{x}$};

        \node [uplink] (f) at (-3.0, 2.5) {$U_{0}$};
        \node [storage] (i) at (-1.5, 1.0) {$N_{0}$};
        \node [storage] (j) at (-2.0,3.75) {$N_{1}$};
        \node [storage] (m) at (1.25, 3.75) {$N_{2}$};

        \node [storage] (p) at (2.0, 0.0) {$N_{3}$};
        \node [uplink] (u) at (3.0, 2.5) {$U_{1}$};
        \node [satellite] (z) at (0, 2.0) {$S_{0}$};

        \path[thick,dashed] (f) edge (i);
        \path[thick,dashed] (f) edge (j);
        \path[thick,dashed] (f) edge (m);
        \path[thick,dashed] (f) edge (p);

        \path[thick,dashed] (u) edge (i);
        \path[thick,dashed] (u) edge (j);
        \path[thick,dashed] (u) edge (m);
        \path[thick,dashed] (u) edge (p);

        \path[thick] (z) edge (f);
        \path[thick] (z) edge (u);
        \path[thick] (z) edge (i);
        \path[thick] (z) edge (j);
        \path[thick] (z) edge (m);
        \path[thick] (z) edge (p);

    \end{tikzpicture}}};

    \node[alice, minimum height=10mm, right=of net] (cauth) {};
    \node[rotate=270,anchor=south] at (cauth.east) {Reviewer};
    \node[minimum height=10mm, below=1mm of cauth] (f2) {\includegraphics[height=10mm]{survey}};

    \node (seg) [below=10mm of f1,xshift=5mm,label={[yshift=0mm,align=center]below:Encrypted\\Segments}] {
        \resizebox{!}{!}{\begin{tikzpicture}[node distance=4mm]
        \node [draw, fill, rectangle, minimum width=12mm] (s1) {};
        \node [draw, fill, rectangle, minimum width=12mm, below=of s1] (s2) {};
        \node [draw=sron0, fill=sron0, rectangle, opacity=0.5, fit=(s1)(s1)] {};
        \end{tikzpicture}}};

    \node (strip) [right=20mm of seg,label={[yshift=0mm]below:Stripes}] {
        \resizebox{!}{!}{\begin{tikzpicture}[node distance=3mm]
        \node [draw, fill, rectangle, minimum width=4mm] (s1) {};
        \node [draw, fill, rectangle, minimum width=4mm, right=of s1] (s2) {};
        \node [draw, fill, rectangle, minimum width=4mm, right=of s2] (s3) {};
        \node [draw=sron0, fill=sron0, rectangle, opacity=0.5, fit=(s1)(s3)] {};
        \node [draw=sron1, fill=sron1, rectangle, opacity=0.5, fit=(s1)(s1),inner sep=4pt] {};
        \node [draw, fill, rectangle, minimum width=4mm, below=of s1] (s4) {};
        \node [draw, fill, rectangle, minimum width=4mm, right=of s4] (s5) {};
        \node [draw, fill, rectangle, minimum width=4mm, right=of s5] (s6) {};
        \end{tikzpicture}}};

    \node (es) [right=10mm of strip,label={[yshift=0mm]below:(2,4) Erasure Coding}] {
        \resizebox{!}{!}{\begin{tikzpicture}[node distance= 2mm]
        \node [draw, fill, rectangle, minimum width=2mm] (s1) {};
        \node [draw, fill, rectangle, minimum width=2mm, below=of s1] (s2) {};
        \node [draw, fill, rectangle, minimum width=2mm, below=of s2] (s3) {};
        \node [draw, fill, rectangle, minimum width=2mm, below=of s3] (s4) {};
        \node [draw, fill, rectangle, minimum width=2mm, right=of s1] (s5) {};
        \node [draw, fill, rectangle, minimum width=2mm, below=of s5] (s6) {};
        \node [draw, fill, rectangle, minimum width=2mm, below=of s6] (s7) {};
        \node [draw, fill, rectangle, minimum width=2mm, below=of s7] (s8) {};
        \node [draw, fill, rectangle, minimum width=2mm, right=of s5] (s9) {};
        \node [draw, fill, rectangle, minimum width=2mm, below=of s9] (s10) {};
        \node [draw, fill, rectangle, minimum width=2mm, below=of s10] (s11) {};
        \node [draw, fill, rectangle, minimum width=2mm, below=of s11] (s12) {};
        \node [draw=sron1, fill=sron1, opacity=0.5, rectangle, fit=(s1)(s4)] {};
        \node [draw=sron2, fill=sron2, opacity=0.5, rectangle, fit=(s1)(s9)] {};
        \end{tikzpicture}}};

    \node (pieces) [right=10mmof es,label={[yshift=0mm]below:Pieces}] {
        \resizebox{!}{!}{\begin{tikzpicture}[node distance= 0mm]
        \node [draw=white, fill, rectangle, minimum width=2mm] (s1) {};
        \node [draw=white, fill, rectangle, minimum width=2mm, right=of s1] (s2) {};
        \node [draw=white, fill, rectangle, minimum width=2mm, right=of s2] (s3) {};
        \node [draw=white, fill, rectangle, minimum width=2mm, below=2mm of s1] (s4) {};
        \node [draw=white, fill, rectangle, minimum width=2mm, right=of s4] (s5) {};
        \node [draw=white, fill, rectangle, minimum width=2mm, right=of s5] (s6) {};
        \node [draw=white, fill, rectangle, minimum width=2mm, below=2mm of s4] (s7) {};
        \node [draw=white, fill, rectangle, minimum width=2mm, right=of s7] (s8) {};
        \node [draw=white, fill, rectangle, minimum width=2mm, right=of s8] (s9) {};
        \node [draw=white, fill, rectangle, minimum width=2mm, below=2mm of s7] (s10) {};
        \node [draw=white, fill, rectangle, minimum width=2mm, right=of s10] (s11) {};
        \node [draw=white, fill, rectangle, minimum width=2mm, right=of s11] (s12) {};
        \node [draw=sron2, fill=sron2, opacity=0.5, rectangle, fit=(s1)(s3)] {};
        \end{tikzpicture}}};

    \draw[->] ([yshift=0mm]seg.east) -- ([yshift=0mm]strip.west) node[left, midway] {};
    \draw[->] ([yshift=0mm]strip.east) -- ([yshift=0mm]es.west) node[left, midway] {};
    \draw[->] ([yshift=0mm]es.east) -- ([yshift=0mm]pieces.west) node[left, midway] {};

    \draw[->] (f1.south) -- (seg.north -| f1) node[right,midway,align=left] {1. Split into segments\\and encrypt};
    \draw[->] ([yshift=7.5mm,xshift=1mm]auth.east) -- ([yshift=7.5mm]net.west) node[above,pos=0.5] {2. Storage request};
    \draw[<-] ([yshift=2.5mm,xshift=1mm]auth.east) -- ([yshift=2.5mm]net.west) node[above,pos=0.5] {3. Storage nodes};
    \draw[->] ([yshift=0mm]seg.east) -- ([yshift=0mm]strip.west) node[above,pos=0.5,align=left] {4. Piece \\ preparation};
    \draw[->] ([yshift=-2.5mm,xshift=1mm]auth.east) -- ([yshift=-2.5mm]net.west) node[above,pos=0.5] {5. Pieces};
    \draw[->] ([yshift=-7.5mm,xshift=1mm]auth.east) -- ([yshift=-7.5mm]net.west) node[above,pos=0.5] {6. Pointer};

    \draw[dashed,->] ([yshift=0.5mm]auth.north) |- ([yshift=1mm]net.north) -| ([yshift=0.5mm]cauth.north) node[above,pos=0] {7. grant access};

    \draw[->] ([yshift=2.5mm,xshift=-1mm]cauth.west) -- ([yshift=2.5mm,xshift=-3mm]net.east) node[above,pos=0.5] {8. Request Pointer};
    \draw[<-] ([yshift=-2.5mm,xshift=-1mm]cauth.west) -- ([yshift=-2.5mm,xshift=-3mm]net.east) node[above,pos=0.5] {9. Pointer};
    \draw[->] ([yshift=-7.5mm,xshift=-3mm]net.east) -| ([xshift=-5mm]f2.west) node[above,pos=0.3] {10. Pieces} -- ([xshift=0mm]f2.west);
\end{tikzpicture}}
  \caption{Conceptual overview of Storj using (2,4) erasure coding.}
  \label{fig:storj}
  \vspace{-1.5em}
\end{figure}

A conceptual overview of Storj can be seen in~\cref{fig:storj}.
To upload the survey paper, the author needs to split it into segments,
which are then encrypted.
The author requests the satellite to store a segment.
The satellite checks capacity of the storage nodes
and returns a lists of adequate storage candidates.
The segment is then split into stripes, which are erasure encoded and arranged into pieces.
The pieces are then uploaded to the storage nodes in parallel.

For the erasure encoding,
Storj uses Reed-Solomon erasure codes~\cite{reed1960polynomial}.
For erasure codes the data is encoded in a $(k,n)$ erasure code.
This means, that an object is encoded into $n$ pieces,
in such a way that only $k$ pieces are necessary to recreate the object.
Storj chooses four values for each object: $k$, $m$, $o$, and $n$.
$k$ represents the minimum of required pieces to reconstruct the data,
$m$ is a buffer for repair, $o$ is a buffer for churn and
$n$ is the total number of pieces.
Erasure codes provide a higher redundancy with less overhead compared to
storing the pieces multiple times.
Furthermore, since only $k$ pieces are required to retrieve the file,
the latency till the file is available can be reduced.

After the upload, a pointer containing meta data of the segment (\eg, hash of pieces,
storage location, erasure encoding scheme) is returned to the satellite.
This is repeated for each segment and the last segment contains additional meta data about the survey.
For downloading the survey paper, the pointer for the segments are requested.
The pieces are requested in parallel from the storage nodes.
Once enough pieces to assemble the segments are gathered, the survey can be read.

To ensure the cooperation of the rational nodes, Storj
provides an incentive system.
The incentive system rewards storage nodes for storing and providing content.
Nodes are monitored with audits and evaluated via a reputation system.
A goal of Storj is low latency, which lead to avoiding
a blockchain dependent incentive mechanism.

\paragraph*{Discussion}
Storj employs some concepts that are unique when compared to other P2P data networks.
In particular, the Amazon S3 compatibility might promote
Storj as decentralized storage system.
The erasure codes add overhead to storing files, but during a file retrieval
only the necessary amount of pieces need to be downloaded.
The decentralization of storage, through the erasure codes, with adequate
storage node selection and the help of a reputation system
increases the protection against data breaches.

The satellite nodes are important parts of the network and partition the network,
since files available at one satellite are not available at another satellite.
This promotes centralization in form of the satellite.
While satellites cannot share the meta data with possible third parties due to the encryption,
it is still possible to leak access patterns.

While Storj is deployed and can indeed be used,
applications and research on the topic is rather rare.
\citeauthor{de2021exploring}~\cite{de2021exploring} analyzed the Storj network and
identified the satellite nodes as possible vectors for Denial-of-Service attacks.
They modified the implementation of storage node's connection handling and
successfully took down a satellite node in the test environment,
rendering payment and file retrieval impossible for some time.
However, the production system should be resistant to such an attack.
Another study also showed a different attack on data networks.
\citeauthor{zhang2019frameup}~\cite{zhang2019frameup} showed, in Storj v2.0,
the possibility to upload unencrypted data to storage nodes,
which can be used to frame owner's of storage nodes.
Nonetheless, Storj's provided privacy guarantees, resilience, acquirable meta data or
the possibility to deploy the different nodes by everyone
could provide valuable insights for cloud storage.

\subsection{Arweave}\label{sec:arweave}
The Arweave protocol~\cite{arweave2019williams} utilizes a blockchain-like structure, called blockweave,
to provide a mechanism for permanent on-chain data storage
as well as payment for storage.
In the blockweave, a block points to the direct preceding block and a recall block,
which is deterministically chosen based on the information of the previous block.
While the weave is immutable and provides censorship-resistance of its data,
every node can decide to refuse accepting content.
Refusing content by a sufficiently large amount of nodes prevents
inclusion of unwanted content.

Arweave utilizes Wildfire, a protocol similar to BitTorrent's tit-for-tat to rank peers.
Through Wildfire each node maintains a list of peers and scores and subsequently ranks the peers
based on their responsiveness, \eg, answering requests or send transactions.
The score is basically determined by received bytes per second averaged over recent requests.
High ranking and therefore best-performing peers receive messages first in parallel,
sequentially followed by the rest.
Connections to low ranking peers are periodically pruned.
This incentivizes nodes to be highly responsive themselves to receive messages as fast as possible.
Furthermore, it should optimize resource utilization of the nodes and reduce communication time.

\begin{figure}
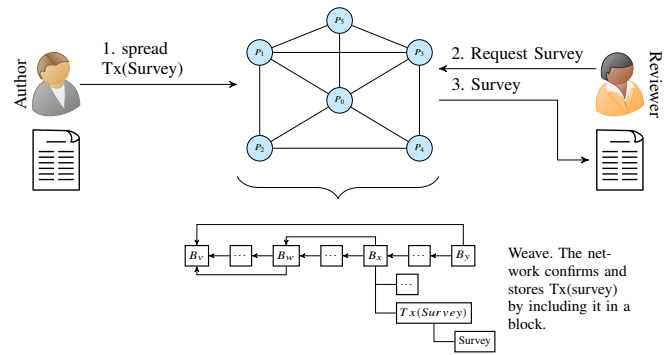

\footnotesize
    \resizebox{\columnwidth}{!}{
    \begin{tikzpicture}[node distance=25mm,auto,>=stealth']

    \node[bob, minimum height=10mm] (auth) {};
    \node[rotate=90,anchor=south] at (auth.west) {Author};
    \node[minimum height=10mm, below=1mm of auth] (f1) {\includegraphics[height=10mm]{survey}};

    \node (net) [minimum height=5mm, right=of auth] {
        \resizebox{30mm}{!}{
        \begin{tikzpicture}
            \tikzstyle{peer}=[circle, draw, thin,fill=cyan!20, minimum width=8mm]
            \node[peer] (p0) at (0,0) {$P_{0}$};
            \node[peer] (p1) at (-2.5,1.5) {$P_{1}$};
            \node[peer] (p2) at (-2.5,-1.5) {$P_{2}$};
            \node[peer] (p3) at (2.5,1.5) {$P_{3}$};
            \node[peer] (p4) at (2.5,-1.5) {$P_{4}$};
            \node[peer] (p5) at (0,2.5) {$P_{5}$};

            \path (p0) edge (p1);
            \path (p0) edge (p2);
            \path (p0) edge (p3);
            \path (p0) edge (p4);
            \path (p0) edge (p5);
            \path (p1) edge (p2);
            \path (p1) edge (p5);
            \path (p1) edge (p3);
            \path (p2) edge (p4);
            \path (p3) edge (p4);
            \path (p3) edge (p5);
        \end{tikzpicture}}};

    \node (chain) [minimum height=5mm, below=8mm of net,label={[yshift=0mm,align=left,text width=20mm,font=\scriptsize]right:Weave. The network confirms and stores Tx(survey) by including it in a block.}] {
    \resizebox{50mm}{!}{
    \begin{tikzpicture}[node distance=5mm,%
    grow via three points={one child at (0.5,-0.7) and
    two children at (0.5,-0.7) and (0.5,-1.4)},
    edge from parent path={(\tikzparentnode.south) |- (\tikzchildnode.west)}]
    \tikzstyle{every node}=[draw=black,thick,anchor=west]
    \node (u) {$B_{v}$};
    \node[right= of u] (v) {\dots};
    \node[right= of v] (w) {$B_{w}$};
    \node[right= of w] (x) {\dots};
    \node[right=of x] (d) {$B_{x}$}
        child { node {\dots}}
        child { node {$Tx(Survey)$}
            child { node {Survey}}};
    \node[right=of d] (y) {\dots};
    \node[right=of y] (f) {$B_{y}$};

    \draw[<-] (u.north) -- ([yshift=5.0mm]u.north) -| (f.north);
    \draw[<-] (w.north) -- ([yshift=2.0mm]w.north) -| (d.north);

    \draw[<-] (u.south) -- ([yshift=-2.0mm]u.south) -| (w.south);

    \draw[<-] (u.east) -- (v.west);
    \draw[<-] (v.east) -- (w.west);
    \draw[<-] (w.east) -- (x.west);
    \draw[<-] (x.east) -- (d.west);
    \draw[<-] (d.east) -- (y.west);
    \draw[<-] (y.east) -- (f.west);
    \end{tikzpicture}}};

    \node[alice, minimum height=10mm, right=of net] (cauth) {};
    \node[rotate=270,anchor=south] at (cauth.east) {Reviewer};
    \node[minimum height=10mm, below=1mm of cauth] (f2) {\includegraphics[height=10mm]{survey}};

    \draw [decorate,decoration={brace,amplitude=10pt,mirror,raise=4pt}]
        (net.south west) -- (net.south east);

    \draw[->] (auth.east) -- (net.west) node[above,pos=0.4,align=left] {1. spread \\ Tx(Survey)};
    \draw[->] ([yshift=2.5mm]cauth.west) -- ([yshift=2.5mm]net.east) node[above,pos=0.5, align=left] {2. Request Survey};
    \draw[->] ([yshift=-2.5mm]net.east) -| ([xshift=-5mm]f2.west) node[above,pos=0.2,align=left] {3. Survey} -- (f2.west);
    \end{tikzpicture}}
    \caption{Conceptual overview of Arweave.}
    \label{fig:arweave}
\end{figure}

At the heart, Arweave is a blockchain-based network.
While Wildfire introduces a ranking that favors certain connections,
it remains an unstructured P2P network.
A conceptual overview of Arweave and how to archive/retrieve a file
can be found in \cref{fig:arweave}.
To archive the survey paper in Arweave, it is necessary to send a transaction to the network.
The peers confirm the transaction by including it in a block.
If someone wants to read the survey the network is asked.
If a peer stores the block containing the survey,
it can be returned and the survey can be read.

Arweave's goal is to provide eternal permanent storage of data,
preserving and time-stamping information in an immutable way.
The data is stored on-chain on the blockweave, therefore, immutable and
only removable through forking the weave.
The blockweave provides decentralized storage for the permaweb.

Storage and maintenance of the blockweave and its data
is ensured through Arweave's cryptocurrency: Arweave tokens.
The tokens are used for rewarding miners and payment for sending transactions.

\paragraph*{Discussion}
The Arweave protocol provides on-chain storage on a blockchain-like structure.
This gives the storage similar advantages and disadvantages of a blockchain.
Arweave provides time-stamping, transparency, incentives, and immutable storage.
The data is stored through transactions providing pseudonymous authors of data.

One of the biggest problems of blockchains is the scalability.
Arweave tries to reduce these problems by utilizing blockshadows,
a mechanism similar to compact blocks,
explained in Bitcoin Improvement Proposal 152~\cite{bip152}, and Wildfire
for fast block propagation reducing fork probability.
Furthermore, the usage of Block Hash List and Wallet List
should reduce the initial cost of participation.
With version 2.0 Arweave introduced a hard fork to improve scalability,
decoupling data from transactions.
Instead of including the data in the transaction,
a Merkle root of the data is included.
This improves transaction propagation speed,
since the data is no longer necessary to forward the transaction.

Due to the pseudo-random recall block,
nodes are incentivized to store many blocks to maximize their mining reward.
This increases the replication of data.
However, not every node necessarily stores every block or content,
every node decides for itself based on content filter which data it stores.
Requesting content might become complicated,
since nodes are requested opportunistically in hope they store the content.

Research about Arweave directly is at most sparse.
However, this can be explained by the broad range of emerging blockchain-based protocols
and research about blockchains can be at least partly applied to Arweave as well.

\subsection{Honorable Mentions and Related Concepts}\label{sec:honorable}
Next to our detailed overview of select P2P data networks,
we provide additional literature on other systems and concepts
concerning the current generation of P2P data networks.
In particular, there are some paper concepts providing different and
sophisticated ideas for P2P content sharing.

Sia~\cite{vorick2014sia} aims to be a decentralized cloud storage platform.
A file is split into chunks, which are encrypted and then stored
via erasure coding on multiple storage nodes.
The location of chunks is stored as metadata.
Sia uses a blockchain to incentivize storage and retrieval of data.
The conditions for and duration of storing the data is fixed in storage contracts.
The data owner is responsible for file health.

The Open Storage Network~(OSN)~\cite{kirkpatrick2021open} is a distributed network for
transferring and sharing research data.
It is comparable to a distributed cloud service
dedicated to large amount of research data.
Data is stored in centrally monitored and maintained pods.
These OSN pods are specially configured server racks
and require a high bandwidth Internet connection.
Institutions that want to contribute to the network can house pods.
As a consequence, researchers can store and share their research data in the OSN network.
The connectivity of the OSN pods shall ensure fast access to the data.
Data can be shared with selected participants or via open access.
The central management and strict conditions for pods
differentiates the OSN from the rest of the presented data networks,
where decentralization and arbitrary participation is a key feature.

\begin{table*}
\footnotesize
\centering
\caption{Summary of the building blocks.}
\newcolumntype{R}{>{\raggedright\arraybackslash}X}
\begin{tabularx}{\textwidth}{llRRRRRRR}
\toprule
 & \textbf{Category} & \textbf{BitTorrent} & \textbf{IPFS/Filecoin} & \textbf{Swarm} & \textbf{Hypercore} & \textbf{SAFE} & \textbf{Storj} & \textbf{Arweave} \\
\midrule
%%%%%%%%%%%%%%%%%%%%%%%%%%%%%%%%%%%%%%%%%%%%%%%%%%%%%%%%%%%%%%%
\multicolumn{2}{l}{\textbf{Network}}\\
& Topology & Unstructured & Hybrid & Kademlia & Unstructured & Kademlia & Unstructured & Unstructured \\
\midrule
%%%%%%%%%%%%%%%%%%%%%%%%%%%%%%%%%%%%%%%%%%%%%%%%%%%%%%%%%%%%%%%
\multicolumn{2}{l}{\textbf{File Handling}}\\
& File Look-up & DHT, Central & DHT, Opportunistic & DHT & DHT & DHT & Central & Opportunistic \\
& Storage & File & Blocks & Chunks & Files & Chunks & Segments & Files \\
& Storage Location & Random & Random & Addressed & Random & Addressed & Random \\
& File Replication & Passive & Passive, Caching & Active/Passive, Caching & Passive & Active, Caching & -- & Passive \\
\midrule
%%%%%%%%%%%%%%%%%%%%%%%%%%%%%%%%%%%%%%%%%%%%%%%%%%%%%%%%%%%%%%%
\multicolumn{2}{l}{\textbf{Information Security}}\\
& Confidentiality & -- & -- & Manifests & Public-key & Self-authentication & Satellite nodes & -- \\
& Integrity & Meta-data file & Content-addressing & Content-addressing & Meta-data file & Content-addressing, self-encryption & Satellite nodes & Blockweave \\
& Availability & Replication, Incentives & Replication, Incentives & Replication, Erasure Codes, Incentives & Replication & Replication, Incentives & Erasure Codes, Incentives & Replication, Incentives \\
\midrule
%%%%%%%%%%%%%%%%%%%%%%%%%%%%%%%%%%%%%%%%%%%%%%%%%%%%%%%%%%%%%%%
\multicolumn{2}{l}{\textbf{Incentivization}} \\
& Upload & Free & Free & Charge & Free & Charge & Free & Charge \\
& Reward (Storing) & -- & For Time & For/Over Time & -- & -- & For Time & Over Time \\
& Punish (Storer) & -- & Misbehavior & Misbehavior & -- & -- & Misbehavior & -- \\
& Chunk/File Trade & Monitor & Monitor & Monitor & -- & -- & Monitor & Monitor \\
& Retrieval Only & Charge (optional) & Charge (optional) & Charge imbalance & -- & Reward & Charge & -- \\
\bottomrule
\end{tabularx}
\label{tab:summary}
\end{table*}

\citeauthor{fukumitsu2017proposal}~\cite{fukumitsu2017proposal} propose
a peer-to-peer-type storage system,
where even meta-data, necessary for reconstructing the stored files,
is stored in the network and can be retrieved
with an ID, a password, and a timestamp.
The authors assume an unstructured P2P network
where each node can offer different services.
Nodes broadcast regularly necessary information about themselves,
\eg, offered services and its IP address.
An important component of the scheme are storage node lists stored on a blockchain.
The storage node list is a randomly ordered list
of selected nodes offering storage services.
Data is stored in parts and the storage process is split into two phases:
storing user data and storing data necessary for reconstructing user data.
User data is encrypted, divided into parts and the parts are stored on
nodes selected from the currently available storage nodes.
The parts can be requested using restore keys.
For reconstructing user data the decryption key and
pairs of storage node and restore keys are necessary.
Therefore, data is replicated on other nodes.
A user creates an ID, password pair, and selects a storage list.
The data is encrypted with the hash of ID, password and storage list.
Storage nodes are chosen deterministically from the storage list.
The restore key for the parts is the hash of the storage list and
the hash of a piece index, the ID and password.
This scheme allows fetching data without storing information on the user device.

\citeauthor{jia2016oblivp2p}~\cite{jia2016oblivp2p}, propose \emph{OblivP2P}
a mechanism implementing ideas from oblivious RAM to hide data access patterns.
While the authors mention that their mechanism is applicable to other peer-to-peer systems,
they focus on a BitTorrent like system with a tracker.

\citeauthor{qian2016garlic}~\cite{qian2016garlic} propose Garlic Cast,
a mechanism for improving anonymity in an overlay network.
Peers do not request and search content directly.
Instead, a peer searches for proxies and the proxies exchange and request the content.
Messages between a peer and its proxy are exchanged
via a security-enhanced information dispersal algorithm (IDA).
An IDA is a form of erasure coding where $k$ of $n$ pieces
are sufficient to reconstruct the object.
The security-enhanced IDA first encrypts a message,
splits the message and key into $n$ fragments with a $k$-threshold IDA,
and sends cloves, messages containing a key and message fragment.
Proxies are discovered via random walks:
Cloves are send to its neighbors,
requesting peers to be a proxy with a random clove sequence number,
each neighbor randomly forwards the clove and maintains the state of successor and predecessor.
A peer with two cloves with the same sequence number can recover the request,
and if it volunteers to be a peer, it returns a reply to the requester.

Other paper concepts utilize a blockchain for access control and
to store data locations instead of a supplement as an incentive mechanism,
\eg Blockstack~\cite{ali2016blockstack}, which maintains meta-data on the blockchain
and relies on external data stores for actual storage of data.
There are also concepts using distributed ledger technologies for
access control \eg Calypso~\cite{kokoris2018calypso},
which uses a skipchain-based identity and access management
allowing auditable data sharing.
However, these systems and systems concentrating only on selling data via the blockchain
are outside of the scope of this survey.

\section{Discussion of Building Blocks}\label{sec:buildingblocks}
After gaining an initial understanding of each system,
we take a closer look at all systems, identifying similarities and distinct differences.
In this discussion, we also include BitTorrent
as prominent example from a previous generation of data networks.
By comparing these systems and reviewing literature on the topic,
we identify building blocks and open challenges in P2P data networks.
In particular, we identified the areas,
network architectures, file handling, information security, and
incentivization as most relevant technical aspects.
We take these building blocks and derive a taxonomy.
In \cref{tab:summary}, we provide a summary of building blocks.

\subsection{Network Architecture}\label{sec:netarchitecture}
Each of the considered data network builds an overlay network
to communicate with other peers.
An overlay network is a logical network of nodes on top of the real network.
While many ways exist to organize an overlay network~\cite{stoica2001chord,rowstron2001pastry},
we clearly see a dominance of Kademlia~\cite{maymounkov2002kademlia}.
Each network uses a Kademlia-based DHT one way or another;
this can result in two different overlay networks,
one for peer discovery and one for the data exchange.

Therefore, it is necessary that nodes can be identified.
Especially, in structured networks, where the identity can determine its neighbors.
In most cases, the networks use a self-determined key pair for identification.
The public key (or the hash of the public key) is then the node identifier.
In case of BitTorrent the node ID is a random string.
SAFE uses many different identities, however the network ID is given by the network.
In Swarm the identity is created from the network ID and an Ethereum address.

Despite using Kademlia, the networks are organized differently upon closer inspection.
IPFS, Swarm, and SAFE use the DHT also to structure the network.
SAFE, however, separates the network additionally in sections,
where each section organizes itself with so-called elders.
Swarm creates a Kademlia topology, where
the identity directly decides the neighbors.
SAFE and Swarm can therefore be classified as structured overlay networks.
While IPFS also uses a DHT, a peer connects to every peer it encounters,
which leads to an unstructured network.
If the number of connections exceeds a certain limit,
connections are pruned with the exception of
actively used or DHT required connections~\cite{henningsen2020mapping}.
This structures the IPFS network to some degree.
Storj used a DHT for peer discovery in the past.
Storage nodes decide how much resources are provided to a satellite and
with which satellite it cooperates.
Therefore, Storj replaced the initial peer discovery with
a direct communication of satellite and storage nodes.
Storage and satellite nodes maintain their own peer list.
Furthermore, cooperation between satellites and storage nodes,
is controlled with a reputation system.
In BitTorrent and Hypercore, the DHT is used for peer discovery only.
Once peers are discovered, a separate unstructured overlay network
is responsible for data exchange.
In BitTorrent, the connection between the peers
are decided based on tit-for-tat.

Arweave is an exception as it does not use a DHT at all.
Arweave uses a gossip protocol similar to Bitcoin,
where peers announce their neighbors and known addresses.
Concerning network organization, Arweave
has no strict structure for its neighbor selection,
although it uses Wildfire, a tit-for-tat based mechanism to rank peers
and drop connections from unresponsive/unpopular peers.

\begin{figure}
    \centering
    \resizebox{0.8\columnwidth}{!}{
    \begin{forest}, forked edges,
    for tree={grow=east,parent anchor=east,child anchor=west,
    draw, align=center, if n children=0 {fill=sron2!30}{fill=sron1!30}}
    [Network Architectures \\ (\Cref{sec:netarchitecture})
        [Unstructured
         [Arweave]
         [Storj]
         [Hypercore]
         [IPFS]
         [BitTorrent]
        ]
        [Structured
         [Chord, fill=sron1!30]
         [Kademlia
          [IPFS]
          [Swarm]
          [SAFE]
         ]
         [Pastry, fill=sron1!30]
        ]
     ]
  \end{forest}
  }

  \caption{Overview of the different network architectures.}
  \label{fig:network}
\end{figure}
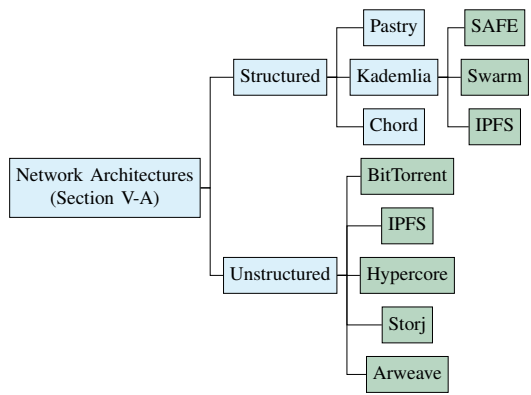

An overview of the presented categorization
with respect to the network architecture
is provided in \cref{fig:network}.

\subsection{File Handling and File Size}\label{sec:file}
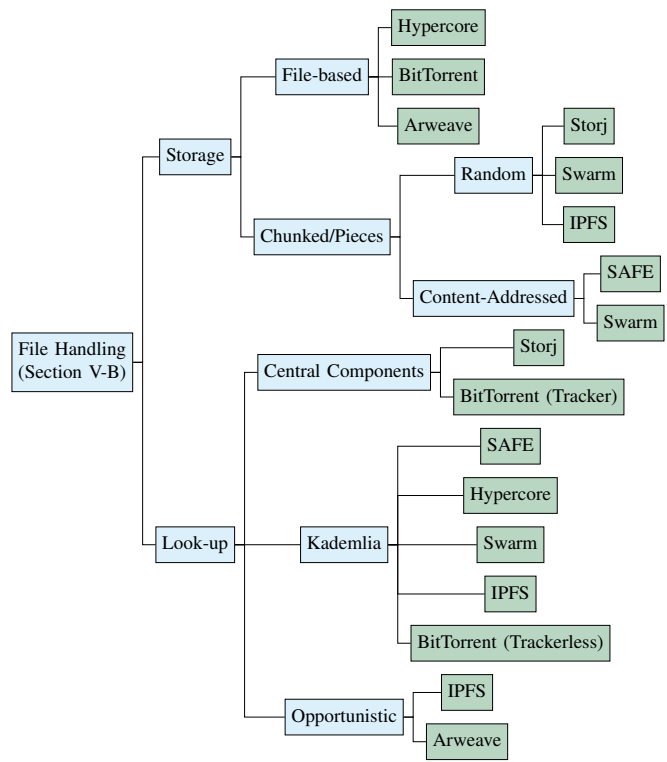
\begin{figure}
    \resizebox{\columnwidth}{!}{
    \begin{forest},forked edges,
    for tree={grow=east,parent anchor=east,child anchor=west,
    draw, align=center, if n children=0 {fill=sron2!30}{fill=sron1!30}}
    [File Handling \\ (\Cref{sec:file})
        [Look-up
         [Opportunistic
          [Arweave]
          [IPFS]
         ]
         [Kademlia
          [BitTorrent (Trackerless)]
          [IPFS]
          [Swarm]
          [Hypercore]
          [SAFE]
         ]
         [Central Components
           [BitTorrent (Tracker)]
           [Storj]
         ]
        ]
        [Storage
         [Chunked/Pieces
          [Content-Addressed
           [Swarm]
           [SAFE]
          ]
          [Random
           [IPFS]
           [Swarm]
           [Storj]
          ]
         ]
         [File-based
          [Arweave]
          [BitTorrent]
          [Hypercore]
         ]
        ]
     ]
  \end{forest}
  }
  \caption{Overview of file storage and look-up mechanisms.}
  \vspace{-1.5em}
  \label{fig:file}
\end{figure}

The file handling is another core component of a data network
and clearly more diverse than the network organization.
We provide an overview of our taxonomy in \cref{fig:file},
which we divide in storage and file look-up mechanisms.

A common pattern with respect to storage is that in each data network,
immutable files or at least immutable data blobs are preferred.
Mutability and intentional deletion of files
is rather a feature than the default.

Due to the respective protocol, the files are split into pieces
either during the exchange (BitTorrent, Hypercore) or
the file is stored in pieces located on potentially different devices.
Splitting files into pieces increases the storage overhead
due to additional meta data.
At the same time, though,
it improves the retrieval process in case of large files.
Arweave does not split files into pieces.
Instead, it uses transactions to store files,
which become part of a block in the blockweave.

While chunking is in general a common feature,
the storage is irregular.
BitTorrent and Hypercore concentrate
more on exchanging data than using the network
to store data on their behalf.
This results in a high probability
of all chunks being present on one device.
The storage is rather file-based since the aim
is the possession of all chunks to possess the file.

IPFS and Swarm split the files into pieces
and build a Merkle Tree/DAG.
The root is then sufficient to retrieve the file.
Each piece can be addressed and retrieved by itself
and individually stored on separate nodes.
In IPFS, the location of chunks is \enquote{random}
in the sense that each node can determine by itself,
if it stores a certain chunk.
In Swarm a chunk's storage location is tied to its address.
However, similar to IPFS other nodes can also decide to additionally store chunks.

SAFE splits the chunks into pieces and
encrypts the chunks with each other.
Similar to Swarm a chunk is content addressed and
the content decides the storage location.

Storj splits the files in erasure encoded pieces,
reducing the required trust in single nodes.
The storage location of the pieces is decided randomly
and distributed on the available storage nodes,
cooperating with the responsible satellite node.

The chunking of files also influences the look-up process.
The request is either referencing a chunk/file directly or
a chunk pointing to other chunks.
The chunks are in general retrieved from neighbors.
The request to neighbors can be directed or random via a broadcast.
In case of Arweave and IPFS, the file look-up can be considered opportunistic
as peers are queried without knowledge about the peers' possession of the chunks/file.
In Storj a central component is available to send direct requests.
In the other data networks, however, peers utilize a DHT for the look-up.
In IPFS the DHT is used as a backup look-up, if the opportunistic request fails.
Since in BitTorrent and Hypercore the exchange overlay network
deals with specific data,
we have to differentiate here:
a neighbor is expected to possess at least part of a file.
Therefore, the peer discovery can be considered as a directed request.
To this end, BitTorrent uses either a central component (\ie, a tracker)
or a DHT (\ie, trackerless).
Hypercore uses a DHT.

Due to the increasing amount of collected data,
it is also appropriate to consider
limits of the data networks concerning file sizes.
In BitTorrent, IPFS, and Hypercore data is stored on the data source's node first
and shared later,
while Swarm, SAFE, and Storj store chunks in the network directly.
This limits possible data sizes in the first case to a node's owned storage capacities and
in the latter case to the network's and peer's willingness to provide storage.
Furthermore, Storj assumes an object size of $4\,MB$ or more and files are erasure encoded.
Swarm splits data into $4\,kB$ chunks which are distributed in the network based on their hash,
which could make it difficult to retrieve large files.
In Arweave, the data is stored on-chain similarly limiting size to the network.
Arweave had a file size limit of $3\,MB$.
However, it was removed with the upgrade to version 2.1.

Hypercore is designed for large data sets and partial data sharing.
Partial data sharing can improve possible file sizes.
IPFS's Merkle DAG allows retrieval of partial data as well.
In contrast, BitTorrent's chunking, SAFE's self-encryption, and Storj's erasure codes
prevent partial data sharing.

For the reasons outlined before,
we believe that most data networks
are unsuitable for large, single datasets in the range of Petabytes
and are rather designed for data in the range of Megabytes and Gigabytes.
However, future performance measurements are necessary
to confirm or deny our reasoning.

\subsection{Information Security}
Confidentiality, integrity, and availability~(CIA)
are important aspects of information security.
These aspects provide additional challenges and
gain additional importance in the distributed setting of data networks.
In a distributed system where data
is potentially stored on different unsupervised devices,
it is hard to protect the data or control access to data.
Since the data comes from many untrusted devices,
the integrity needs to be guaranteed.
We can generally expect improved availability,
\eg, due to the redundant storage and distribution of data.
However, considering availability as long term file persistence remains a challenge.
Any node could delete content and arbitrary
join or leave the network,
which results in files becoming unavailable.

\subsubsection{Confidentiality}
To keep content and meta-data of data confidential from other participants
is difficult in a distributed environment.
Even nodes storing data are possible information leaks.
Encryption is the main instrument
to protect the data in distributed systems.
The encryption prevents other parties from reading
the content of files despite fetching or storing the data.
An additional protection against storage nodes is chunking of files.
By chunking the file and ideally distributing the chunks
on different nodes a storage node is unable to identify content.
Swarm, SAFE, and Storj distribute the chunks
during the storage process.
In the other data networks, the distribution is less prominent,
or in case of Arweave not present at all.

Another aspect which protects the content of data is access control.
Access control in the presented data networks
is mostly realized through distributing decryption keys.
The exchange of the decryption key is mainly handled
by the concerned parties directly outside of the data network.
BitTorrent, IPFS, and Arweave employ no additional access control.
However, some data networks also provide additional mechanisms.
In Storj, satellite nodes verify and authorize access requests.
Data access is additionally restricted by satellites,
where another satellite cannot grant access
to data submitted to another satellite.
SAFE uses self-authentication to authenticate access to private data.
Swarm provides access control through so-called manifests.
Swarm manifests provide mappings between paths and files to represent collections.
The manifest contains additionally to Swarm root hash, meta data like media mime type and access control parameter.
In Hypercore, it is necessary to know
the public key of the directory
for discovering peers and decrypting the communication.
This provides an additional distinction between write and read access.

\subsubsection{Integrity}
For the integrity of data, it is possible
to rely on and trust the data provider.
However, in a distributed system it is hard to trust all peers.
The presented data networks utilize
hash functions to ensure integrity.
The hash value has to be known in advance
and therefore might require out-of-band communication.
Given a hash and the algorithm used for the hash,
content can be verified by regenerating the hash
and comparing it with a given hash.

The usage of hash functions is different.
In BitTorrent and Hypercore,
the hash is provided by a file containing meta-data.
IPFS, Swarm, and SAFE use the hash for content-addressing,
meaning the content decides the address and
content is retrieved by their address.
Therefore, the acquired data can be directly verified.
Additionally, SAFE uses self-encryption,
where data is only restorable if it is the right data.
Storj relies on the satellite nodes,
which perform random audits on storage nodes utilizing hashes.
Furthermore, satellite and storage nodes are evaluated
with a reputation system to increase their credibility.
In Arweave, data is stored in a blockweave,
which is similar to a blockchain.
Each block confirms its predecessor by including a hash pointer
and therefore provides data integrity.

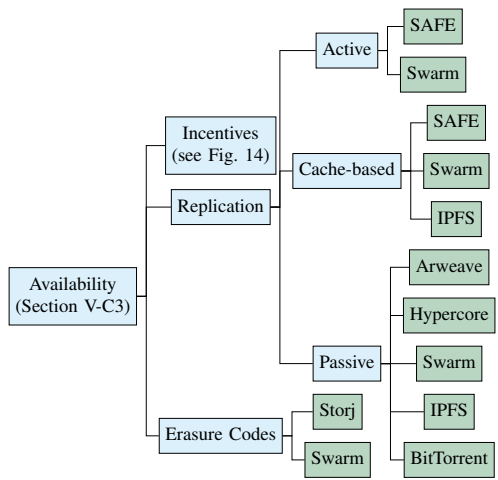
\begin{figure}
    \centering
    \resizebox{0.75\columnwidth}{!}{
    \begin{forest},forked edges,
    for tree={grow=east,parent anchor=east,child anchor=west,
    draw, align=center, if n children=0 {fill=sron2!30}{fill=sron1!30}}
    [Availability \\ (\Cref{sec:availability})
     [Erasure Codes
        [Swarm]
        [Storj]
     ]
     [Replication
      [Passive
       [BitTorrent]
       [IPFS]
       [Swarm]
       [Hypercore]
       [Arweave]
      ]
      [Cache-based
        [IPFS]
        [Swarm]
        [SAFE]
      ]
      [Active
        [Swarm]
        [SAFE]
      ]
     ]
     [Incentives\\(see \cref{fig:incentives}), fill=sron1!30]
    ]
  \end{forest}
  }
  \caption{Overview of availability mechanisms.}
  \label{fig:availability}
\end{figure}

\subsubsection{Availability}\label{sec:availability}
Due to node failure or maintenance, nodes can become unavailable,
eventually decreasing the availability of stored chunks.
Furthermore, node churn, \eg, due to unstable connections or Denial of Service attacks
can render service unavailable over a certain time period.
Therefore to improve availability,
multiple copies of chunks might be required.
Long term availability is a serious problem
of P2P systems in general.
The availability of content can be increased through
active, passive, and cache-based replication.
In \cref{fig:availability}, we provide an overview
of the different availability mechanisms used by data networks.
Popular content profits from cache-based replication,
which can happen naturally through requests and as an optimization.
Next to replication erasure codes can also increase the availability.
While they introduce a per chunk storage overhead,
files and missing chunks can be reconstructed without acquiring all chunks.
Incentive mechanisms can improve replication mechanisms and
ensure redundancy through monetary means.
Note, that we discuss incentivization in a separate section.

BitTorrent and Hypercore rely only on passive replication
and therefore volunteers hosting files.
Arweave's blockweave is utilizing passive replication,
ensuring replicas of blocks on the participants
and therefore the content.
However, every node can decide which content it stores
based on its content policies.
This means that not all content is available on all nodes.
IPFS uses cache-based replication,
additionally to the passive replication through pinning of chunks.
SAFE uses cache-based replication and
has data managers which are responsible
to actively maintain a few redundant copies of chunks.
Storj uses erasure codes instead of replication
providing a certain safety margin against segment loss.
Furthermore, the satellite nodes are responsible for auditing
storage nodes repairing files as necessary.
Swarm utilizes four methods:
erasure codes, passive replication through pinning,
cache-based replication, and
active replication with the nearest neighbor set.

Replication is a simple method to increase availability,
especially cache-based and passive replication rely on
volunteers and require no coordination.
Active replication requires some degree of coordination and communication
to ensure that if a peer leaves the network, a handover of the data is ensured.
Erasure codes are an alternative
(or as shown by Swarm)
an additional method to ensure replication.
Erasure codes introduce less overhead while ensuring a similar degree of availability.
However, erasure codes require coordination for sustaining the required amount of pieces
and are applied on a file level, which also removes some possibility
considering multiple files or file versioning.
In IPFS, blocks are possibly used by different files, which adds additional replication
and reduces overall storage requirements.
This would not be possible with erasure code as they work as a set.
Both methods, erasure coding and replication, have their advantages and disadvantages.
The presented data networks seem to prefer replication.

\begin{figure}
    \resizebox{\columnwidth}{!}{
    \begin{forest}, forked edges,
    for tree={grow=east,parent anchor=east,child anchor=west,
    draw, align=center, if n children=0 {fill=sron2!30}{fill=sron1!30}}
    [Incentives \\ (\Cref{sec:incentives})
        [Exchange
         [Retrieval
          [BitTorrent Token]
          [Filecoin]
          [SAFE]
          [Storj]
          [Arweave*]
         ]
         [Trade
          [BitTorrent*]
          [IPFS*]
          [Swarm]
         ]
        ]
        [Storage
         [Reward (Storer)
          [Continuous Storage
           [Arweave]
           [Storj]
           [Swarm, name=swarm0]
          ]
          [Storage Time Period, name=stp
           [Swarm]
           [Filecoin]
          ]
         ]
         [Punish (Storer)
           [Storj]
           [Filecoin]
           [Swarm]
         ]
         [Charge (Upload)
           [Swarm]
           [SAFE]
           [Arweave]
         ]
        ]
     ]
  \end{forest}

  }
  \caption{Overview of different incentive mechanisms
  (data networks marked with an asterisk do not use monetary incentivization in this category).}
  \label{fig:incentives}
\end{figure}
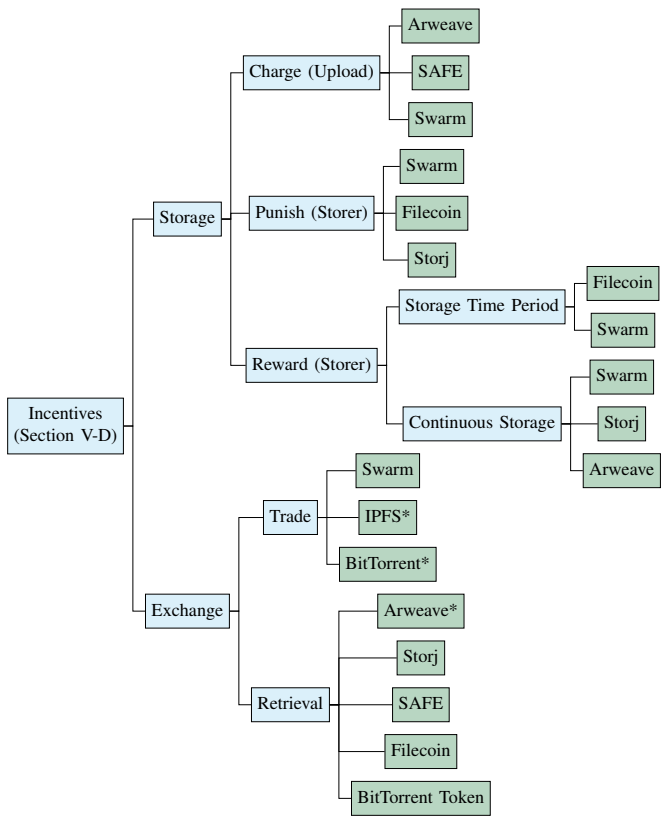

\subsection{Incentivization}\label{sec:incentives}
Incentives are crucial in open/public P2P networks to motivate compliant behavior.
Otherwise, we have to rely on altruism and benign peers.
In the presence of \enquote{selfish} or malicious peers,
this however might lead to an deteriorated data network.
Most of the presented data networks employ some kind of incentive mechanism.
An exception is Hypercore,
which does not employ an incentive mechanism and
is excluded from the following observation.
An overview of the different incentive mechanisms is provided in \cref{fig:incentives}.

One aspect of the incentive mechanism is compensation.
While actions can be rewarded or punished
with preferential treatment or depriving services,
the data networks employ their own additional compensation methods.
The compensation can be considered as a monetary incentive.
The data networks use cryptocurrencies or crypto-tokens,
which can be earned by or used to pay for services.
In BitTorrent, BitTorrent Token supplements the service.
The BitTorrent Token~\cite{bittorrentfoundation2019bittorrentbtt} is a TRC-10 utility token
of the TRON blockchain~\cite{tronfoundation2018tron}.
IPFS itself does not employ a currency.
But it uses Filecoin~\cite{protocollabs2017filecoin} to complement its protocol
to incentivize data reliability/availability.
Likewise, the other data networks use a cryptocurrency or token
one way or the other to compensate services.
Specifically, Swarm uses Ethereum (ether)~\cite{wood2014ethereum,tron2016swapswearswindle},
SAFE uses Safecoins~\cite{safeprimer},
Storj uses ERC-20 STORJ tokens~\cite{erc20token,storjlabs2018storj},
and Arweave~\cite{arweave2019williams} uses its own cryptocurrency.

Another aspect is the purpose of the incentive mechanism.
We observe two different incentive purposes:
promoting participation and increasing availability.
Participation is stimulated by regulating content retrieval.
In all presented data networks, peers
keep track of the exchanged data.
They can be further differentiated by
a trade relationship, where the received and send data are compared
and one sided observations, where peers are evaluated
based on retrieved data.

Except for SAFE, all presented data networks use reputation or monetary incentive
to prevent free-riding and promote active cooperation.
SAFE has a reputation system and
a certain reputation is necessary to actively participate in decisions.
However, concerning the exchange of file,
while SAFE rewards peers for answering request
it does not punish peers for slow responses or charge clients for reading/consuming bandwidth.
BitTorrent, IPFS, and Swarm compare send and received data.
BitTorrent punishes unresponsive, free-riding peers
by disconnecting from these peers, refusing further service.
Additionally, the BitTorrent Token can be used
to compensate peers which offer chunks.
Swarm similarly punishes uncooperative peers by disconnecting them,
however, Swarm also allows rebalancing the scale by
issuing cheques to peers compensating a lack of send pieces.
In IPFS, the Bitswap protocol ranks peers
based on send and received data.
Additionally, in Filecoin content retrieval is charged
and peers providing the content are compensated with filecoin.
Arweave monitors the responsiveness of peers, ranking the peers,
rewarding high ranking peers with preferential treatment.
In Storj, satellite nodes compensate storage nodes for the provided bandwidth.
Storj does not directly compensate the storage node
and instead cumulates the used bandwidth.

It is noticeable that the compensation of file retrievals,
in Filecoin, Swarm, and Storj is similar to a payment channel~\cite{poon2016bitcoin,mccorry2016towards},
\ie, a bilateral channel between two peers
used to exchange (micro-)payments instantaneously.
Payment channels are backed by a cryptocurrency but do not require
to commit every update to the blockchain and therefore
promise improved scalability.
Filecoin uses payment channels for the retrieval process,
files are retrieved in small pieces and each piece is compensated.
Swarm's chequebook contract behaves similar to a payment channel,
where off-chain payment can be cashed in at any point in time.
In Storj the bandwidth is monitored by allocating
a pre-determined amount of bandwidth.
The compensation of pieces in BitTorrent with BitTorrent Token also follows a payment channel.

The availability of files also benefits from the participation.
By compensating file retrieval, nodes gain an incentive
to cache files and answer requests.
However, long-term availability is also important.
Additionally, storing data on other devices might require
an additional incentive for peers to accept the content.
Therefore, the incentive mechanism sometimes
focuses on rewarding and punishing storage nodes.

IPFS's Filecoin, Swarm, Storj, and Arweave reward nodes storing data.
The reward is either for storing the data
over time or for a specific time period.
The time period is defined and nodes are pre- or postpaid,
misbehaving storage nodes are then punished or not compensated.
In IPFS's Filecoin, users rent specific storage for a time period.
In Swarm, storage guarantees are sold.
Swarm, Storj and Arweave reward nodes for
storing data over a long time without defined time constraints.
In Swarm, storage nodes can participate in a lottery,
if they store certain chunks and might be rewarded for the continued storage.
In Storj, storage nodes are compensated in time intervals for the data they stored during the interval,
in case of storage failures the reward is instead used for file repair compensating the new nodes.
In Arweave, the network is paid to store data for a long term.
When a node creates a new block,
proving storage of data, the node is compensated
for its continued provision of storage capacity.

Punishment of nodes is used to guarantee storage in case of prepaid storage.
If a node breaks its storage promises, it looses funds.
A missed audit in Filecoin or failing to proof storage in Swarm reduces
an escrow deposit of the storage node.
In Storj part of the payment to new storage nodes is used as an escrow
until the storage nodes gained enough reputation.
The escrow will be kept if the node leaves the network too early.
In Arweave, instead of punishing nodes, nodes can no longer be rewarded,
if they stop storing blocks.

SAFE and Swarm charge the initial upload of data.
However, this is a protection against arbitrary uploads
rather than an increase in availability.
Swarm finances the lottery with the upload fee.
In Arweave, the upload of data is paid with transaction fees.
Part of the fees go to the miner and part is kept by the network.

\begin{table}
\footnotesize
\centering
\caption{Overview of research on data networks.}
\begin{tabular}{llll}
\toprule
 & \textbf{Paper} & \textbf{System} & \textbf{Short Description} \\
\midrule
%%%%%%%%%%%%%%%%%%%%%%%%%%%%%%%%%%%%%%%%%%%%%%%%%%%%%%%%%%%%
\multicolumn{4}{l}{\textbf{Performance and Structure}}\\
& \cite{shen2019understanding} & IPFS & Read and write performance \\
& \cite{muralidharan2019interplanetary} & IPFS & Cluster IoT data sharing \\
& \cite{ascigil2019towards} & IPFS & Enhancing with ICN \\
& \cite{nyaletey2019blockipfs} & IPFS & Meta-Data storage on blockchain \\
& \cite{heinisuo2019asterism} & IPFS & On mobile devices \\
& \cite{henningsen2020mapping} & IPFS & Network mapping \\
& \cite{henningsen2020crawling} & IPFS & Network crawler \\
& \cite{de2021accelerating} & IPFS & Improving Bitswap \\
& \cite{guidi2021data} & IPFS & Node identity, data availability \\
\midrule
%%%%%%%%%%%%%%%%%%%%%%%%%%%%%%%%%%%%%%%%%%%%%%%%%%%%%%%%%%%%
\multicolumn{4}{l}{\textbf{Confidentiality and Access Control}}\\
& \cite{wang2018blockchain} & IPFS & Blockchain-based, encryption \\
& \cite{steichen2018blockchain} & IPFS & Blockchain-based, modified client \\
& \cite{khatal2020fileshare} & IPFS & Blockchain-based, modified application \\
& \cite{hoang2020privacy} & IPFS & Blockchain-based, encryption \\
& \cite{battah2020blockchain} & IPFS & Blockchain-based, encryption \\
& \cite{politou2020delegated} & IPFS & Delegated content erasure \\
\midrule
%%%%%%%%%%%%%%%%%%%%%%%%%%%%%%%%%%%%%%%%%%%%%%%%%%%%%%%%%%%%
\multicolumn{4}{l}{\textbf{Security and Privacy}}\\
& \cite{patsakis2019hydras} & IPFS & Botnet coordination \\
& \cite{prunster2020total} & IPFS & Eclipse attack \\
& \cite{karapapas2020ransomware} & IPFS & Ransomware \\
& \cite{balduf2021monitoring} & IPFS & Monitor data request \\
& \cite{paul2014security} & SAFE & CIA and possible attacks \\
& \cite{jacob2015security} & SAFE & Security analysis \\
& \cite{de2021exploring} & Storj & Denial-of-Service attack \\
& \cite{zhang2019frameup} & Storj & Storing unencrypted data \\
\bottomrule
\end{tabular}
\label{tab:research}
\end{table}

\section{Research Areas and Open challenges}\label{sec:goals}
Previous generation of data networks had different network architectures, structured and unstructured,
and used an incentive mechanism mainly to promote cooperation and prevent
uncooperative behavior, \eg, free-riders,
mainly with reputation systems~\cite{androutsellis2004survey}.
Other incentive structures were also explored.
The next generation uses mainly Kademlia-based architectures,
and employs an incentive structure to increase availability and long term persistence.

The previous generation already faced some challenges,
which still apply to the next generation data networks.
In 2005, \citeauthor{hasan2005survey}~\cite{hasan2005survey} identified certain challenges
that peer-to-peer systems have to overcome to gain acceptance for real-life scenarios.
This includes deployment, naming, access control, DDoS attack protections,
preventing junk data, and churn protection.
We observe that the next generation data networks address these problems
and provide possible solutions.
However, the degree of maturity,
the interaction with other mechanism, and the adoption rate need more consideration.
In the literature review for the search of current generation data networks,
we found a large body of literature utilizing or analyzing IPFS.
Analyses of other systems are at most sparse.
One reason could be lack of actual deployment, small user base or lack of implementation.
Another reason, which this survey tries to address,
is in our opinion a lack of concise and structured documentation.
Some of the presented systems make it hard
to get into the system, understand the concepts and show that the system is valid.

We observe five main challenges of data networks,
which provide new opportunities for research:
performance, confidentiality and access control, security, anonymity, and naming.
An overview of existing research can be found in \cref{tab:research}.

\subsection{Performance}
A research direction which is already pursued by some researchers is the performance of the systems.
Investigating the performance, read/write times, storage overhead, file look-up, churn resistance
through simulations or tests,
can be used to identify new use cases and fortify claims that a system might replace centralized counterparts.
The IPFS developers developed \enquote{Testground}~\cite{gittestground} for testing and benchmarking P2P systems at scale.
In that sense, the performance of Testground and its ability to replicate real systems,
is also an area worthy to be researched.
There exists other research analyzing the performance of IPFS,
\eg, the read and write latency~\cite{shen2019understanding,nyaletey2019blockipfs},
using IPFS cluster for Internet of Things data sharing~\cite{muralidharan2019interplanetary},
improving the system~\cite{ascigil2019towards,de2021accelerating},
or analyzing the network~\cite{henningsen2020mapping,henningsen2020crawling,guidi2021data}.
\citeauthor{heinisuo2019asterism}~\cite{heinisuo2019asterism} showed that IPFS
needed improvement to be used on mobile devices due to
high network traffic draining the battery.
Research concerning IPFS's competitors is lacking.
Additionally, \citeauthor{naik2020next}~\cite{naik2020next} focus on the topic of churn in P2P networks.
Furthermore, considering the increase of research data,
the feasibility of data networks for large, single datasets in the range of Tera- and Petabytes should be investigated.

\subsection{Confidentiality and Access Control}
The past and present generation of data networks
provide some confidentiality and access control,
but the systems are rather designed for public data than private data.
The knowledge gained of nodes while storing data needs to be researched.
This concerns not only information about the content of data but also meta-data like access patterns.
The security of the existing access control needs to be investigated.
There are research proposals for access control
with blockchains~\cite{wang2018blockchain,steichen2018blockchain,battah2020blockchain,khatal2020fileshare,hoang2020privacy},
however the immutability of blockchains makes this questionable for private and personal data.
Another aspect concerning private data is deleting data.
While it is useful for censorship-resistance to prevent deletion of data,
the possibility to delete personal, malicious or illegal data might raise acceptance of data networks.
For example, \citeauthor{politou2020delegated}~\cite{politou2020delegated} propose a mechanism for deleting content in IPFS.
Investigating and improving the existing systems increases the trust in data networks.
An increased trust in the confidentiality and the protection from unwarranted access
can open these systems for storing private and personal data.

\subsection{Security}
As typical for security research,
work in this area is in a constant back and forth between
finding and fixing new vulnerabilities.
In addition, research is also concerned with malicious activities
using P2P data networks to exchange data with malware~\cite{patsakis2019hydras,karapapas2020ransomware}.

With respect to security vulnerabilities,
\citeauthor{prunster2020total}~\cite{prunster2020total} disclose an eclipse attack on IPFS and
\citeauthor{de2021exploring}~\cite{de2021exploring} showed a Denial-of-Service attack on Storj's test network.
Furthermore, it is not only necessary to investigate known attack vectors,
but also to investigate the existence of new attack vectors.
For example, Storj acknowledges the possibility of an \enquote{Honest Geppetto} attack,
where an attacker operates many storage nodes (honestly) for a long time,
effectively controlling a large part of the storage capabilities.
This control allows taking data \enquote{hostage} or taking down the data in general,
rendering the data network inoperable.
Another example is Frameup~\cite{zhang2019frameup}, where unencrypted data
is stored on storage nodes, which could lead to legal issues.
Storing arbitrary data might also pose a risk to the storage device.
Security is the research area where we observe research beyond IPFS.

\subsection{Anonymity}
Next to confidentiality, which concerns data security and privacy,
protecting the privacy of individuals is another relevant aspect;
in particular, anonymity, which describes the inability
to identify an individual in a group of individuals,
\ie, unlinkability~\cite{pfitzmann2000anon}.

With respect to anonymity, various entities can be protected in data networks:
the content creator, the storage node, and the user requesting content.
Of the previous generation data networks,
especially Freenet~\cite{clarke2000freenet}
and GNUnet~\cite{bennett2002gnunet} focused on protecting
the identity of the different entities.
\citeauthor{balduf2021monitoring}~\cite{balduf2021monitoring} already showed for IPFS
the continued existance of privacy problems by identifying
content requesters through monitoring data requests.

Due to the incentive mechanisms and the resulting charge of individuals
it is hard to guarantee anonymity as at least pseudonyms are required.
As soon as the incentive mechanism is used,
information about the requester is gained.
A distributed ledger recording transactions,
\eg, Filecoin, Ethereum Swarm, Arweave,
can reveal additional information and as a result participants are pseudonymous.
When a central component authorizes requests and deals with incentivization,
\eg, satellite nodes in Storj, requester, storage node and central component know each other.
In case of incentivizing requests, the requesting node and storage nodes are revealed.
The identity of requesters can be partly secured via forwarding strategies or proxies, \eg, Swarm, SAFE.

The first generation had systems like Freenet
which aimed for anonymity and censorship-resistance.
The anonymity of the current generation seems to fall behind the first generation.
Despite advances in anonymous communication with mixnets or Tor~\cite{dingledine2004tor},
there are no data networks providing strong anonymity.
Altogether, the provided anonymity guarantees
and further enhancements need to be investigated.
This includes the anonymity-utility trade-off and an analysis of different attacker models.
Anonymity is not only important to protect the privacy of individuals,
but is also important to guarantee the claimed censorship-resistance.
If the identity of storage nodes can be easily inferred it is possible that,
even though the network protects against deletion, law enforcement can enforce the censorship.
This is a concern especially for systems like Swarm,
where the location of a stored chunk is predetermined
and node identity is linked to Ethereum pseudonyms.

\subsection{Naming}
Naming, in particular providing human-readable names in a distributed system, is a known challenge.
The problem and its adjacent challenges is captured by Zooko's Triangle~\cite{zookotriangle}.
It describes the difficulty of building a distributed namespace,
which is distributed (without a central authority),
secure (clear-cut resolution), and human-readable.

In all systems, the addressing of data
lacks either distribution (tracker-based BitTorrent and Storj) or human-readability
(trackerless BitTorrent, Hypercore, IPFS, Swarm and SAFE).
BitTorrent is a good example where the tracker is a central authority
and in the case of trackerless BitTorrent the human-readable torrent is
addressed with the not so readable infohash (hash of the torrent).
In the v3.0 of Storj, the satellite is a central component.

The lack of human-readability is a result of self-certifying data,
where the data determines the address or the name of the data.
If the data is changed the address changes.
Therefore, human-readability is supported through a different mechanism,
a naming independent of the content.
An exception is Hypercore.
In Hypercore, the data group is bound to the public key and
the mutability inside the group is secured through versioning.

One solution to provide human-readability is name resolution.
Name resolution allows the mapping of keys to self-certifying content.
The name resolution can provide human-readability and
provide support for versioning of files.
However, due to the possibility of updating the value
and delays in propagation one could argue that security is violated,
even if the key is unique.
Independent of Zooko's Triangle, the name resolution
announces content and gives ambiguous character strings meaning
and should only be used for public data,
unless the name resolution provides access control.

To this end, IPFS, Swarm, and SAFE provide some kind of naming service.
In fact, IPFS provides two naming services, IPNS and DNSLink,
which are used for different purposes.
IPNS is used for mapping the hash of a public key to an IPFS CID, allowing mutable data.
DNSLink uses DNS TXT records for mapping domain names to an IPFS address.

Swarm also provides two naming systems: single-owner chunks and ENS~\cite{eip137}.
Single-owner chunks provide a data identification based on an owner and an identifier,
providing a secure, non human-readable key with an updatable value.
The Ethereum Name System is similar to DNS, where a record is mapped to an address.

\citeauthor{squarezooko}~\cite{squarezooko} argued that a blockchain-based
name service provides all three properties of Zooko's triangle.
Anybody can register the name on the blockchain providing decentralization,
the name can be anything providing human-readability, and
the tamperproof ledger ensures unique names providing security.
Following this line of argument, systems like Namecoin,
Blockstack~\cite{ali2016blockstack}, and ENS,
which adopt the idea of a blockchain-based name system, are developed.
Although these systems exist, except for Swarm with ENS,
none of the systems seem to provide a solution for Zooko's triangle.
However, due to the lack of transaction finality and possible blockchain forks,
it could be argued that blockchain-based systems violate strong security aspects
and only provide eventual security.

\section{Conclusion and Lessons Learned}\label{sec:conclusion}
The first generation of P2P data networks taught us that
P2P-driven file exchange works and has some major advantages, \eg, self-scalability.
Another indicator for the persistence of this technology is
BitTorrent's continued existence and wide user base.
The first generation however also taught us weaknesses, \eg, a lack of long term availability.
The next generation data networks builds upon and improved the previous generation,
taking advantage of technological advancements and concepts to address the weaknesses.

In this survey paper, we studied this emerging new generation of P2P data networks.
In particular, we investigated new developments and technical building blocks.
From our qualitative comparison, we can conclude that except for the overlay structure
the various data networks explore different solutions with respect to
file management, availability, and incentivization.
Most notably, explicit incentive mechanisms,
\eg, using a cryptocurrency or some sort of token, seem to be ubiquitous
to ensure long-term availability and the participant's engagement.
We also see different measurements to ensure availability in the face
of Denial-of-Service attacks or churn beyond incentive mechanisms,
\ie, replication, erasure codes, or even a combination of both.
Moreover, since many systems combine naming services and content addressing
in a distributed architecture, they have the potential to reconcile
the system properties of human readability, security, and decentrality
as conjured by Zooko's triangle.

An important open task is now to investigate and evaluate the various building blocks.
Especially, incentive mechanisms are notoriously difficult to design.
To some extent, we can consider the different deployments of P2P data networks
as a large field test where we can observe the effects of certain design decisions.
In general, P2P data networks have become part of the research agenda,
either as a basis for other applications or as research object itself.

Yet, many challenges and open research questions remain,
\eg, investigating anonymity, participant's privacy and access control,
opening P2P data networks to a wider range of possible use cases.
We therefore believe that this new generation of P2P data networks
provide many exciting future research opportunities.

\IEEEtriggeratref{90}
\printbibliography[heading=bibintoc]

@INPROCEEDINGS{hasan2005survey,
  TITLE = {A Survey of Peer-to-Peer Storage Techniques for Distributed File Systems},
  AUTHOR = {Hasan, Ragib and Anwar, Zahid and Yurcik, William and Brumbaugh, Larry and Campbell, Roy},
  VOLUME = {2},
  PAGES = {205--213},
  CROSSREF = {itcc05}
}

@INPROCEEDINGS{thanh2008taxonomy,
  TITLE = {A Taxonomy and Survey on Distributed File Systems},
  AUTHOR = {Thanh, Tran Doan and Mohan, Subaji and Choi, Eunmi and Kim, SangBum and Kim, Pilsung},
  VOLUME = {1},
  PAGES = {144--149},
  CROSSREF = {ncm08}
}

@INPROCEEDINGS{ashraf2019comparative,
  TITLE = {Comparative Analysis of Unstructured {P2P} File Sharing Networks},
  AUTHOR = {Ashraf, Fasiha and Naseer, Ateeqa and Iqbal, Shaukat},
  PAGES = {148--153},
  CROSSREF = {icisdm19}
}

@ARTICLE{huang2020blockchain,
  TITLE = {When Blockchain Meets Distributed File Systems: An Overview, Challenges, and Open Issues},
  AUTHOR = {Huang, Huawei and Lin, Jianru and Zheng, Baichuan and Zheng, Zibin and Bian, Jing},
  JOURNAL = {IEEE Access},
  VOLUME = {8},
  PAGES = {50574--50586},
  YEAR = {2020},
  PUBLISHER = {IEEE}
}

@article{benisi2020blockchain,
  TITLE = {Blockchain-based Decentralized Storage Networks: A Survey},
  AUTHOR = {Benisi, Nazanin Zahed and Aminian, Mehdi and Javadi, Bahman},
  JOURNAL = {Journal of Network and Computer Applications},
  VOLUME = {162},
  PAGES = {102656},
  YEAR = {2020},
  PUBLISHER = {Elsevier}
}

@article{androutsellis2004survey,
  TITLE = {A survey of peer-to-peer content distribution technologies},
  AUTHOR = {Androutsellis-Theotokis, Stephanos and Spinellis, Diomidis},
  JOURNAL ={{ACM} Computing Surveys},
  VOLUME = {36},
  NUMBER = {4},
  PAGES = {335--371},
  YEAR = {2004}
}

@article{casino2019immutability,
  TITLE = {Immutability and Decentralized Storage: An Analysis of Emerging Threats},
  AUTHOR = {Casino, Fran and Politou, Eugenia and Alepis, Efthimios and Patsakis, Constantinos},
  JOURNAL ={{IEEE} Access},
  VOLUME = {8},
  PAGES = {4737--4744},
  YEAR = {2019}
}

@ARTICLE{naik2020next,
  TITLE = {Next level peer-to-peer overlay networks under high churns: a survey},
  AUTHOR = {Naik, Ashika R and Keshavamurthy, Bettahally N},
  JOURNAL = {Peer-to-Peer Networking and Applications},
  VOLUME ={13},
  NUMBER = {3},
  PAGES = {905--931},
  YEAR = {2020}
}

@INPROCEEDINGS{cohen2003incentives,
  TITLE = {Incentives build robustness in BitTorrent},
  AUTHOR = {Cohen, Bram},
  PAGES = {68--72},
  CROSSREF = {p2pecon03}
}

@INPROCEEDINGS{pouwelse2005bittorrent,
  TITLE = {The Bittorrent {P2P} File-Sharing System: Measurements and Analysis},
  AUTHOR = {Pouwelse, Johan and Garbacki, Pawel and Epema, Dick and Sips, Henk},
  PAGES = {205--216},
  CROSSREF = {iptps05}
}

@INPROCEEDINGS{bharambe2006analyzing,
  TITLE = {Analyzing and Improving a Bittorrent Networks Performance Mechanisms},
  AUTHOR = {Bharambe, Ashwin R and Herley, Cormac and Padmanabhan, Venkata N},
  PAGES = {1--12},
  CROSSREF = {infocom06}
}

@ARTICLE{xia2010survey,
  TITLE = {A Survey of BitTorrent Performance},
  author = {Xia, Raymond Lei and Muppala, Jogesh K},
  JOURNAL = {{IEEE} Communications Surveys and Tutorials},
  VOLUME = {12},
  NUMBER = {2},
  PAGES = {140--158},
  YEAR = {2010}
}

@TECHREPORT{bittorrentfoundation2019bittorrentbtt,
 TITLE = {BitTorrent (BTT) White Paper},
 AUTHOR = {{BitTorrent Foundation}},
 INSTITUTION = {BitTorrent Foundation},
 DATE = {2019-02}
}

@INPROCEEDINGS{ramanathan2020quantifying,
  title={Quantifying the Impact of Blocklisting in the Age of Address Reuse},
  author={Ramanathan, Sivaramakrishnan and Hossain, Anushah and Mirkovic, Jelena and Yu, Minlan and Afroz, Sadia},
  PAGES = {360--369},
  CROSSREF = {imc20}
}

@MISC{bep005,
  TITLE={bep\_0005.rst\_post},
  AUTHOR = {Loewenstern, Andrew and Norberg, Arvid},
  howpublished={\url{http://bittorrent.org/beps/bep_0005.html}},
  note={Accessed: 2021-06}
}

@TECHREPORT{benet2014ipfs,
  TITLE = {{IPFS} - Content Addressed, Versioned, {P2P} File System (DRAFT 3)},
  AUTHOR = {Benet, Juan},
  INSTITUTION = {Protocol Labs},
  DATE = {2014-07}
}

@TECHREPORT{protocollabs2017filecoin,
  TITLE = {Filecoin: A Decentralized Storage Network},
  AUTHOR = {{Protocol Labs}},
  INSTITUTION={Protocol Labs},
  DATE = {2017-07}
}

@TECHREPORT{benet2017porep,
  TITLE = {Proof of Replication},
  AUTHOR = {Benet, Juan and Dalrymple, David and Greco, Nicola},
  INSTITUTION={Protocol Labs},
  DATE = {2017-07}
}

@ARTICLE{patsakis2019hydras,
  TITLE = {Hydras and {IPFS:} A Decentralised Playground for Malware},
  AUTHOR = {Patsakis, Constantinos and Casino, Fran},
  JOURNAL = {International Journal of Information Security},
  VOLUME = {18},
  NUMBER = {6},
  PAGES = {787--799},
  YEAR = {2019}
}

@INPROCEEDINGS{karapapas2020ransomware,
  TITLE = {Ransomware as a Service using Smart Contracts and IPFS},
  AUTHOR = {Karapapas, Christos and Pittaras, Iakovos and Fotiou, Nikos and Polyzos, George C},
  PAGES = {1--5},
  CROSSREF = {icbc20}
}

@INPROCEEDINGS{shen2019understanding,
  TITLE = {Understanding {I/O} Performance of {IPFS} storage: a client's perspective},
  AUTHOR = {Shen, Jiajie and Li, Yi and Zhou, Yangfan and Wang, Xin},
  PAGES = {17:1--17:10},
  CROSSREF = {iwqos19}
}

@INPROCEEDINGS{muralidharan2019interplanetary,
  TITLE = {An InterPlanetary File System {(IPFS)} based IoT framework},
  AUTHOR = {Muralidharan, Shapna and Ko, Heedong},
  PAGES = {1--2},
  CROSSREF = {icce19}
}

@INPROCEEDINGS{henningsen2020mapping,
  TITLE = {Mapping the Interplanetary Filesystem},
  AUTHOR = {Henningsen, Sebastian and Florian, Martin and Rust, Sebastian and Scheuermann, Bj{\"o}rn},
  PAGES = {289--297},
  CROSSREF = {networking2020}
}

@INPROCEEDINGS{henningsen2020crawling,
  TITLE = {Crawling the {IPFS} Network},
  AUTHOR = {Henningsen, Sebastian and Rust, Sebastian and Florian, Martin and Scheuermann, Bj{\"o}rn},
  PAGES = {679--680},
  CROSSREF = {networking2020}
}

@ARTICLE{balduf2021monitoring,
  TITLE = {Monitoring Data Requests in Decentralized Data Storage Systems: A Case Study of {IPFS}},
  author={Balduf, Leonhard and Henningsen, Sebastian and Florian, Martin and Rust, Sebastian and Scheuermann, Bj{\"o}rn},
  journal={arXiv preprint arXiv:2104.09202},
  year={2021}
}

@INPROCEEDINGS{guidi2021data,
  TITLE = {Data Persistence in Decentralized Social Applications: The IPFS approach},
  AUTHOR = {Guidi, Barbara and Michienzi, Andrea and Ricci, Laura},
  PAGES = {1--4},
  CROSSREF = {ccnc21}
}

@INPROCEEDINGS{confais2017object,
  TITLE = {An Object Store Service for a Fog/Edge Computing Infrastructure based on {IPFS} and a Scale-out {NAS}},
  AUTHOR = {Confais, Bastien and Lebre, Adrien and Parrein, Beno{\^\i}t},
  PAGES={41--50},
  CROSSREF = {icfec17}
}

@INPROCEEDINGS{ali2017iot,
  TITLE = {IoT Data Privacy via Blockchains and {IPFS}},
  AUTHOR = {Ali, Muhammad Salek and Dolui, Koustabh and Antonelli, Fabio},
  PAGES={14:1--14:7},
  CROSSREF = {iot17}
}

@INPROCEEDINGS{norvill2018ipfs,
  TITLE = {{IPFS} for Reduction of Chain Size in Ethereum},
  AUTHOR = {Norvill, Robert and Pontiveros, Beltran Borja Fiz and State, Radu and Cullen, Andrea},
  PAGES = {1121--1128},
  CROSSREF = {ithingsgreenvomcpscomsmartdata18}
}

@INPROCEEDINGS{ascigil2019towards,
  TITLE = {Towards Peer-to-Peer Content Retrieval Markets: Enhancing {IPFS} with {ICN}},
  AUTHOR = {Ascigil, Onur and Re{\~{n}}{\'e}, Sergi and Kr{\'{o}}l, Michal and Pavlou, George and Zhang, Lixia and Hasegawa, Toru and Koizumi, Yuki and Kita, Kentaro},
  PAGES = {78--88},
  CROSSREF = {icn2019}
}

@INPROCEEDINGS{heinisuo2019asterism,
  TITLE = {Asterism: Decentralized File Sharing Application for Mobile Devices},
  AUTHOR = {Heinisuo, Olli-Pekka and Lenarduzzi, Valentina and Taibi, Davide},
  PAGES = {38--47},
  CROSSREF = {mobilecloud19}
}

@ARTICLE{xu2019healthchain,
  TITLE = {Healthchain: A Blockchain-Based Privacy Preserving Scheme for Large-Scale Health Data},
  AUTHOR = {Xu, Jie and Xue, Kaiping and Li, Shaohua and Tian, Hangyu and Hong, Jianan and Hong, Peilin and Yu, Nenghai},
  JOURNAL = {{IEEE} Internet of Things Journal},
  VOLUME = {6},
  NUMBER = {5},
  PAGES = {8770--8781},
  YEAR = {2019}
}

@INPROCEEDINGS{kumar2020distributed,
  TITLE = {Distributed Off-Chain Storage of Patient Diagnostic Reports in Healthcare System using {IPFS} and Blockchain},
  AUTHOR = {Kumar, Randhir and Marchang, Ningrinla and Tripathi, Rakesh},
  PAGES = {1--5},
  CROSSREF = {comsnets20}
}

@ARTICLE{hao2018safe,
  TITLE = {A Safe and Efficient Storage Scheme Based on Blockchain and IPFS for Agricultural Products Tracking},
  AUTHOR = {Hao, JinTao and Sun, Yan and Luo, Hong},
  JOURNAL = {Journal of Computers},
  VOLUME = {29},
  NUMBER = {6},
  PAGES = {158--167},
  YEAR = {2018}
}

@INPROCEEDINGS{alam2016interplanetary,
  TITLE = {Interplanetary Wayback: The Permanent Web Archive},
  AUTHOR = {Alam, Sawood and Kelly, Mat and Nelson, Michael L},
  PAGES = {273--274},
  CROSSREF = {jcdl16}
}

@INPROCEEDINGS{tenorio2019towards,
  TITLE = {Towards a Decentralized Process for Scientific Publication and Peer Review using Blockchain and {IPFS}},
  AUTHOR = {Tenorio-Forn{\'e}s, Antonio and Jacynycz, Viktor and Llop-Vila, David and S{\'a}nchez-Ruiz, Antonio and Hassan, Samer},
  PAGES = {1--10},
  CROSSREF = {hicss19}
}

@INPROCEEDINGS{steichen2018blockchain,
  TITLE = {Blockchain-Based, Decentralized Access Control for {IPFS}},
  AUTHOR = {Steichen, Mathis and Fiz, Beltran and Norvill, Robert and Shbair, Wazen and State, Radu},
  PAGES = {1499--1506},
  CROSSREF = {ithingsgreenvomcpscomsmartdata18}
}

@ARTICLE{battah2020blockchain,
  TITLE = {Blockchain-based Multi-Party Authorization for Accessing IPFS Encrypted Data},
  AUTHOR = {Battah, Ammar Ayman and Madine, Mohammad Moussa and Alzaabi, Hamad and Yaqoob, Ibrar and Salah, Khaled and Jayaraman, Raja},
  JOURNAL = {{IEEE} Access},
  VOLUME={8},
  PAGES={196813--196825},
  YEAR={2020}
}

@inproceedings{khatal2020fileshare,
  TITLE = {FileShare: A Blockchain and IPFS framework for Secure File Sharing and Data Provenance},
  AUTHOR = {Khatal, Shreya and Rane, Jayant and Patel, Dhiren and Patel, Pearl and Busnel, Yann},
  booktitle={MoSICom~'20: International Conference on Modelling, Simulation \& Intelligent Computing},
  DATE = {2020-01},
  LOCATION = {Dubai, United Arab Emirates}
}

@ARTICLE{wang2018blockchain,
  TITLE = {A Blockchain-Based Framework for Data Sharing With Fine-Grained Access Control in Decentralized Storage Systems},
  AUTHOR = {Wang, Shangping and Zhang, Yinglong and Zhang, Yaling},
  JOURNAL = {{IEEE} Access},
  VOLUME = {6},
  PAGES = {38437--38450},
  YEAR = {2018}
}

@article{prunster2020total,
  TITLE = {Total Eclipse of the Heart -- Disrupting the InterPlanetary File System},
  AUTHOR={Pr{\"u}nster, Bernd and Marsalek, Alexander and Zefferer, Thomas},
  YEAR={2020}
}

@INPROCEEDINGS{nyaletey2019blockipfs,
  TITLE = {BlockIPFS - Blockchain-Enabled Interplanetary File System for Forensic and Trusted Data Traceability},
  AUTHOR = {Nyaletey, Emmanuel and Parizi, Reza M and Zhang, Qi and Choo, Kim-Kwang Raymond},
  PAGES = {18--25},
  CROSSREF = {blockchain19}
}

@ARTICLE{politou2020delegated,
  TITLE = {Delegated content erasure in IPFS},
  author={Politou, Eugenia and Alepis, Efthimios and Patsakis, Constantinos and Casino, Fran and Alazab, Mamoun},
  JOURNAL = {Future Generation Computer Systems},
  VOLUME = {112},
  pages={956--964},
  year={2020}
}

@INPROCEEDINGS{hoang2020privacy,
  TITLE = {Privacy-Preserving Blockchain-Based Data Sharing Platform for Decentralized Storage Systems},
  AUTHOR = {Hoang, Van-Hoan and Lehtihet, Elyes and Ghamri-Doudane, Yacine},
  PAGES={280--288},
  CROSSREF = {networking2020}
}

@article{de2021accelerating,
  TITLE = {Accelerating Content Routing with Bitswap: A Multi-Path File Transfer Protocol in IPFS and Filecoin},
  AUTHOR = {De la Rocha, Alfonso and Dias, David and Psaras, Yiannis},
  YEAR = {2021},
  PAGES = {11}
}

@BOOKLET{tron2020swarmbook,
  TITLE = {The Book of Swarm},
  AUTHOR = {Tr\'{o}n, Viktor},
  HOWPUBLISHED = {online},
  DATE = {2020-06},
  NOTE = {v1.0 pre-release}
}

@TECHREPORT{tron2016swapswearswindle,
  TITLE = {Swap, Swear and Swindle incentive System for Swarm},
  AUTHOR = {Tr\'{o}n, Viktor and Fischer, Aron and Nagy, Daniel A. and Felf\"oldi, Zsolt and Johnson, Nick},
  INSTITUTION = {Ethereum Foundation},
  DATE = {2016-05}
}

@ONLINE{wood2014ethereum,
  AUTHOR = {Wood, Gavin},
  URL = {http://gavwood.com/Paper.pdf},
  DATE = {2014},
  TITLE = {Ethereum: A Secure Decentralised Generalised Transaction Ledger},
}

@TECHREPORT{ogden2018dat,
  TITLE = {Dat - Distributed Dataset Synchronization And Versioning},
  AUTHOR = {Ogden, Maxwell and McKelvey, Karissa and Buus Madsen, Mathias and {Code for Science}},
  INSTITUTION = {Dat Foundation},
  DATE = {2018-01}
}

@MISC{datproto,
  TITLE = {How Dat Works},
  AUTHOR = {Keall, Duncan},
  howpublished={\url{https://datprotocol.github.io/how-dat-works/}},
  note={Accessed: 2021-01}
}

@TECHREPORT{irvine2010self,
  TITLE = {Self-authentication},
  AUTHOR = {Irvine, David},
  DATE = {2010-09}
}

@TECHREPORT{irvine2010selfencrypt,
  TITLE = {Self encrypting data},
  AUTHOR = {Irvine, David},
  DATE = {2015-06}
}

@ARTICLE{lambert2014safe,
  TITLE = {The SAFE Network a New, Decentralised Internet},
  AUTHOR = {Lambert, Nick and Bollen, Benjamin},
  YEAR = {2014}
}

@INPROCEEDINGS{paul2014security,
  TITLE = {Security of the MaidSafe Vault Network},
  AUTHOR = {Paul, Greig and Hutchison, Fraser and Irvine, James},
  CROSSREF = {wwrf14}
}

@INPROCEEDINGS{jacob2015security,
  TITLE = {A Security Analysis of the Emerging P2P-Based Personal Cloud Platform MaidSafe},
  AUTHOR = {Jacob, Florian and Mittag, Jens and Hartenstein, Hannes},
  VOLUME = {1},
  PAGES = {1403--1410},
  CROSSREF = {trustcombigdataseispa15}
}

@MISC{safeprimer,
  TITLE={The SAFE Network Primer},
  AUTHOR = {MaidSafe},
  howpublished={\url{https://primer.safenetwork.org/}},
  note={Accessed: 2021-01}
}

@TECHREPORT{storjlabs2018storj,
  TITLE = {Storj: A Decentralized Cloud Storage Network Framework v3.0},
  AUTHOR = {{Storj Labs Inc.}},
  INSTITUTION = {Storj Labs, Inc.},
  DATE = {2018-10}
}

@INPROCEEDINGS{de2021exploring,
  TITLE = {Exploring the Storj Network: A Security Analysis},
  AUTHOR = {De Figueiredo, Sammy and Madhusudan, Akash and Reniers, Vincent and Nikova, Svetla and Preneel, Bart},
  PAGES = {257--264},
  CROSSREF = {sac21}
}

@ARTICLE{zhang2019frameup,
  TITLE = {Frameup: An Incriminatory Attack on Storj: A Peer to Peer Blockchain Enabled Distributed Storage System},
  AUTHOR = {Zhang, Xiaolu and Grannis, Justin and Baggili, Ibrahim and Beebe, Nicole Lang},
  JOURNAL = {Digital Investigation},
  VOLUME = {29},
  PAGES = {28--42},
  YEAR = {2019}
}

@INPROCEEDINGS{clarke2000freenet,
  TITLE = {Freenet: A Distributed Anonymous Information Storage and Retrieval System},
  AUTHOR = {Clarke, Ian and Sandberg, Oskar and Wiley, Brandon and Hong, Theodore W.},
  PAGES = {46--66},
  CROSSREF = {pet00}
}

@TECHREPORT{bennett2002gnunet,
  TITLE = {GNUnet - A truly anonymous networking infrastructure},
  AUTHOR = {Bennett, Krista and Grothoff, Christian and Horozov, Tzvetan and Patrascu, Ioana and Stef, Tiberiu},
  INSTITUTION = {In: Proc. Privacy Enhancing Technologies Workshop (PET},
  YEAR = {2002}
}

@inproceedings{saroiu2001measurement,
  AUTHOR = {Saroiu, Stefan and Gummadi, P. Krishna and Gribble, Steven D.},
  TITLE = {{Measurement study of peer-to-peer file sharing systems}},
  VOLUME = {4673},
  booktitle = {Multimedia Computing and Networking 2002},
  ORGANIZATION = {International Society for Optics and Photonics},
  PUBLISHER = {SPIE},
  DATE = {2001-12},
  PAGES = {156 -- 170},
}

@INPROCEEDINGS{jia2016oblivp2p,
  TITLE = {Oblivp2p: An Oblivious Peer-to-Peer Content Sharing System},
  AUTHOR = {Jia, Yaoqi and Moataz, Tarik and Tople, Shruti and Saxena, Prateek},
  PAGES = {945--962},
  CROSSREF = {usenixsecurity16}
}

@INPROCEEDINGS{qian2016garlic,
  TITLE = {Garlic Cast: Lightweight and Decentralized Anonymous Content Sharing},
  AUTHOR = {Qian, Chen and Shi, Junjie and Yu, Zihao and Yu, Ye and Zhong, Sheng},
  PAGES = {216--223},
  CROSSREF = {icpads16}
}

@ARTICLE{tschorsch2016bitcoin,
  TITLE = {Bitcoin and Beyond: A Technical Survey on Decentralized Digital Currencies},
  AUTHOR = {Tschorsch, Florian and Scheuermann, Bj{\"o}rn},
  JOURNAL = {IEEE Communications Surveys \& Tutorials},
  VOLUME = {18},
  NUMBER = {3},
  PAGES = {2084--2123},
  YEAR = {2016},
  PUBLISHER = {IEEE}
}

@TECHREPORT{arweave2019williams,
  TITLE = {Arweave: A Protocol for Economically Sustainable Information Permanence},
  AUTHOR = {Williams, Sam and Diordiiev, Viktor and Berman, Lev and Raybould, India and Uemlianin, Ivan},
  INSTITUTION = {arweave.org},
  DATE = {2019-11}
}

@TECHREPORT{tronfoundation2018tron,
  TITLE = {{TRON} Advanced Decentralized Blockchain Platform - Whitepaper Version 2.0},
  AUTHOR = {{{TRON} Foundation}},
  INSTITUTION = {{TRON} Foundation},
  DATE = {2018-12}
}

@TECHREPORT{vorick2014sia,
  TITLE = {Sia: Simple decentralized storage},
  AUTHOR = {Vorick, David and Champine, Luke},
  INSTITUTION = {Nebulous Inc.},
  DATE = {2014-11}
}

@INPROCEEDINGS{ali2016blockstack,
  TITLE = {Blockstack: A Global Naming and Storage System Secured by Blockchains},
  AUTHOR = {Ali, Muneeb and Nelson, Jude and Shea, Ryan and Freedman, Michael J},
  PAGES = {181--194},
  CROSSREF = {usenixatc16}
}

@INPROCEEDINGS{chen2017improved,
  TITLE = {An improved P2P File System Scheme based on {IPFS} and Blockchain},
  AUTHOR = {Chen, Yongle and Li, Hui and Li, Kejiao and Zhang, Jiyang},
  PAGES = {2652--2657},
  CROSSREF = {bigdata17}
}

@INPROCEEDINGS{fukumitsu2017proposal,
  TITLE={A Proposal of a Secure P2P-Type Storage Scheme by Using the Secret Sharing and the Blockchain},
  AUTHOR = {Fukumitsu, Masayuki and Hasegawa, Shingo and Iwazaki, Jun-ya and Sakai, Masao and Takahashi, Daiki},
  PAGES = {803--810},
  CROSSREF = {aina17}
}

@ARTICLE{ahlgren2012survey,
  TITLE = {A Survey of Information-Centric Networking},
  AUTHOR = {Ahlgren, Bengt and Dannewitz, Christian and Imbrenda, Claudio and Kutscher, Dirk and Ohlman, B{\"{o}}rje},
  JOURNAL = {{IEEE} Communications Magazine},
  VOLUME = {50},
  NUMBER = {7},
  PAGES = {26--36},
  YEAR = {2012}
}

@ARTICLE{tourani2017security,
  TITLE = {Security, Privacy, and Access Control in Information-Centric Networking: A Survey},
  author={Tourani, Reza and Misra, Satyajayant and Mick, Travis and Panwar, Gaurav},
  journal={IEEE communications Surveys \& Tutorials},
  volume={20},
  number={1},
  pages={566--600},
  year={2017},
  publisher={IEEE}
}

@INPROCEEDINGS{jacobson2009networking,
  TITLE = {Networking Named Content},
  AUTHOR = {Jacobson, Van and Smetters, Diana K. and Thornton, James D. and Plass, Michael F. and Briggs, Nicholas H. and Braynard, Rebecca L.},
  PAGES = {1--12},
  CROSSREF = {conext09}
}

@ARTICLE{zhang2014named,
  TITLE = {Named Data Networking},
  AUTHOR = {Zhang, Lixia and Afanasyev, Alexander and Burke, Jeffrey and Jacobson, Van and Crowley, Patrick and Papadopoulos, Christos and Wang, Lan and Zhang, Beichuan},
  JOURNAL = {Computer Communication Review},
  VOLUME = {44},
  NUMBER = {3},
  PAGES = {66--73},
  YEAR = {2014},
  PUBLISHER = {ACM}
}

@INPROCEEDINGS{mastorakis2017ntorrent,
  TITLE = {nTorrent: Peer-to-Peer File Sharing in Named Data Networking},
  AUTHOR = {Mastorakis, Spyridon and Afanasyev, Alexander and Yu, Yingdi and Zhang, Lixia},
  PAGES = {1--10},
  CROSSREF = {icccn17}
}

@MISC{rfc7927,
  AUTHOR = {Kutscher, D. and Eum, S. and Pentikousis, K. and Psaras, I. and Corujo, D. and Saucez, D. and Schmidt, T. and Waehlisch, M.},
  INSTITUTION = {RFC Editor},
  LOCATION = {Fremont, CA, USA},
  ORGANIZATION = {RFC Editor},
  PUBLISHER = {RFC Editor},
  URL = {https://www.rfc-editor.org/rfc/rfc7927.txt},
  DATE = {2016-06},
  DOI = {10.17487/RFC7927},
  HOWPUBLISHED = {RFC 7927 (Informational)},
  NUMBER = {7927},
  PAGES = {1--38},
  SERIES = {Internet Request for Comments},
  TITLE = {{Information-centric networking (ICN) research challenges}},
  TYPE = {RFC},
}

@MISC{rfc7945,
  AUTHOR = {Pentikousis, Kostas and Ohlman, Borje and Davies, Elwyn B and Spirou, Spiros and Boggia, Gennaro},
  INSTITUTION = {RFC Editor},
  LOCATION = {Fremont, CA, USA},
  ORGANIZATION = {RFC Editor},
  PUBLISHER = {RFC Editor},
  URL = {https://www.rfc-editor.org/rfc/rfc7945.txt},
  DATE = {2016-09},
  DOI = {10.17487/RFC7945},
  HOWPUBLISHED = {RFC 7945 (Informational)},
  NUMBER = {7945},
  PAGES = {1--38},
  SERIES = {Internet Request for Comments},
  TITLE = {Information-Centric Networking: Evaluation and Security Considerations},
  TYPE = {RFC},
}

@INPROCEEDINGS{rowstron2001pastry,
  TITLE = {Pastry: Scalable, Decentralized Object Location, and Routing for Large-Scale Peer-to-Peer Systems},
  AUTHOR = {Rowstron, Antony and Druschel, Peter},
  PAGES = {329--350},
  CROSSREF = {ifip01}
}

@inproceedings{stoica2001chord,
  TITLE = {Chord: A Scalable Peer-to-Peer lookup Service for Internet Applications},
  AUTHOR = {Stoica, Ion and Morris, Robert and Karger, David and Kaashoek, M. Frans and Balakrishnan, Hari},
  PAGES = {149--160},
  CROSSREF = {sigcomm01}
}

@INPROCEEDINGS{ratnasamy2001scalable,
  TITLE = {A Scalable Content-Addressable Network},
  AUTHOR = {Ratnasamy, Sylvia and Francis, Paul and Handley, Mark and Karp, Richard and Shenker, Scott},
  PAGES={161--172},
  CROSSREF={sigcomm01}
}

@INPROCEEDINGS{maymounkov2002kademlia,
  TITLE = {Kademlia: A Peer-to-Peer Information System Based on the {XOR} Metric},
  AUTHOR = {Maymounkov, Petar and Mazi\`{e}res, David},
  PAGES = {53--65},
  CROSSREF = {iptps02}
}

@misc{nakamoto2009bitcoin,
  author = {Nakamoto, Satoshi},
  title = {Bitcoin: A peer-to-peer electronic cash system},
  url = {http://www.bitcoin.org/bitcoin.pdf},
  year = 2009
}

@ARTICLE{kokoris2018calypso,
  TITLE={Calypso: Auditable sharing of private data over blockchains},
  AUTHOR={Kokoris-Kogias, Eleftherios and Alp, Enis Ceyhun and Siby, Sandra Deepthy and Gailly, Nicolas and Gasser, Linus and Jovanovic, Philipp and Syta, Ewa and Ford, Bryan},
  JOURNAL={Cryptology ePrint Archive, 2018/209, Tech. Rep.},
  YEAR={2018}
}

@INPROCEEDINGS{pfitzmann2000anon,
  AUTHOR = {Pfitzmann, Andreas and Köhntopp, Marit},
  CROSSREF = {pet00},
  DATE = {2000},
  PAGES = {1--9},
  TITLE = {Anonymity, Unobservability, and Pseudonymity - A Proposal for Terminology},
}

@article{narayanan17bitcoinpedigree,
  TITLE = {Bitcoin's Academic Pedigree},
  AUTHOR = {Narayanan, Arvind and Clark, Jeremy},
  JOURNAL = {{ACM} Queue},
  VOLUME = {15},
  NUMBER = {4},
  PAGES = {20},
  YEAR = {2017},
}

@INPROCEEDINGS{bonneau15challenges,
  TITLE = {SoK: Research Perspectives and Challenges for Bitcoin and Cryptocurrencies},
  AUTHOR = {Bonneau. Joseph and Miller, Andrew and Clark, Jeremy and Narayanan, Arvind and Kroll, Joshua A. and Felten, Edward W.},
  PAGES = {104--121},
  CROSSREF = {sp15}
}

@INPROCEEDINGS{gervais16blockchain-sec-perf,
  TITLE = {On the Security and Performance of Proof of Work Blockchains},
  AUTHOR = {Gervais, Arthur and Karame, Ghassan O. and W{\"{u}}st, Karl and Glykantzis, Vasileios and Ritzdorf, Hubert and Capkun, Srdjan},
  PAGES = {3--16},
  CROSSREF = {ccs16}
}

@article{reed1960polynomial,
  title={Polynomial codes over certain finite fields},
  author={Reed, Irving S and Solomon, Gustave},
  journal={Journal of the society for industrial and applied mathematics},
  volume={8},
  number={2},
  pages={300--304},
  year={1960},
  publisher={SIAM}
}

@MISC{zookotriangle,
  TITLE={Names: Distributed, Secure, Human-Readable: Choose Two},
  AUTHOR = {Zooko Wilcox-O'Hearn},
  howpublished={\url{https://web.archive.org/web/20011020191610/http://zooko.com/distnames.html}},
  note={Accessed: 2021-01}
}

@MISC{erc20token,
  TITLE={EIP-20 - ERC-20 Token Standard},
  AUTHOR = {Vogelsteller, Fabian and Buterin, Vitalik},
  howpublished={\url{https://github.com/ethereum/EIPs/blob/master/EIPS/eip-20.md}},
  note={Accessed: 2021-01}
}

@MISC{eip137,
  TITLE={EIP-137 - Ethereum Domain Name Service - Specification},
  AUTHOR = {Johnson, Nick},
  howpublished={\url{https://eips.ethereum.org/EIPS/eip-137}},
  note={Accessed: 2021-01}
}

@ONLINE{bip152,
  AUTHOR = {Corallo, Matt},
  URL = {https://github.com/bitcoin/bips/blob/master/bip-0152.mediawiki},
  DATE = {2016-04},
  TITLE = {BIP~152: Compact Block Relay},
  note = {Accessed: 2021-02}
}

@MISC{squarezooko,
  TITLE={Squaring the Triangle: Secure, Decentralized, Human-Readable Names},
  AUTHOR = {Aaron Swartz},
  howpublished={\url{http://www.aaronsw.com/weblog/squarezooko}},
  note={Accessed: 2021-01}
}

@ARTICLE{poon2016bitcoin,
  TITLE = {The bitcoin lightning network: Scalable off-chain instant payments},
  AUTHOR = {Poon, Joseph and Dryja, Thaddeus},
  DATE = {2016-01}
}

@INPROCEEDINGS{mccorry2016towards,
  TITLE = {Towards Bitcoin Payment Networks},
  AUTHOR = {McCorry, Patrick and M{ö}ser, Malte and Shahandashti, Siamak Fayyaz and Hao, Feng},
  CROSSREF = {acisp16},
  DATE = {2016},
  PAGES = {57--76}
}

@INPROCEEDINGS{dingledine2004tor,
  AUTHOR = {Dingledine, Roger and Mathewson, Nick and Syverson, Paul F.},
  CROSSREF = {usenixsecurity04},
  DATE = {2004},
  PAGES = {303--320},
  TITLE = {Tor: The Second-Generation Onion Router},
}

@MISC{dworkin2015sha,
  TITLE = {{SHA-3} Standard: Permutation-Based Hash and Extendable-Output Functions},
  AUTHOR = {Dworkin, Morris J},
  PUBLISHER = {Federal Inf. Process. Stds. (NIST FIPS), National Institute of Standards and Technology, Gaithersburg, MD},
  DATE = {2015-08}
}

@TECHREPORT{kirkpatrick2021open,
  TITLE = {Open Storage Network Retrospective and the Future of Distributed Storage for eInfrastructure},
  AUTHOR = {Kirkpatrick, Christine and Coakley, Kevin and Cragin, Melissa and Cramer, Catherine and McHenry, Kenton and Marini, Luigi and Kooper, Rob and Glasgow, Jim and Foster, Ian and Szalay, Alex and Simmel, Derek},
  INSTITUTION = {{The Open Storage Network}},
  DATE = {2021-08},
}

@MISC{web3foundation,
  TITLE = {About - Web3 Foundation},
  AUTHOR = {{Web3 Foundation}},
  howpublished={\url{https://web3.foundation/about/}},
  note={Accessed: 2021-09}
}

@MISC{gitipfs,
  TITLE = {{IPFS} - Github},
  AUTHOR = {{Protocol Labs}},
  PUBLISHER = {GitHub},
  JOURNAL = {GitHub repository},
  howpublished={\url{https://github.com/ipfs}},
  note={Accessed: 2021-01}
}

@MISC{gitlibp2p,
  TITLE = {libp2p - Github},
  AUTHOR = {{Protocol Labs}},
  PUBLISHER = {Github},
  JOURNAL = {GitHub repository},
  howpublished={\url{https://github.com/libp2p}},
  note={Acessed: 2021-07}
}

@MISC{gittestground,
  TITLE = {Testground - Github},
  AUTHOR = {{Testground team}},
  PUBLISHER = {GitHub},
  JOURNAL = {GitHub repository},
  howpublished={\url{https://github.com/testground/}},
  note={Accessed: 2021-09}
}

@MISC{gitswarm,
  TITLE = {Ethersphere - Github},
  AUTHOR = {Ethersphere},
  PUBLISHER = {GitHub},
  JOURNAL = {GitHub repository},
  howpublished={\url{https://github.com/ethersphere}},
  note={Accessed: 2021-01}
}

@MISC{gitdat,
  TITLE = {Hypercore Protocol - Github},
  AUTHOR = {{Hypercore Protocol developers}},
  PUBLISHER = {GitHub},
  JOURNAL = {GitHub repository},
  howpublished={\url{https://github.com/hypercore-protocol}},
  note={Accessed: 2021-01}
}

@MISC{gitsafe,
  TITLE = {SAFE Network - Github},
  AUTHOR = {MaidSafe},
  PUBLISHER = {GitHub},
  JOURNAL = {GitHub repository},
  howpublished={\url{https://github.com/safenetwork}},
  note={Accessed: 2021-01}
}

@MISC{gitstorj,
  TITLE = {Storj Labs - Github},
  AUTHOR = {{Storj Labs}},
  PUBLISHER = {GitHub},
  JOURNAL = {GitHub repository},
  howpublished={\url{https://github.com/Storj/}},
  note={Accessed: 2021-01}
}

@MISC{gitarweave,
  TITLE = {Arweave - Github},
  AUTHOR = {ArweaveTeam},
  PUBLISHER = {GitHub},
  JOURNAL = {GitHub repository},
  howpublished={\url{https://github.com/ArweaveTeam}},
  note={Accessed: 2021-01}
}

@PROCEEDINGS{sac21,
  LOCATION = {Gwangju, Korea},
  BOOKTITLE = {SAC~'21: Proceedings of the 36th {ACM/SIGAPP} Symposium On Applied Computing},
  DATE = {2021-03}
}

@PROCEEDINGS{ccnc21,
  LOCATION = {Las Vegas, NV, USA},
  BOOKTITLE = {CCNC~'21: Proceedings of the 18th {IEEE} Annual Consumer Communications {\&} Networking Conference},
  DATE = {2021-01}
}

@PROCEEDINGS{imc20,
  LOCATION = {Virtual Event, USA},
  BOOKTITLE = {IMC~'20: Proceedings of the {ACM} Internet Measurement Conference},
  DATE = {2020-10}
}

@PROCEEDINGS{networking2020,
  LOCATION = {Paris, France},
  BOOKTITLE = {Networking~'20: Proceedings of the 19th IFIP Networking Conference},
  DATE = {2020-06}
}

@PROCEEDINGS{icbc20,
  LOCATION = {Toronto, ON, Canada},
  BOOKTITLE = {ICBC~'20: Proceedings of the 2020 {IEEE} International Conference on Blockchain and Cryptocurrency},
  DATE = {2020-05}
}

@PROCEEDINGS{comsnets20,
  LOCATION = {Bengaluru, India},
  BOOKTITLE = {COMSNETS~'20: Proceedings of 2020 International Conference on COMmunication Systems {\&} NETworkS},
  DATE = {2020-01}
}

@PROCEEDINGS{icn2019,
  LOCATION = {Macao, SAR, China},
  BOOKTITLE = {ICN~'19: Proceedings of the 6th {ACM} Conference on Information-Centric Networking},
  DATE = {2019-09} 
}

@PROCEEDINGS{blockchain19,
  LOCATION = {Atlanta, GA, USA},
  BOOKTITLE = {BLOCKCHAIN~'19: Proceedings of the 2019 International Conference on Blockchain},
  DATE = {2019-07},
}

@PROCEEDINGS{iwqos19,
  LOCATION = {Phoenix, AZ, USA},
  BOOKTITLE = {IWQoS~'19: Proceedings of the International Symposium on Quality of Service},
  DATE = {2019-06}
}

@PROCEEDINGS{mobilecloud19,
  LOCATION = {Newark, CA, USA},
  BOOKTITLE = {MobileCloud~'19: Proceedings of the 7th {IEEE} International Conference on Mobile Cloud Computing, Services, and Engineering},
  DATE = {2019-04}
}

@PROCEEDINGS{icisdm19,
  LOCATION = {Houston, TX, USA},
  BOOKTITLE = {ICISDM~'19: Proceedings of the 3rd International Conference on Information System and Data Mining},
  DATE = {2019-04}
}

@PROCEEDINGS{icce19,
  LOCATION = {Las Vegas, NV, USA},
  BOOKTITLE = {ICCE~'19: Proceedings of the 37th {IEEE} International Conference on Consumer Electronics},
  DATE = {2019-01}
}

@PROCEEDINGS{hicss19,
  LOCATION = {Grand Wailea Maui, Hawaii, USA},
  BOOKTITLE = {HICCS~'19: Proceedings of the 52nd Hawaii International Conference on System Sciences},
  DATE = {2019-01}
}

@PROCEEDINGS{ithingsgreenvomcpscomsmartdata18,
  LOCATION = {Halifax, NS, Canada},
  BOOKTITLE = {iThings/GreenCom/CPSCom/SmartData~'18: Proceedings of the 2018 {IEEE} International Conference on Internet of Things and {IEEE} Green Computing and Communications and {IEEE} Cyber, Physical and Social Computing and {IEEE} Smart Data},
  DATE = {2018-08}
}

@PROCEEDINGS{iot17,
  LOCATION = {Linz, Austria},
  BOOKTITLE = {IOT~'17: Proceedings of the 7th International Conference on the Internet of Things},
  DATE = {2017-10}
}

@PROCEEDINGS{icccn17,
  LOCATION = {Vancouver, BC, Canada},
  booktitle={ICCCN~'17: Proceedings of the 26th International Conference on Computer Communication and Networks},
  DATE = {2017-07}
}

@PROCEEDINGS{icfec17,
  LOCATION = {Madrid, Spain},
  BOOKTITLE = {ICFEC~'17: Proceedings of the 2017 {IEEE} 1st International Conference on Fog and Edge Computing},
  DATE = {2017-05},
}

@PROCEEDINGS{bigdata17,
  LOCATION = {Honolulu, HI, USA},
  BOOKTITLE = {BigData Congress~'17: 2017 {IEEE} International Congress on Big Data},
  DATE = {2017-06}
}

@PROCEEDINGS{aina17,
  LOCATION = {Taipei, Taiwan},
  BOOKTITLE = {AINA~'17: Proceedings of the 31st {IEEE} International Conference on Advanced Information Networking and Applications},
  DATE = {2017-03}
}

@PROCEEDINGS{icpads16,
  LOCATION = {Wuhan, China},
  booktitle={ICPADS~'16: Proceedings of the 22nd {IEEE} International Conference on Parallel and Distributed Systems},
  DATE = {2016-12}
}

@PROCEEDINGS{ccs16,
  LOCATION = {Vienna, Austria},
  BOOKTITLE = {CCS~'16: Proceedings of the 23rd {ACM} {SIGSAC} Conference on Computer and Communications Security},
  DATE = {2016-10},
}

@PROCEEDINGS{usenixsecurity16,
  LOCATION = {Austin, TX, USA},
  BOOKTITLE = {USENIX Security~'16: Proceedings of the 25th {USENIX} Security Symposium},
  DATE = {2016-08}
}

@PROCEEDINGS{acisp16,
  LOCATION = {Melbourne, VIC, Australia},
  BOOKTITLE = {ACISP~'2016: Proceedings of the 21st Australasian Conference on Information Security and Privacy},
  DATE = {2016-07},
}

@PROCEEDINGS{jcdl16,
  LOCATION = {Newark, NJ, USA},
  BOOKTITLE = {JCDL~'16: Proceedings of the 16th {ACM/IEEE-CS} on Joint Conference on Digital Libraries},
  DATE = {2016-06},
}

@PROCEEDINGS{usenixatc16,
  LOCATION = {Denver, CO, USA},
  BOOKTITLE = {{USENIX} {ATC} '16: Proceedings of the 2016 {USENIX} Annual Technical Conference},
  DATE = {2016-06}
}

@PROCEEDINGS{trustcombigdataseispa15,
  LOCATION = {Helsinki, Finland},
  BOOKTITLE = {{IEEE} Trustcom/BigDataSE/ISPA~'15: Proceedings of the 2015 {IEEE} Trustcom/BigDataSE/ISPA},
  DATE = {2015-08}
}

@PROCEEDINGS{sp15,
  LOCATION = {San Jose, CA, USA},
  BOOKTITLE = {SP~'15: Proceedings of the 36th {IEEE} Symposium on Security and Privacy},
  DATE = {2015-05},
}

@PROCEEDINGS{wwrf14,
  LOCATION = {Morocco},
  BOOKTITLE = {WWRF~'14: Wireless World Research Forum Meeting 32},
  DATE = {2014-05}
}

@PROCEEDINGS{conext09,
  LOCATION = {Rome, Italy},
  BOOKTITLE = {CoNext~'09: Proceedings of the 5th {ACM} International Conference on Emerging Networking Experiments and Technologies},
  DATE = {2009-12}
}

@PROCEEDINGS{ncm08,
  LOCATION = {Gyeongju, South Korea},
  BOOKTITLE = {NCM~'08: Proceedings of the 4th International Conference on Networked Computing and Advanced Information Management},
  DATE = {2008-09}
}

@PROCEEDINGS{infocom06,
  LOCATION = {Barcelona, Catalunya, Spain},
  BOOKTITLE = {INFOCOM~'06: Proceedings of the 25th {IEEE} International Conference on Computer Communications},
  DATE = {}
}

@PROCEEDINGS{itcc05,
  LOCATION = {Las Vegas, NV, USA},
  BOOKTITLE = {ITCC~'05: Proceedings of the 2005 International Conference on Information Technology: Coding and Computing},
  DATE = {2005-04}
}

@PROCEEDINGS{iptps05,
  LOCATION = {Ithaca, NY, USA},
  BOOKTITLE = {IPTPS~'05: Proceedings of the 4th International Workshop on Peer-To-Peer Systems},
  DATE = {2005-02}
}

@PROCEEDINGS{usenixsecurity04,
  LOCATION = {San Diego, CA, USA},
  BOOKTITLE = {USENIX Security~'04: Proceedings of the 13th USENIX Security Symposium},
  DATE = {2004-08},
}

@PROCEEDINGS{p2pecon03,
  LOCATION = {Berkeley, CA, USA},
  BOOKTITLE = {P2PEcon~'03: Proceedings of the 1st Workshop on Economics of Peer-to-Peer Systems},
  DATE = {2003-06},
}

@PROCEEDINGS{iptps02,
  LOCATION = {Cambridge, MA, USA},
  BOOKTITLE = {IPTPS'02: Proceedings of the 1st International Workshop on Peer-to-Peer Systems},
  DATE = {2002-03}
}

@PROCEEDINGS{ifip01,
  LOCATION = {Heidelberg, Germany},
  BOOKTITLE = {Middleware~'01: Proceedings of the 2001 {IFIP}/{ACM} International Conference on Distributed Systems Platforms},
  DATE = {2001-11}
}

@PROCEEDINGS{sigcomm01,
  LOCATION = {San Diego, CA, USA},
  BOOKTITLE = {SIGCOMM'01: Proceedings of the 2001 {ACM} Conference on Applications, Technologies, Architectures, and Protocols for Computer Communication},
  DATE = {2001-08}
}

@PROCEEDINGS{pet00,
  LOCATION = {Berkeley, CA, USA},
  BOOKTITLE = {PET~'00: Proceedings of the International Workshop on Designing Privacy Enhancing Technologies: Design Issues in Anonymity and Unobservability},
  DATE = {2000-07},
}

\end{document}